\documentclass[aps,prb,twocolumn,superscriptaddress,longbibliography]{revtex4-2}

\usepackage{amsmath,amssymb,amsfonts,bm,graphicx,mathtools,braket}

\usepackage{graphicx}
\usepackage{placeins}
\usepackage{dcolumn}
\usepackage{bm}
\usepackage{hyperref}
\hypersetup{colorlinks=true,linkcolor = blue,
            urlcolor  = blue,
            citecolor = blue,
            anchorcolor = blue, linktocpage}

\usepackage{subfigure}
\usepackage{times}{\Large }

\newcommand{\revise}[1]{\textcolor{black}{ {#1}}\color{black}} 

\newcommand{\equa}[1]{Eq.~\eqref{#1}} 
\newcommand{\fig}[1]{Fig.~\ref{#1}}

\begin{document}

\title{Fate of transient order parameter domain walls in ultrafast experiments}
\author{Lingxian Kong} 
\affiliation{International Center for Quantum Materials, School of Physics, Peking University, Beijing 100871, China}

\author{Ryuichi Shindou}
\affiliation{International Center for Quantum Materials, School of Physics, Peking University, Beijing 100871, China}%

\author{Zhiyuan Sun}%
\email{Corresponding author. zysun@tsinghua.edu.cn}
\affiliation{State Key Laboratory of Low-Dimensional Quantum Physics and Department of Physics, Tsinghua University, Beijing 100084, China}%

\begin{abstract}
In ultrafast experiments, an optical pump pulse often generates transient domain walls of the order parameter in materials with spontaneous symmetry breaking, due to either a finite penetration depth of light on a three-dimensional (3D) material, or a finite spot size on a two-dimensional (2D) material. We show that the domain wall decays due to unstable order parameter fluctuations. We study a generic system with $U(1)$-symmetric order, and those with an additional weak $Z_2$ ($U(1)$-symmetry-breaking) term, representing the charge-density-wave (CDW) orders in recent experiments. During the first stage of the decay dynamics, exponentially growing thermal fluctuations convert the domain wall into an interface with randomly distributed topological defects. In the second stage, the topological defects undergo a coarsening dynamics within the interface. For a 2D interface in a 3D system, the coarsening dynamics leads to a diffusive growth of the correlation length. For a one-dimensional (1D) interface in a 2D system with the weak $Z_2$ term, the correlation-length growth shows a crossover from diffusive to sub-diffusive behavior. 
\end{abstract}

\maketitle
\section*{Introduction}
    Transient manipulation of order parameters by an ultrafast laser pulse has been one of the major topics in condensed matter 
physics
~\cite{RevModPhys.82.2731, RevModPhys.93.041002, Fausti2011, Mitrano:2016aa,Sentef2017, Kennes2017, PhysRevLett.119.167203, vogelgesang2018phase, Zhang2018a, Cremin2019, Zong2019, doi:10.1021/acs.nanolett.9b01865, Claassen:2019aa, demsar2020non, Dolgirev.2020,Dolgirev.2020_amplitude, 
Sun.2020_Metastable, sun2020topological, Sun.2021_Josephson, Gao.2022,  bao2022light, PhysRevB.106.L121109, PhysRevLett.131.196401, Sun.2023_ponderomotive, Cheng2024, zhou2024transient}. 
Recent 
experiments~\cite{yusupov2010coherent,duan2021optical,PhysRevB.103.054109,PhysRevLett.113.026401, duan2025identification} reported that in several CDW materials, 
a laser pulse (pump) created domain walls of the CDW order parameter. 
There the pump pulse 
heats the surface of the system so that the CDW order 
parameter $\psi$ in the surface region is suppressed 
in time from its initial equilibrium value $\psi_{\text{eq}}$ 
to zero, following a classical equation of motion. 
From the principle of inertia, 
$\psi$ may cross zero, and stabilize in the energy minimum at $-\psi_{\text{eq}}$ after the system is cooled down~\cite{yusupov2010coherent,duan2021optical,PhysRevB.103.054109, PhysRevLett.113.026401}, getting `inverted'. 
Due to the finite penetration depth of the pump pulse, 
only the order parameter in the surface region gets inverted, 
yielding a domain wall (the white plane in the left 
panel of Fig.~\ref{punchline}) between the
surface region and the bulk at time $t=t_0$, see the Supplementary Note 1. 
Within a 
mean-field theory of 
the time-dependent Ginzburg-Landau (TDGL) dynamics 
that neglects 
fluctuations, the domain wall undergoes no further evolution~\cite{yusupov2010coherent,duan2021optical, PhysRevLett.113.026401, PhysRevB.103.054109}.

In this work, we propose that due to fluctuations, the domain wall will decay following a two-stage 
dynamics. In the first stage, the domain wall decays by 
the exponential growth of 
unstable fluctuations~\cite{bray2002theory, Sun.2020_Metastable} 
of $\psi$ on the wall, which happens both in the $U(1)$ symmetric case,  
and in the system with a weak $Z_2$ term. 
The dynamics are characterized by increasing magnitude and correlation length of the fluctuations.
At the end of the first stage, around a timescale $t\simeq t_{\rm c}$, the wall is transformed into an interface containing randomly distributed topological defects, 
as shown by the vortex strings in Fig.~\ref{punchline} for a 3D system. In 2D systems, the interface is 1D and the defects are vortices and antivortices. 


In the second stage, the motion and annihilation of topological defects on the interface dominate the dynamics. It happens under driving forces originating from the tension of the vortex strings in 3D systems and the attraction between vortices and antivortices in 2D systems ~\cite{PhysRevE.47.1525, PhysRevA.46.7765,PhysRevA.42.5865,10.1143/PTP.78.237, PhysRevA.45.657}. 
Universal features of the second-stage dynamics are characterized by 
a mean distance $L(t)$ between the defects on the interface. 
The system undergoes a self-similar evolution with a global 
increase of $L(t)$, 
leading to a coarsening phenomenon~\cite{bray2002theory, PhysRevA.46.7765, PhysRevE.47.1525, PhysRevE.47.1525, 10.1143/PTP.78.237}, as illustrated in Fig.~\ref{punchline}. We show that in 3D systems, the coarsening dynamics exhibits a diffusive behavior in both the $U(1)$-symmetric case and the case with the additional weak $Z_2$ term, i.e., the length scale increases as $L(t) \propto t^{1/2}$. In 2D systems with 
the weak $Z_2$ term, $L(t) \propto t^{p}$ exhibits a crossover from $p\approx 1/2$ to $p\approx 0$ as $L$ gets larger, namely a diffusive-to-subdiffusive crossover. The length scale $L(t)$ can be measured from the structure factor in time-resolved X-ray diffraction experiments~\cite{orenstein2023subdiffusive, PhysRevB.99.104111}. 

\begin{figure}[h]
	\centering
	\includegraphics[width= 3.2in]{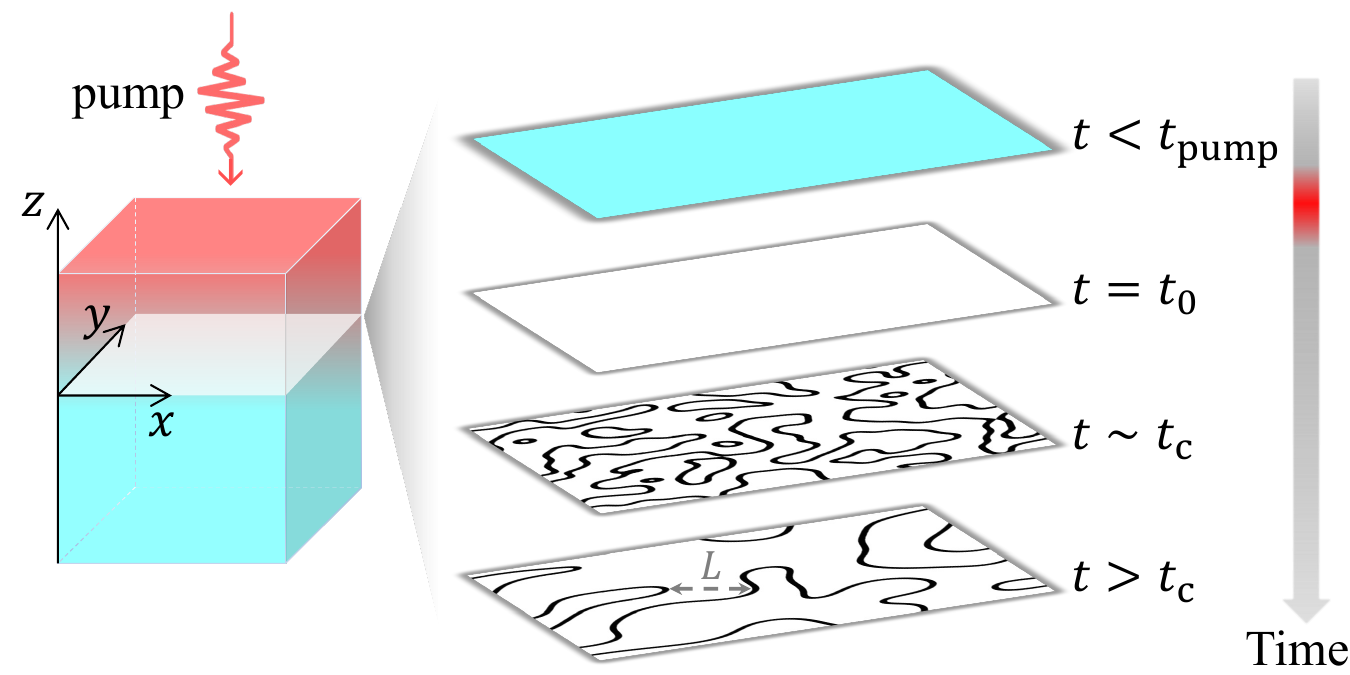} 
	\caption {{\bf Illustration of the temporal evolution of a 3D sample.} At $t=t_{\text{pump}}$, the surface region is pumped by an ultrafast pulse represented by the red segment on the time axis. An order parameter domain wall is created at $t=t_0$ on the white interface. The four planes on the right show the order parameter configuration of the interface at different times. The black curves denote the vortex strings. 
	}
	\label{punchline}
\end{figure}

\section*{Results}
\subsection*{$U(1)$-symmetric model}
We consider the ``model A" dynamics~\cite{Hohenberg1977, Kopnin}, whose mean-field version is also called the `TDGL equation', together with a $U(1)$-invariant free energy: 
\begin{align}
    &\frac{1}{\gamma} \partial_t \psi = -\frac{1}{E_{\text{c}}} \frac{\delta F}{\delta \psi^*} + \eta  \label{model_A} 
    \,, 
    \\
    & 
   F = E_{\text{c}} \int f\ \ d^{D}  \bm{r} 
     \,, \quad 
     f = 
        \left(|\psi|^2 + \alpha \right)^2 +  \xi_0 ^2  |\nabla \psi|^2
    \label{eqn:F_U1}
    \,.
\end{align} 
Here $\psi(\bm{r},t)=\psi_1+i\psi_2$ is the dimensionless complex order parameter, $\gamma$ is the relaxation rate, $E_{\text{c}}$ is the condensation energy density, $D$ is the space dimension,
$\alpha \sim (T-T_{\text{c}})/T_{\text{c}}$ is negative at a temperature $T$ below the critical temperature $T_\text{c}$, and $\xi_0$ is the bare coherence length.  The thermal noise field $\eta = \eta_1 + i\eta_2$ arises from a 
thermal bath made of the other degrees of freedom, 
including 
high-energy electrons and phonons. 
We characterize the noise by a correlator $\langle \eta_{i}(\bm{r},t) \eta_{i^{\prime}}(\bm{r^{\prime}},t^{\prime}) \rangle = 2T(\gamma E_{\text{c}})^{-1} \delta_{ii^{\prime}}\delta(\bm{r}-\bm{r^{\prime}})\delta(t-t^{\prime})$ with 
an effective temperature $T$ of the bath and $i/i^{\prime} = 1,2$. 
Here the Ginzburg parameter $\zeta = |\alpha|^{D/2 - 2}T/(E_{\text{c}}\xi_0^D )$
plays the role of a 
dimensionless measure  
of the fluctuations in the non-equilibrium process as well as 
the equilibrium thermal fluctuations
~\cite{Sun.2020_Metastable, Mitra.2018}. 
Eqs.~\eqref{model_A}\eqref{eqn:F_U1} are reasonable approximations to the dynamics of 
the mean-field order parameter and its fluctuations at temperatures close to $T_{\text{c}}$~\cite{Hohenberg1977, Mitra.2018} in 
incommensurate CDW~\cite{Dolgirev.2020, Dolgirev.2020_amplitude}, spin density waves (SDW)~\cite{RevModPhys.66.1}, excitonic insulators (EI), and 
Bardeen–Cooper–Schrieffer (BCS) superconductors~\cite{Kopnin, Mitra.2017, Mitra.2018, Sun.2020_Metastable}. \revise{A microscopic derivation of the free energy model from an electron-phonon coupling Hamiltonian is presented in Supplementary Note 7, which shows $E_{\rm c} \sim \nu T_{\rm c}^2$ with $\nu \sim p_{\rm F}^D/E_{\rm F} $ being the density of states of electrons contributing to the Fermi surface nesting. Here $p_{\rm F}$ and $E_{\rm F}$ denote the Fermi momentum and Fermi energy of the metallic state without the CDW formation. Note that the CDW often has different coherence lengths in different directions while Eq.~(\ref{eqn:F_U1}) is written in a rescaled spatial coordinate so that it appears isotropic. Around $T_{\rm c}$, one has $\xi_0  \sim v_{\rm F}/T_{\rm c} $ where $v_{\rm F} = E_{\rm F}/p_{\rm F}$ denotes the Fermi velocity~\cite{PhysRevLett.31.462, Gruner.1988_RMP}. This leads to $\zeta \sim T_{\rm c}^{D-1}/E_{\rm F}^{D-1}|\alpha|^{D/2-2}$~\cite{ Sun.2020_Metastable} which is similar to SDW, EI, and BCS superconductors~\cite{RevModPhys.66.1, nagaosa2013quantum,  PhysRev.158.462, kozlov1965metal}.
}

\begin{figure}[h]
	\centering
	\includegraphics[width= 3.3in]{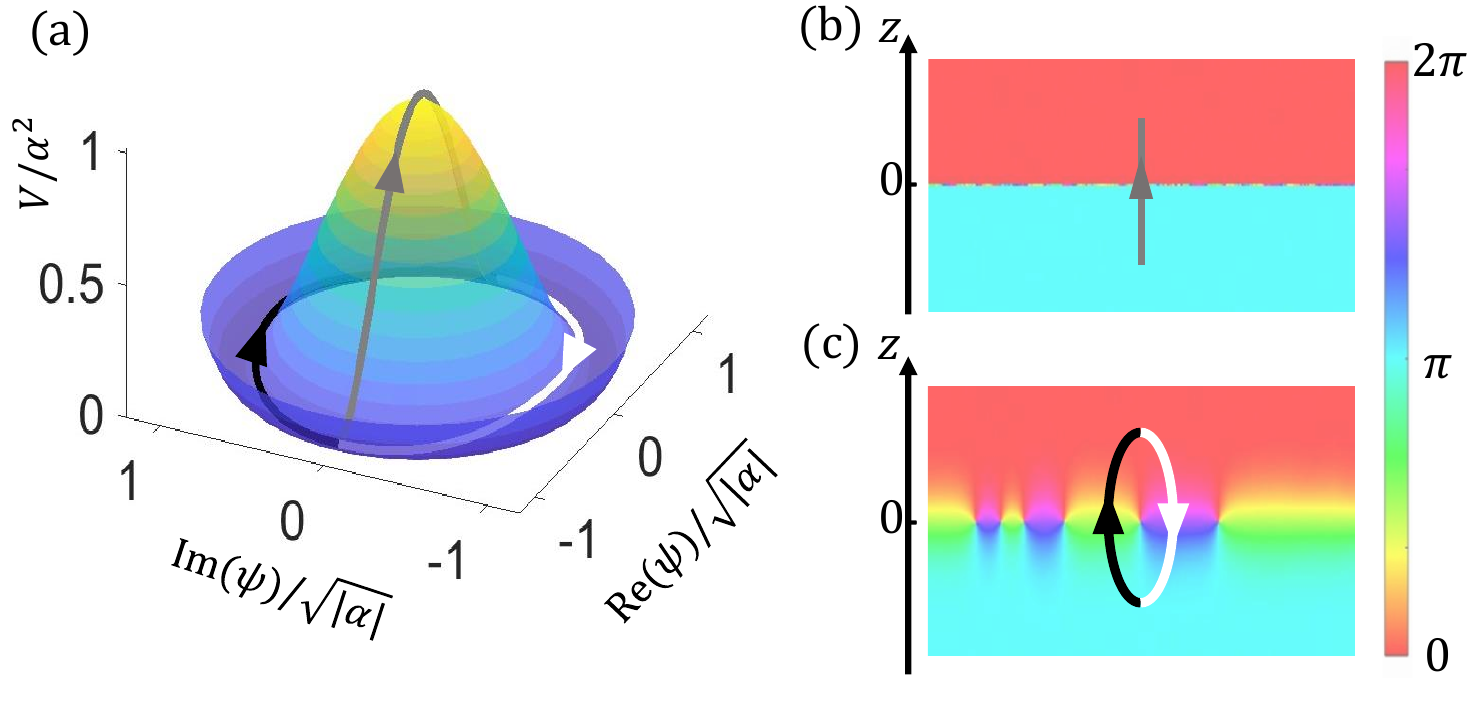} 
	\caption{
		{\bf Illustration of the decay of the domain wall.} (a) The `Mexican hat'  free energy landscape (the potential part) of Eq.~(\ref{eqn:F_U1}) plotted on the complex plane of $\psi$. 
		(b,c) The phase of $\psi$ plotted on the $y-z$ plane at (b) $t=t_0$ and (c) $t\geq t_{\text{c}}$, showing the domain wall at $z=0$.
		The $\psi$ and free energy along the gray, black, and white lines are also plotted in (a).
		In the initial domain-wall configuration,
		$\psi = \psi_0(z)$ changes from $-\sqrt{|\alpha|}$ to $\sqrt{|\alpha|}$ crossing the potential peak along the gray line. To lower the energy, $\psi$ fluctuates in the imaginary direction.
		After $t_{\text{c}}$,   random phase domains appear on the interface.   The loop denoted by black and white lines in (c) encloses a vortex with a phase winding $2\pi$. In a 2D interface of a 3D system, the interface becomes 
		full of vortex-strings as 
		shown in Fig.~\ref{punchline}. }
	\label{stageI}
\end{figure}

We first consider the 2D domain wall in the 3D systems
($D=3$), where 
the domain wall in Fig.~\ref{punchline} is set as the $z=0$ plane. 
An order parameter configuration 
of the domain wall satisfies the saddle-point condition $\delta F/\delta \psi = 0$. Without loss of generality, we choose the real solution $\psi_0(\bm{r}) = \sqrt{|\alpha|}\tanh(z/\xi)$ as shown by the gray line in Fig.~\ref{stageI}(a), 
where $\xi = \xi_0/\sqrt{|\alpha|}$ is the coherence length. 

To study the first stage of dynamics, we set $t_0 = 0$ and write  $\psi(\bm{r},t) = \psi_{0}(\bm{r}) + \delta\psi(\bm{r},t) $ where $\delta \psi = \phi_1 + i\phi_2$ is the fluctuation field with the real part 
$\phi_1$ and imaginary part $\phi_2$.
The linearized dynamics of $\delta \psi$ is expanded from Eq.~(\ref{model_A}) as
\begin{align}
     \frac{1}{\gamma} \partial_t \phi_{a} &
     = \left(-\hat{L}_{a} + \xi_0 ^2 \nabla_{\bm{R}}^2
     \right) \phi_{a} 
     + \eta_{a}
    \label{linear_dynamics}
\end{align}
where $a = 1,2$, $\hat{L}_{1} = -\xi_0 ^2 \partial_z^2 + 2\alpha + 6\psi_0^2  $, $\hat{L}_{2} = -\xi_0 ^2 \partial_z^2 + 2\alpha + 2\psi_0^2 $, and $\nabla_{\bm{R}}^2 = \partial_x^2 + \partial_y^2$ is the Laplacian in $D-1$ dimensions. The eigen-function of $\hat{L}_{a} - \xi_0 ^2 \nabla_{\bm{R}}^2$ can be written as $\phi_{a}^{\bm{K}}(\bm{r}) = u_{a}(z)e^{i\bm{K}\cdot\bm{R}}$ with eigenvalue $\epsilon_{a} + \xi_0^2 K^2$, where $u_{a}(z)$ is the eigen-function of $\hat{L}_{a}$ with eigenvalue $\epsilon_{a}$. Here $\bm{R}$($\bm{K}$) denotes the spatial coordinate (momentum) in the $(D-1)$-dimensional plane normal to $z$. Eigenvalues of $\hat{L}_1$ are all non-negative, while $\hat{L}_2$ has 
one negative eigenvalue, 
$\epsilon_2 = \alpha$, corresponding to a 
`bound-state' eigen-function $u_2(z) = \text{sech}(z/\xi)$, see the Supplementary Note 2. It
implies that the long-wavelength fluctuations in the imaginary part
direction, written as $\phi_{2}^{\bm{K}}(\bm{r}) \propto \text{sech}(z/\xi)e^{i\bm{K}\cdot\bm{R}}$ with $K<\sqrt{|\alpha|}/\xi_0$, will grow exponentially in time with the exponent $\left(|\alpha|-\xi_0^2 K^2\right) \gamma t$~\cite{Sun.2020_Metastable}. 

With $\phi_2 (\bm{r}) = \varphi(\bm{R},t)\text{sech}(z/\xi)$, 
the correlation function $C(\bm{R},t) = \langle \varphi(\bm{0},t)\varphi(\bm{R},t) \rangle$ 
depicts the real space structure of the fluctuation within 
the interface. From Eq.~(\ref{linear_dynamics}), it has the following analytical expression for $\gamma t \gg 1/|\alpha|$ (see the Supplementary Note 2):
\begin{align}
    C\left(\bm{R},t \right) \approx  \zeta|\alpha|  \frac{e^{2|\alpha|\gamma t}}{\sqrt{8\pi|\alpha|\gamma t}} 
    I_0\left(\frac{R^2}{16\xi_0^2 \gamma t}\right) e^{-\frac{R^2}{16\xi_0^2 \gamma t}}
    \label{grt}
\end{align}
with the modified Bessel function $I_0$ of the first kind. 
Therefore, 
the fluctuation magnitude grows exponentially with the correlation length growing as $L(t)= 4\xi_0 \sqrt{\gamma t}$, a diffusive behavior.
Note that due to the initial fluctuations $\langle \phi_{2}(\bm{r}) \phi_{2}(\bm{r^{\prime}}) \rangle \sim 1/|\bm{r} - \bm{r^{\prime}}|$ at long distances, the spatial correlation \revise{crosses over } from an exponential decay at $R\ll L(t)$ to a power law decay $C\left(\bm{R},t \right) \propto L(t)/R$ at $R\gg L(t)$.
The first-stage dynamics ends at a crossover time $t_{\text{c}}$ when the fluctuation becomes so large that the linear approximation in \equa{linear_dynamics} fails, and
the $|\psi|^4$ term in \equa{eqn:F_U1} starts to stop its growth. One may estimate this timescale by $\langle |\psi(\bm{r},t_{\text{c}})|^2 \rangle \sim |\alpha|$, which gives $C(\bm{0},t_{\text{c}}) \sim |\alpha|$ and  $\gamma t_{\text{c}} \sim (2 |\alpha|)^{-1} \ln(\zeta^{-1}) $. Note that $\gamma t_{\text{c}}$ thus obtained is much larger than 
$1/|\alpha|$ given that $\zeta$ is small enough.

At $t\simeq t_{\text{c}}$, %
the interface 
consists of randomly distributed positive-$\phi_2$ and negative-$\phi_2$ domains with the
typical domain size of  
$4\xi_0 \sqrt{\gamma t_{\text{c}}}$~\cite{Sun.2020_Metastable}.
Boundaries between the domains quickly relax into stable topological defects with the phase winding $\pm 2\pi$, as illustrated in \fig{stageI}(c), meaning that 
the interface becomes full of randomly distributed 2D vortex strings 
as in Fig.~\ref{punchline}.

At $t \gtrsim t_{\text{c}}$, the second stage of dynamics takes over, which has been argued to be 
a coarsening dynamics~\cite{bray2002theory} of the vortex strings. Specifically, distributions of the vortex strings at different times show a statistical self-similarity, which is 
characterized by a growing correlation length $L(t)$, or the
typical inter-string distance.   
The coarsening dynamics with $L(t)\propto t^p$ and 
$p>1/2$, $p=1/2$, and $p<1/2$ are referred to as 
superdiffusive, diffusive, and sub-diffusive dynamics, respectively. Eq.~(\ref{model_A}) implies the equation of motion for the length scale:
\begin{align}
    \lambda(L) \frac{dL}{dt} = F_{\text{d}} (L)
    \ ,
    \label{estimate}
\end{align}
see Refs.~\cite{bray2002theory, PhysRevA.46.7765, PhysRevE.47.1525, PhysRevE.47.1525, 10.1143/PTP.78.237} and the Supplementary Note 3.
Here $\lambda(L) \frac{dL}{dt}$ and $F_{\text{d}} (L)$ are the frictional force with a friction coefficient $\lambda(L)$ and the driving force acting on a unit length of the vortex string, respectively. \equa{estimate} is nothing but a balance equation between the two forces.
Consider a circular vortex-string loop with the diameter $L$, a typical topological defect, its energy
is $E\sim |\alpha|E_{\text{c}} \xi_0^2 L\ln(L/\xi)$~\cite{bray2002theory}.
Therefore, the driving force that contracts the loop is $F_{\text{d}}(L) \sim L^{-1}dE/dL \sim  |\alpha| E_{\text{c}} \xi_0^2 L^{-1}\ln(L/\xi) $ in the large $L$ limit. The friction coefficient $\lambda(L) \sim \gamma^{-1}E_{\text{c}} |\alpha| \ln (L/\xi)$ is derived from the damping term~\cite{bray2002theory}, i.e. first-order time derivative term in \equa{model_A}. 
Combining these two, we obtain a solution of 
Eq.~(\ref{estimate}) in the large-$L$ limit as 
$L(t) \sim \xi_0 \sqrt{\gamma t}$, which manifests 
a diffusive coarsening dynamics.

\begin{figure}[h]
	\centering
	\includegraphics[width= 3.4in]{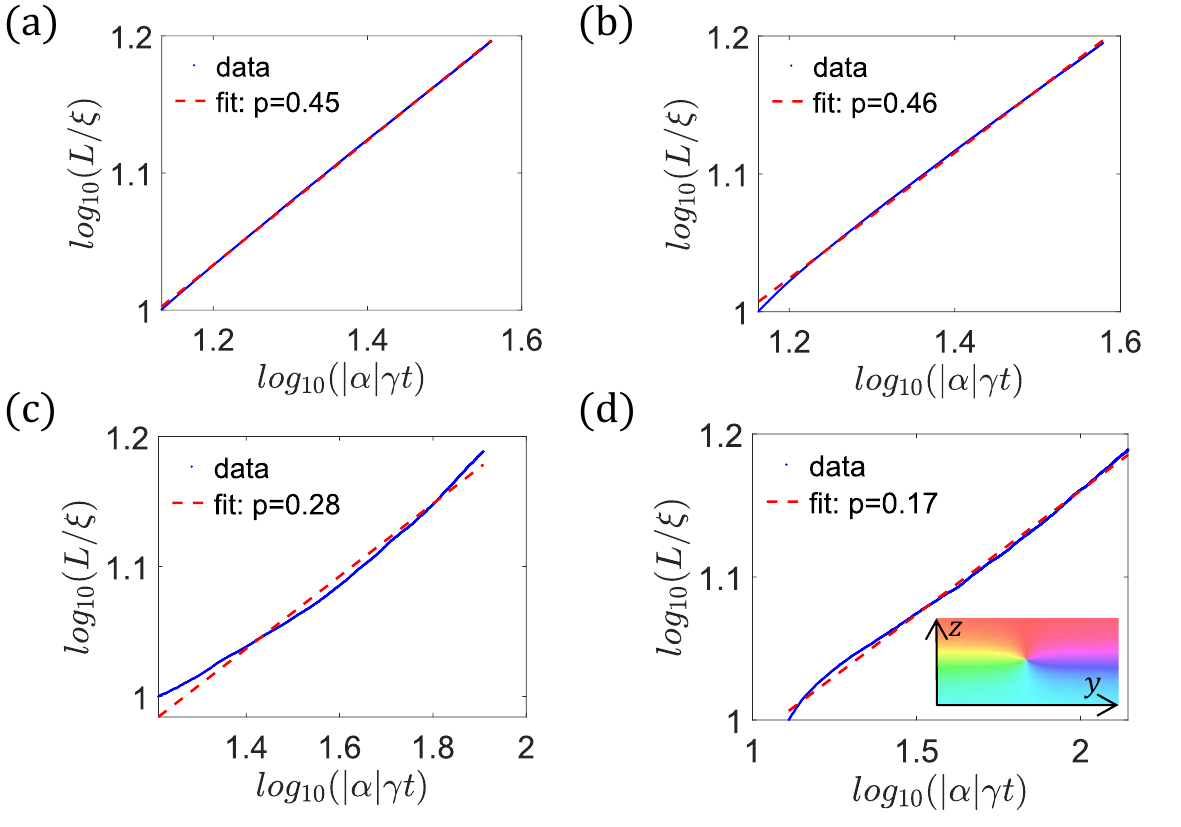} 
	\caption {{\bf Numerically computed $\log_{10}\left(L/\xi \right)$ as a function of $\log_{10} (|\alpha|\gamma t)$  for $2\zeta = 0.0001$, and the linear fitting.} In the numerical simulation, space and time are discretized by step sizes $0.5 \xi$ and $0.03 (|\alpha|\gamma)^{-1} $. Data points are selected within $\log_{10} (L/\xi) \in (1,1.2)$. \revise{The results are statistical averages over ten independent simulations. } (a) The 3D $U(1)$ symmetric case.   (b) The 3D $U(1)$ breaking case with $\beta/|\alpha| = 0.1$. Both (a) and (b) are simulated in a $L_x \times L_y \times L_z$ grid with $L_z /\xi = 60 $, $L_x /\xi = L_y / \xi = 1280$.    
		(c)(d) The 2D case with the $Z_2$ term $\beta/|\alpha|$ being $0.001$ (c)  and $0.1$ (d).  The phase distribution of a topological defect is sketched in the inset of (d) with the same colorbar as Fig.~\ref{stageI}(b,c).
		Both (c) and (d) are simulated in a $ L_y \times L_z$ grid with $L_z /\xi = 120 $, $L_y /\xi = 38400$.   } 
	\label{data}
\end{figure}

We numerically simulated Eq.~(\ref{model_A}) with the domain-wall initial condition and observed the coarsening dynamics. To evaluate the length scale $L(t)$ in the simulation, we count a 
number $N$ of vortices in the $y$-$z$ planes for each discrete $x$, 
average $N$ over \revise{each $x$ in ten independent simulations}, and define $L = L_y/\langle N\rangle$, where $L_y$ is the system size along $y$. Fig.~\ref{data}(a) shows $L$ versus time by a log-log plot in the interval of $\log_{10} (L/\xi) \in (1,1.2)$, from which we estimate the exponent $p$ by a linear fit $\log_{10} \left(L/\xi \right)=p \log_{10} (|\alpha|\gamma t) + c_0$. The estimate gives $p=0.45$ for the 2D interface of a 3D system.
The discrepancy from the theoretical result may be because the simulation has not entered the long-time region. 

\vspace{0.5cm}
\subsection*{Effects of a $Z_2$ term}
To model the case without the $U(1)$ symmetry, we add a term $2\beta \psi_2^2$ (`$Z_2$ term') with $\beta > 0$ to the free energy density in Eq.~(\ref{eqn:F_U1}), 
reducing the $U(1)$ symmetry to a $Z_2 \otimes Z_2$ symmetry under the transformation $\psi_{1(2)} \rightarrow \pm\psi_{1(2)}$. 
Due to the $Z_2$ term,  the low-temperature equilibrium state is long-range ordered even for a 2D system so that a well-defined 1D domain wall could be generated.
This model describes commensurate CDW/SDW~\cite{Bak:1982aa, Gruner.1988_RMP} and certain excitonic insulators~\cite{Sun.2021_Josephson}, etc. As shown below, the decay process of the domain wall for the small $\beta$ case is characterized by a similar 
two-stage dynamics as the $U(1)$ symmetric case, while the scaling 
behavior may be different. 

For the first-stage dynamics, the $Z_2$ term modifies $\hat{L}_{2}$ in Eq.~(\ref{linear_dynamics}),  $\hat{L}_{2} = -\xi_0 ^2 \partial_z^2 + 2(\alpha + \beta) +  2\psi_0^2$, so that the lowest eigenvalue of $\hat{L}_2$ 
becomes $\alpha+2\beta$. For $\beta > |\alpha|/2$, 
the eigenvalue is positive, and the domain wall is stable against any fluctuation of the imaginary part of $\psi$. For $\beta < |\alpha|/2 $, the mode $\phi_{2}^{\bm{K}}(\bm{r}) \propto \text{sech}(z/\xi)e^{i\bm{K}\cdot\bm{R}}$ with $K<\sqrt{|\alpha|-2\beta}/\xi_0$ 
grows exponentially in time. The growth is characterized by the correlation function $C(\bm{R},t)$. The correlation function in the long-time regime with $\gamma t \gg 1/\beta$ and $\gamma t \gg 1/(|\alpha|-2\beta)$ is computed from 
Eq.~(\ref{linear_dynamics}) as
\begin{align}
    C(\bm{R},t) \approx c_{1} \zeta|\alpha|  \frac{e^{2\left(|\alpha|-2\beta \right)\gamma t}}{\sqrt{8\pi|\alpha|\gamma t}^{D-1}}  e^{-\frac{R^2}{8\xi_0^2 \gamma t}}
    \label{g_beta}
\end{align}
where $c_{1} \approx \sqrt{|\alpha|/\beta} $ if $\beta \ll |\alpha|$. Similar to \equa{grt}, one has a growing correlation length $L(t) \sim \xi_0 \sqrt{8\gamma t}$. The crossover time to the second-stage dynamics is $\gamma t_{\text{c}} \sim (2|\alpha|)^{-1} \ln(\zeta^{-1})$ for $\beta \ll |\alpha|$, given that $\ln(\zeta^{-1}) \gg |\alpha|/\beta$ such that $t_\text{c}$ is in the long-time regime. 
\revise{Note that although the domain wall is locally stable for $\beta > |\alpha|/2$ in our calculation, it will finally decay in pump-probe experiments from other effects such as the nonzero curvature of the domain wall due to the finite spot size of lasers, see the Supplementary Note 8. It may also tunnel into another type of domain wall by nucleation processes, which we leave for future study.}

The coarsening dynamics of the topological defects 
with the $Z_2$ term are also described by the balance equation 
Eq.~(\ref{estimate}), while
the friction coefficient $\lambda(L)$ and the driving force $F_{\text{d}}(L)$ depends on 
a spatial profile of the topological defect that is deformed 
by the $Z_2$ term. In 
the case of $\beta \ll |\alpha|$,  
the free energy density $f$ is dominated by its phase 
part, 
\begin{align}
    f \approx |\alpha| 
    \left(\xi_0^2 |\nabla \theta|^2 + 2\beta \sin^2 \theta
    \right)
    \,. \label{f_beta}
\end{align}
Even with the small $\beta$, 
the topological defect is no longer a perfectly circular
vortex, but connects two different sine-Gordon solitons 
$\theta_0^{\pm} = \text{arg} \left(\sinh (\sqrt{2\beta}z/\xi_0) \pm i \right)$ ~\cite{ Eto2023-ky}, which is sketched in the inset of Fig.~\ref{data}(d). 
When the distance $r$ from the vortex core in the 2D plane 
is much smaller than a length scale $\xi_0 /\sqrt{2\beta}$, the spatial profile approaches the form of a circular symmetric vortex. 
For $r \gg \xi_0 /\sqrt{2\beta}$, the profile approaches the sine-Gordon solitons 
as $\theta \approx \theta_0^{\pm} + s(\bm{r})$, where $s(\bm{r}) \sim e^{-r\sqrt{2\beta}/\xi_0} $ decays exponentially. 

As a result, the coarsening dynamics \revise{crosses over } between two asymptotic time domains. In the early-time domain with $L(t) \ll \xi_0 /\sqrt{2\beta}$, adjacent defects see each other as circular symmetric vortices. Therefore, the driving force and frictional force are the same as the $U(1)$ symmetric case. In 2D systems, 
they are $F_{\text{d}}(L) \sim E_{\text{c}} |\alpha| \xi_0^2 /L$, and 
$\lambda(L) \sim E_{\text{c}} |\alpha| \ln (L/\xi)/\gamma$, respectively. From Eq.~(\ref{estimate}), one has $L(t) \sim \xi_0 \sqrt {\gamma t/\ln(|\alpha|\gamma t)}$, which is nearly diffusive in the large-$\gamma t$ limit. In 3D systems, the calculation is explained below Eq.~(\ref{estimate}). In the late-time domain with $L(t) \gg \xi_0 /\sqrt{2\beta}$, 
the adjacent vortices see each other as the sine-Gordon solitons. 
Thereby, the friction coefficient becomes a constant $\lambda \sim E_{\text{c}}|\alpha|/\gamma$. The driving force in the 2D system becomes short-range attraction between vortices and antivortices with $F_{\text{d}}(L) \sim  \xi_0 E_{\text{c}} |\alpha| \sqrt{2\beta} e^{-L\sqrt{2\beta}/\xi_0}$, while the driving force in the 3D system is dominated by the string tension with $F_{\text{d}}(L) \sim  \xi_0^2 E_{\text{c}} |\alpha|L^{-1}$, 
see the Supplementary Note 3. 
This results in the following scaling for 
the late-time domain of the coarsening dynamics, 
 \begin{align}
    L(t) \sim 
    \begin{cases}
    \frac{\xi_0}{\sqrt{2\beta}} \ln(2\beta\gamma t)   & \text{in 2D systems,} \\
    \xi_0\sqrt{\gamma t}              & \text{in 3D systems.}
\end{cases}
\label{Lt_beta}
\end{align}
Thus, the coarsening dynamics in the 2D systems exhibit a 
diffusive-to-subdiffusive crossover as $L$ goes from $ L \ll \xi_0/\sqrt{2\beta}$ 
to $L \gg \xi_0/\sqrt{2\beta}$, while that in the 3D 
systems are always diffusive.

Numerical simulations of Eq.~(\ref{model_A}) with the 
$Z_2$ term shows consistent coarsening dynamics.   
In Fig.~\ref{data}(b,c,d), 
we fit the value of $p$ in the interval of $\log_{10} (L/\xi) \in (1,1.2)$ for several $\beta/|\alpha|$. We get $p=0.46$ in the 3D systems with $\beta/|\alpha| = 0.1$ 
(Fig.~\ref{data}(b)), which is consistent with 
the diffusive dynamics.  
In the 2D systems, we obtain $p=0.28$ and $0.17$  with  
$\beta/|\alpha| = 0.001$ ($L < \xi_0 /\sqrt{2\beta}$) and $0.1$ ($L > \xi_0 /\sqrt{2\beta}$), respectively 
(Fig.~\ref{data}(c,d)), which is consistent with 
the diffusive-to-subdiffusive crossover.

\subsection*{Stability of the interface}
In addition to the motion within the interface, the defects also move in the $z$-direction  
due to the thermal noise $\eta$. From Eq.~(\ref{model_A}), 
the $z$-direction motion of a 
defect 
is a Brownian motion given by 
$d(z/\xi_0)/d(\gamma t) = H_z$ with a random force 
$H_z$ dependent on the defect size. For a typical defect with size $L$, 
the random force for a vortex satisfies 
 \begin{align}
    \langle H_z (t) H_z(t^{\prime})\rangle = 
    \begin{cases}
    \frac{2\zeta}{ h_z(L)} \delta(\gamma t -  \gamma t^{\prime})   & \text{in 2D systems,} \\
    \frac{2\zeta}{ h_z(L)L/\xi}  \delta(\gamma t - \gamma t^{\prime})              & \text{in 3D systems.}
\end{cases}
\label{H_z}
\end{align}
Here $h_z(L) = \int d^2\bm{r} \ (\partial_z\theta)^2$ where $\theta(\bm{r})$ is the profile of a vortex with size $L$ discussed below Eq.~(\ref{f_beta}), see the Supplementary Note 3. For the $U(1)$ symmetric case as well as 
for the $L \ll \xi_0/\sqrt{2\beta}$ case with a nonzero $\beta$, one has $h_z(L) \sim \ln(L/\xi)$, while it is hard to evaluate $h_z(L)$ for the other cases. 
\equa{H_z} implies that as long as $\zeta$ is small, 
the $z$-direction motion of the defects 
is negligible and the interface is stable.

\subsection*{Effects of the inertia term} 
So far, we have clarified the decaying process of the domain wall in terms of the ``model A" dynamics, Eq.~(\ref{model_A}). The ``model A" dynamics neglects the inertia of the order parameter, the $\partial_t^2 \psi$ term, while the 
inertia is crucial for the inversion of the order parameter by the pump pulse that generates the domain wall~\cite{yusupov2010coherent,duan2021optical,PhysRevB.103.054109,PhysRevLett.113.026401}. 
Thus, we now proceed to add the inertia term $\gamma_0^{-2}\partial_t^2 \psi$ to the left-hand side of Eq.~(\ref{model_A}), and discuss how the decaying processes of the domain wall are modified. 
Note that 
microscopic derivations of the relevant TDGL theory~\cite{Mitra.2018,Sun.2020_SC}
gives $\gamma,\, \gamma_0\sim T$ for 
pure electronic density-wave orders 
arising from Fermi surface nesting, see Ref.~\cite{Mitra.2018} and Appendix~E of Ref.~\cite{Sun.2020_SC}, while 
a mixing with lattice distortions and electron-phonon scattering 
may significantly modify these parameters. 

In the first-stage dynamics, 
a small $|\alpha|$  suppresses the inertia term more effectively than the damping term.  
Thus, as long as the temperature is close to $T_{\text{c}}$ where 
$|\alpha|\ll1$, the inertia term does not change the qualitative picture 
of the first-stage dynamics, modifying only the coefficients. 
Specifically, the time-dependent correlation length in Eqs.~(\ref{grt},\ref{g_beta}) is renormalized by a factor $1/\sqrt{\kappa}$ with $\kappa = \sqrt{1+4(|\alpha|-2\beta)\gamma^2/\gamma_0^2}$, see the Supplementary Note 4. 

In the second-stage dynamics, the $\partial_t^2 \psi$ term gives the topological defect 
an inertial mass $m(L) = \lambda(L)\gamma/\gamma_0^2$~\cite{PhysRevLett.68.1216}, 
which is dependent on the defect size $L$. Thereby, 
Eq.~(\ref{estimate}) is modified into
\begin{align}
   m(L)\frac{d^2 L}{dt^2} + \lambda(L) \frac{dL}{dt} = F_{\text{d}} (L),
   \label{estimate2}
\end{align}
see the Supplementary Note 4. Note that all the results of $L(t)$ satisfying Eq.~(\ref{estimate}) obtained in this work satisfy Eq.~(\ref{estimate2}) in the
long-time limit, when the term $m(L)d^2L /dt^2$ is much smaller than the other two terms.

\subsection*{ Estimation of parameters and timescales.}
\revise{Phenomenological parameters $\gamma, \xi_0, \zeta$, and $\alpha$ determine the timescale of the dynamics. In relevant CDW materials such as RTe$_3$ (R = La, Sm, Tb, Ce), 1T-TaS$_2$, and 1T-TiSe$_2$~\cite{PhysRevB.71.085114, PhysRevB.77.035114, PhysRevB.90.085105}, the relaxation time is roughly $1/\gamma \sim 1$ ps from ultrafast experiments~\cite{Dolgirev.2020_amplitude, Zong2019, PhysRevLett.130.226501, PhysRevB.103.054109, yusupov2010coherent, Gao.2022}. 
From the expression $\xi_0  \sim v_{\rm F}/T_{\rm c}$, $\xi_0$ is estimated as several nanometers for LaTe$_3$~\cite{PhysRevB.106.L121109}, while a similar value was obtained for 1T-TaS$_2$ from the domain-wall width in scanning tunneling microscopy images~\cite{Gao.2022}. Experiments are typically performed in 3D materials, where $T_{\rm c}^{D-1}/E_{\rm F}^{D-1} \sim 10^{-4}$ given that $E_{\rm F} \sim 10^4-10^5$ K for metals and $ T_{\rm c}\sim 10^2 - 10^3$ K. Therefore, the Ginzburg parameter is estimated by $\zeta \sim 10^{-4}/\sqrt{|\alpha|}$ for a certain experimental temperature. For example, in room-temperature ($T\approx 300$ K) experiments for SmTe$_3$ with $T_{\rm c} = 416$ K or LaTe$_3$ with $T_{\rm c} = 670$ K~\cite{PhysRevB.103.054109, Zong2019}
, one has $\sqrt{|\alpha|} \sim 1$ and $\zeta \sim 10^{-4}$. The first-stage dynamics finishes at a crossover time $t_{\rm c} \sim 10 \gamma^{-1} \sim 10$ ps. The second-stage dynamics will continue until $L(t)$ reaches a certain length scale $L_{\rm s}$ when the domain wall starts to interact with other domain walls or the sample surface. $L_{\rm s}$ is the domain-wall depth or the inter-domain-wall distance if multiple domain walls are generated. From the literature~\cite {PhysRevB.103.054109, duan2021optical}, an estimation gives $L_{\rm s}/\xi_0 \sim 10$ in 3D systems. Therefore, the scaling $L(t) \sim \xi_0 \sqrt{\gamma t}$ gives a timescale $\sim 10^2$ ps for the second-stage dynamics.   }

\subsection*{Experimental scheme.}
\revise{ The CDW order induces a lattice distortion $\bm{b}(\bm{r}) \propto \psi(\bm{r},t)e^{i\bm{Q}\cdot \bm{r}}$, where $\bm{Q}$ is the CDW wavevector lying in the $x-y$ plane, see the Supplementary Note 7. One could measure the evolution of the CDW order from the lattice structure factor in ultrafast x-ray scattering~\cite{orenstein2023subdiffusive,doi:10.1080/08940886.2016.1220273, PhysRevB.99.104111}. For example, in a recent ultrafast x-ray scattering experiment for LaTe$_3$ which exhibits an incommensurate CDW, the material was excited by a 1.55 eV, 35 fs optical pump pulse with the fluence of 8 mJ/cm$^2$ and probed by a 50 fs, 10 keV hard x-ray pulse~\cite{orenstein2023subdiffusive}. To investigate the dynamics of the $z=0$ interface parametrized by $\bm{R}$, one could extract a structure factor defined by $S(\bm{R},t) \equiv \langle b^{\dagger}(\bm{0},t) b(\bm{R},t)\rangle \propto \langle \psi^{\dagger}(\bm{0},t) \psi(\bm{R},t)\rangle e^{i\bm{Q}\cdot \bm{R}}$, whose Fourier transform $\tilde{S}(\bm{K},t)$ has a peak at $\bm{K} = \bm{Q}$ with the peak intensity corresponding to the amplitude $|\psi|^2$ and the inverse of the full width at half maximum (IFWHM) corresponding to the correlation length $L(t)$. In the long-time regime of the first-stage dynamics, Eqs.~\eqref{grt}\eqref{g_beta} predict the peak intensity to show an exponential growth and the IFWHM to exhibit a diffusive growth. In the second-stage dynamics, the IFWHM is expected to show a diffusive growth for 3D materials while the peak intensity has no significant growth since the growth of fluctuation magnitude has saturated. }

\section*{Discussion}
Our theory enriches the mechanism of topological defect generation in ultrafast experiments~\cite{Kibble1976,zurek1985cosmological,zurek1996cosmological, Zong2019, Cheng2024, PhysRevLett.110.156401}.  It applies to a wide class of systems with order parameters breaking a (quasi) continuous symmetry, such as charge order~\cite{joe2014emergence, PhysRevLett.118.106405, Gruner.1988_RMP, PhysRevLett.131.196401, PhysRevLett.130.226501}, spin order~\cite{RevModPhys.66.1}, excitonic order~\cite{ PhysRev.158.462, kozlov1965metal, Sun.2021_Josephson}, 
and superconductivity~\cite{Kopnin, Mitra.2017, Mitra.2018, Sun.2020_Metastable}.

In some experiments for 3D systems, a strong pump fluence could generate multiple domain walls at different 
depths in the $z$ direction with a typical inter-wall distance $L_z$~\cite{duan2021optical,yusupov2010coherent,PhysRevB.103.054109}. In this paper,
we focused on the decaying dynamics of an individual domain wall neglecting the interaction between the walls. This is appropriate as long as the inter-wall distance $L_z$ is much larger 
than the typical size $L$ of the topological defects on the walls, such that each wall evolves independently with a diffusive growth of in-plane correlation length $L$ while $L_z$ does not grow. In other words, the measured correlation-length dynamics depend on the direction of the measurement. This might be relevant to a recent pump-probe experiment that reports anomalous sub-diffusive coarsening dynamics of the CDW in $\text{LaTe}_3$~\cite{orenstein2023subdiffusive}.

\revise{Optically generated domain walls may also appear and lead to interesting dynamics in other situations that warrant further investigation~\cite{grandi2024nonequilibrium, gassner2024light, lin2024ultrafast},  e.g., the domain wall between superconducting orders with different chirality~\cite{PhysRevResearch.3.013253}. } Another example is the system with competing orders~\cite{Sun.2020_Metastable}: a real field $\psi$ breaking a $Z_2$ symmetry (e.g., commensurate CDW) competes with a complex field $\Psi = \Psi_1 + i\Psi_2$ breaking a $U(1)$ symmetry (e.g., superconductivity or incommensurate CDW)~\cite{PhysRevLett.118.106405, joe2014emergence} such that $\psi$ dominates 
and $\Psi$ is completely suppressed in equilibrium. If a pump pulse generates a domain wall of $\psi$, the fluctuations of $\Psi$ may become unstable on the domain wall and grow into a random domain 
structure with increasing correlation length. \revise{A preliminary study of a model with a quartic competing term $|\Psi|^2 \psi^2$ is provided in Supplementary Note 5, while it is meaningful to perform a systematical study with various competing terms between multi-component orders. } 

\section*{Methods}
\revise{Both analytical and numerical results in this work are obtained by the ``model A" dynamics, Eq.~(\ref{model_A}). The results of the first-stage dynamics are derived from the linearized dynamical equation, Eq.~(\ref{linear_dynamics}). The analytical results of the second-stage dynamics are obtained from the force-balance equation, Eq.~(\ref{estimate}), which is derived from Eq.~(\ref{model_A}).  In the numerical simulations, the ``model A" dynamical equation was rescaled by
\begin{align}
    \alpha &= \bar{\alpha} b^2,  \ \ \beta = \bar{\beta} b^2,  \ \ E_{\text{c}} = \bar{E_{\text{c}}} b^{-4}, \ \  \psi= \bar{\psi} b, \nonumber \\ \ \ \eta &=  \bar{\eta} b^3, 
    \ \ \xi_0 =  \bar{\xi_0} b, \ \ \gamma = \bar{\gamma} b^{-2},
    \label{rescaling}
\end{align}
with $b=\sqrt{|\alpha|}$ and $\bar{\alpha} = -1$. We further define $\tau = |\alpha|\gamma t$, $\tilde{\bm{r}} = \bm{r}/\xi$ and $\tilde{\nabla} = \partial/\partial \tilde{\bm{r}}$ and update $\bar{\psi}$ after each time step according to the dynamical equation given in the following form:
\begin{widetext}
\begin{align}
    \bar{\psi}(\tilde{\bm{r}}, \tau + d\tau) - \bar{\psi}(\tilde{\bm{r}}, \tau ) =  \Big( \tilde{\nabla}^2 \bar{\psi}(\tilde{\bm{r}}, \tau) + 2\bar{\psi}(\tilde{\bm{r}}, \tau) -2\frac{\beta}{|\alpha|} i \bar{\psi_2}(\tilde{\bm{r}}, \tau) - 2|\bar{\psi}(\tilde{\bm{r}}, \tau)|^2 \bar{\psi}(\tilde{\bm{r}}, \tau) \Big) d\tau + \bar{\eta}(\tilde{\bm{r}},\tau) d\tau.
\end{align}
\end{widetext}
The rescaled noise correaltor reads $\langle \bar{\eta}_{i}(\tilde{\bm{r}},\tau) \bar{\eta}_{i^{\prime}}(\tilde{\bm{r}}^{\prime},\tau^{\prime}) \rangle = 2\zeta \delta_{ii^{\prime}}\delta(\tilde{\bm{r}}-\tilde{\bm{r}}^{\prime})\delta(\tau - \tau^{\prime})$.  $\bar{\eta}(\tilde{\bm{r}},\tau) d\tau$ are generated by $\bar{\eta}_i(\tilde{\bm{r}},\tau) d\tau = \sqrt{\frac{2\zeta d\tau}{\Pi_j dr_j}} n_i(\tilde{r},\tau)$ with $n_i(\tilde{r},\tau)$ being generated from independent standard normal distribution for each discretized $\tilde{\bm{r}}$ and $\tau$. The values of $2\zeta$ and the discretization step sizes, $dr_{j=x,y,z}$ and $d\tau$, are provided in the caption of Fig.~\ref{data}. The second-order derivative, $\tilde{\nabla}^2 \bar{\psi}$, is computed using the central difference method. A free boundary condition is applied along the $z$-direction, while periodic boundary conditions are imposed along the $x$- and $y$-directions. } 

\section*{Data availability statement}
The data that support the findings of this study are available from the corresponding author upon reasonable request.
\section*{Code availability statement}
The code for the numerical simulation of this study is available from the corresponding author upon reasonable request.

\section*{Acknowledgments}
We thank Andrew J. Millis, Mariano Trigo, Alfred Zong, Shaofeng Duan, Qingzheng Qiu, Ke Liu, Yeyang Zhang, and Zhenyu Xiao for helpful discussions. 
Lingxian Kong and Ryuichi Shindou acknowledge the support from the National Basic Research Programs of China (No. 2019YFA0308401) and the National Natural Science Foundation of China (No. 11674011 and No. 12074008). \revise{Zhiyuan Sun acknowledges the support from the National Natural Science Foundation of China (Grants No. 12421004 and No. 12374291) and the State Key Laboratory of Low-Dimensional Quantum Physics at Tsinghua University. } \\

\section*{Author contributions}
Z.S. conceived and supervised the project. \revise{L.K. performed the calculations. L.K., R.S., and Z.S. co-wrote the manuscript. }

\section*{Competing Interests}
The authors declare no competing interests.

\section*{\bf References}
\bibliography{dw}


\end{document}


\title{Supplementary Notes for ``Fate of transient order parameter domain walls in ultrafast experiments"}
\author{Lingxian Kong}
\affiliation{International Center for Quantum Materials, School of Physics, Peking University, Beijing 100871, China}

\author{Ryuichi Shindou}
\affiliation{International Center for Quantum Materials, School of Physics, Peking University, Beijing 100871, China}%

\author{Zhiyuan Sun}%
\email{Corresponding author. zysun@tsinghua.edu.cn}
\affiliation{State Key Laboratory
of Low-Dimensional Quantum Physics and Department of Physics, Tsinghua University, Beijing 100084, China}%


\maketitle

\setcounter{equation}{0}
\setcounter{figure}{0}
\setcounter{table}{0}
\setcounter{page}{1}
\makeatletter
\renewcommand{\theequation}{S\arabic{equation}}
\renewcommand{\thefigure}{S\arabic{figure}}

\tableofcontents

\section*{\textsl{\NoCaseChange{Supplementary Note 1: }}Brief introduction about the formation of the order-parameter domain wall in ultrafast experiments. }
\label{dw_form_intro}
\begin{figure}[h]
    \centering
    \includegraphics[width= 4in]{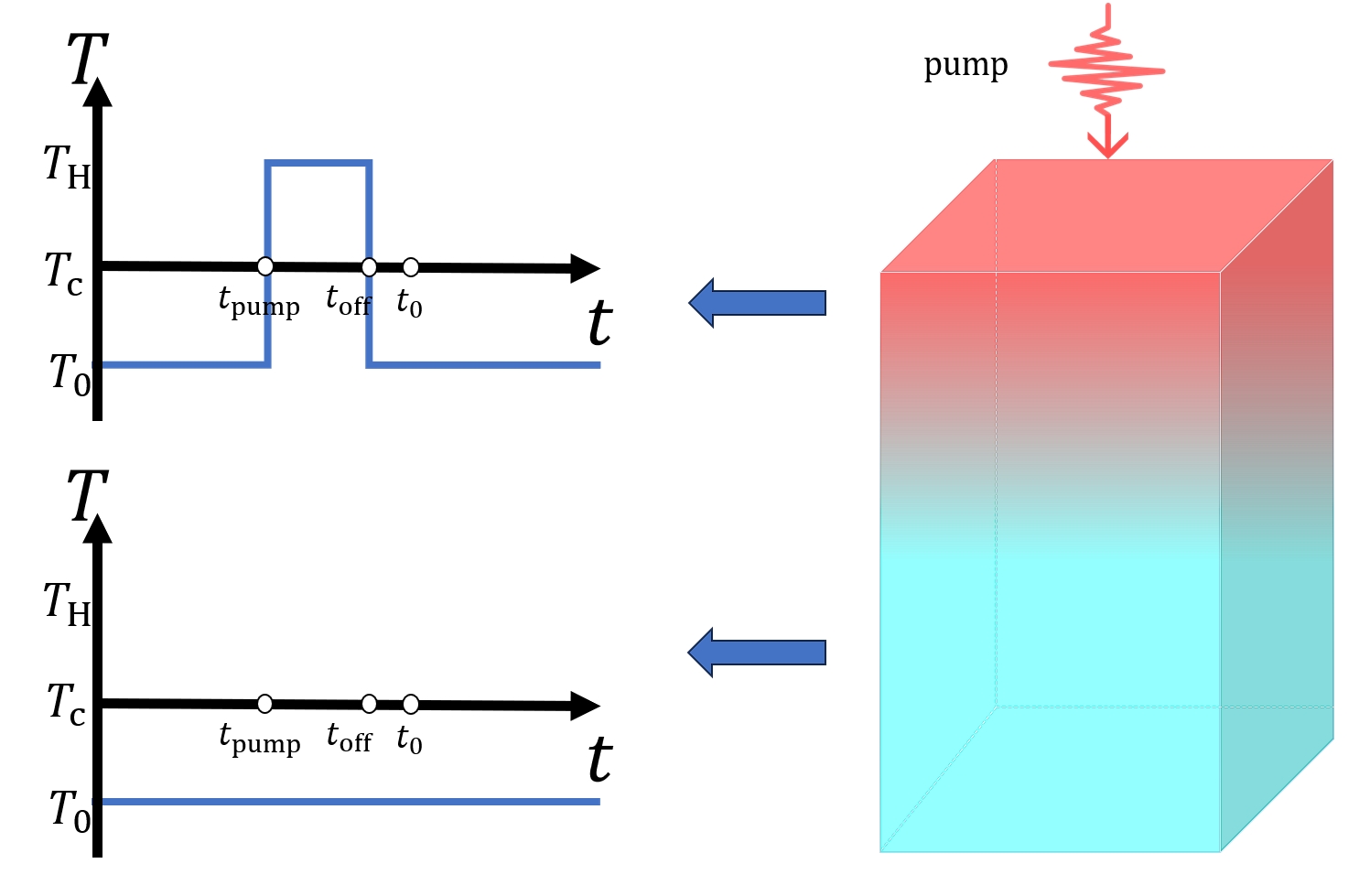} 
    \caption{Schematic pictures of the 3D sample under an optical pump, 
    and time($t$)-dependence of local temperature of a surface region.  
    For $t_{\text{pump}} < t < t_{\text{off}}$, the optical pump acts on the upper surface of the sample, elevating the local temperature of the surface region to $T_{\text{H}}$. For the bulk region away from the surface, the temperature remains $T_0$.}
    \label{Tt_profile}
\end{figure}

In this section, we introduce a mechanism of the domain wall formation~\cite{yusupov2010coherent, duan2021optical, PhysRevB.103.054109}. 
It has been reported both theoretically and experimentally that an optical pump can invert the charge-density-wave (CDW) order parameter in surface regions of samples, generating domain walls of the order parameter. The formation of the domain wall is driven by a 
thermal effect due to the optical pump. When the optical pump acts 
on a sample surface (Fig.~\ref{Tt_profile}), layers near the surface 
get hotter instantly, while layers inside a sample bulk 
remains cold due to a finite penetration depth of the optical pump. 
Suppose that during the optical pump, $t_{\text{pump}} < t < t_{\text{off}}$, 
the temperature in the surface layers is elevated to a temperature with a spatial dependence $T(z,t) = T_0 + \Theta(t-t_{\rm pump}) \Theta(t_{\rm off} -t) (T_{\text{H}} - T_0) e^{-(z_{\text{top}}-z)/ z_{\text{p}} }$. Here $\Theta(t)$ is the Heaviside step function, $z_{\text{top}}$ is a location of the top surface, and $z_{\text{p}}$ is the penetration depth of the optical pump. 
$T_{\text{H}}$ is the temperature on the top surface of the sample, which is above an equilibrium critical temperature $T_{\text{c}}$. $T_0$ is the temperature deep inside the sample, which is below $T_{\rm c}$.

\begin{figure}[h]
    \centering
    \includegraphics[width= 6.5in]{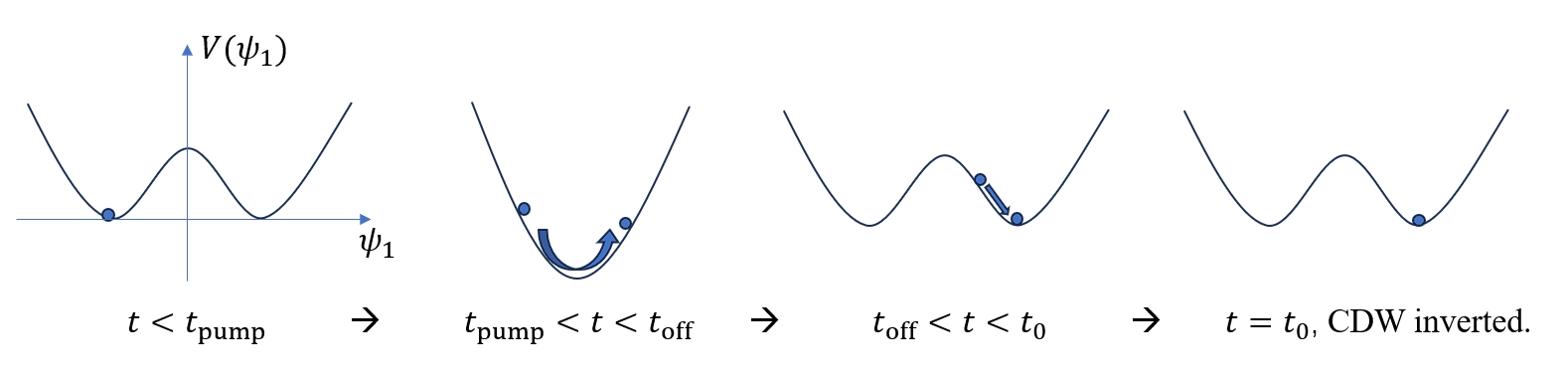} 
    \caption{ Schematic picture of the order parameter inversion in the surface region. The curves shows an evolution of a free-energy potential for $\psi_1$ in the surface region, $V(\psi_1) = E_{\text{c}} [\psi_1^2 + \alpha(t)]^2$. Before the pump ($t<t_{\rm pump}$), and after the pump ($t>t_{\rm off}$), $\alpha(t)$ is negative, and the function has double wells. During the pump ($t_{\rm pump}<t<t_{\rm off}$), $\alpha(t)>0$, and the potential is in the single-minimum form. A ball in the pictures represents the real part $\psi_1$ of the order parameter $\psi$. In the picture, $\psi_1$ is in the minimum with the negative value for $t<t_{\rm pump}$.
    In the duration of the pump, $\psi_1$ `slides down' along the potential, and changes its sign. When the pump is turned off, $V(\psi_1)$ changes into the double-well form again, and $\psi_1$ relaxes into the other well with the positive value.}
    \label{landscape}
\end{figure}

Consider that the CDW free energy 
with the ``Mexican hat" potential $V(|\psi|) = E_{\text{c}} [|\psi|^2 + \alpha(z,t)]^2$ 
depends on time $t$ through the coefficient $\alpha(z,t)$ and
$\alpha(z,t)$ depends on a local temperature $T(z,t)$, i.e. $\alpha(z,t) = \alpha(T(z,t))$. $\alpha(z,t)$ is positive/negative if $T(z,t)$ is above/below a transition 
temperature $T_{\text{c}}$.
For $t < t_{\text{pump}}$, the order parameter 
in the whole sample is in the equilibrium energy minima. Without loss of 
generality, we assume that $\psi$ for $t<t_{\text{pump}}$ is real, e.g. 
$\psi =\psi_{\text{eq}}= -\sqrt{|\alpha|}$. In experiments, 
the duration of the optical pump is typically short, so that dynamics of 
the phase degree of freedom can be ignored~\cite{yusupov2010coherent}. 
Therefore, during such a short duration, $\psi$ continues to be real, 
where the potential depends only on $ \text{Re}\psi \equiv \psi_1$ for 
$t_{\text{pump}} < t <t_{\text{off}}$. 
Fig.~\ref{landscape} shows schematically 
an evolution of the potential $V(\psi_1)$ and how the order parameter 
is inverted in the region of the surface layers. After the pump, 
$\psi$ in the surface layers follows an oscillatory motion, 
while $\psi$ deep inside the bulk remains unchanged with
$\psi = \psi_{\text{eq}} = -\sqrt{|\alpha|}$. When the 
the pump is off at an appropriate time, $\psi$  
in the surface region find its minimum on the other side 
of the double-well potential. Consequently, a domain wall shall be 
formed at a certain time, $t=t_0>t_{\rm off}$. 
A $\psi$-configuration with single domain wall at $z=0$ can given by 
$\psi(\bm{r}) = \sqrt{|\alpha|}\tanh\frac{z}{\xi}$ with 
$\xi\equiv \xi_0/\sqrt{|\alpha|}$ being the coherence length.  

\begin{figure}[h]
    \centering
    \includegraphics[width= 3in]{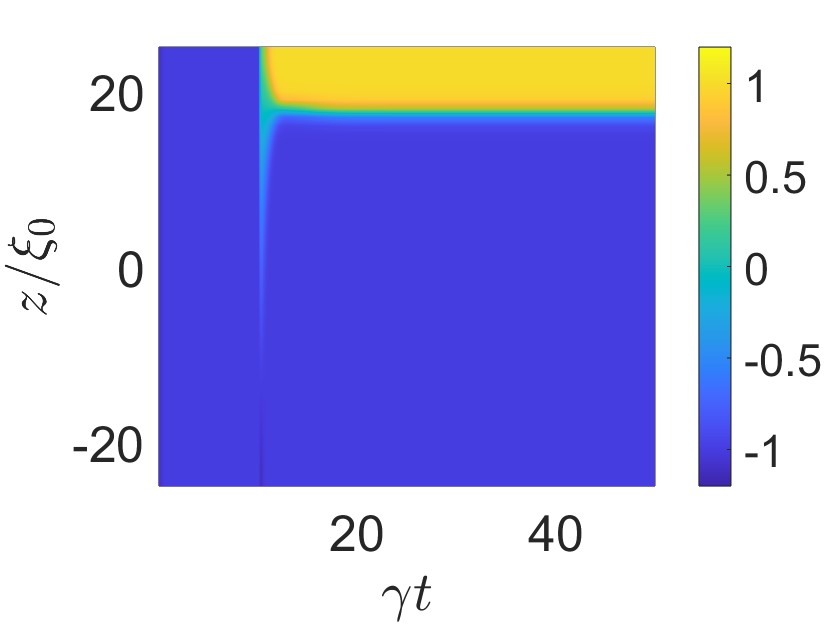} 
    \caption{Plot of $\psi$ in the time evolution of Eq.~(\ref{dw_form}) with the initial condition $\psi = -1$.  The pump is applied in $\gamma t \in [10,10.2]$. i.e. $\alpha(z,t) = \alpha(T_0) + \Theta(t-10) \Theta(10.2 -t) \alpha(T_{\rm H}) e^{-(z_{\text{top}}-z)/ z_{\text{p}} }$. Here $\alpha(T_0)=-1$, $\alpha(T_{\rm H}) = 7$, $z_{\text{p}}/\xi_0 = 20$. $\gamma_0 = 0.1\gamma$.}
    \label{dw_res}
\end{figure}

In the following, we will explain the mechanism of the domain-wall formation 
more formally~\cite {yusupov2010coherent,duan2021optical, PhysRevB.103.054109}. 
To describe the oscillatory motion of $\psi$ in the duration of the optical pump, we 
add an inertia term into the time-dependent Ginzburg-Landau (TDGL) equation, 
\begin{align}
    \frac{1}{\gamma_0^2} \partial_t^2 \psi + \frac{1}{\gamma} \partial_t \psi  = 
    \xi_0^2 \nabla^2 \psi - 2\alpha(z,t) \psi - 2|\psi|^2 \psi. 
    \label{dw_form}
\end{align}
Here we set $\alpha(z,t) = \alpha(T_0) + \Theta(t-t_{\rm pump}) \Theta(t_{\rm off} -t) \alpha(T_{\rm H}) e^{-(z_{\text{top}}-z)/ z_{\text{p}} }$. 
Typically $z_{\text{p}}\approx 20$nm ~\cite{yusupov2010coherent, duan2021optical, PhysRevB.103.054109}. By solving Eq.~(\ref{dw_form}) with the initial condition $\psi=-\sqrt{|\alpha|}$ before the pump, we confirm 
that the domain wall emerges after the duration of the optical 
pump, as shown in Fig.~(\ref{dw_res}).  

Several remarks on Eq.~(\ref{dw_form}) should be noted in order. To begin, 
the dynamics of Eq.~(\ref{dw_form}) is in the mean-field level, 
which does not include the fluctuation effect. 
In the absence of the fluctuation, 
$\psi$ remains to be real with the real $\psi$ at the initial time, and 
the phase dynamics does not develop at all. Such dynamics is approximately valid 
within the short duration $t_{\rm pump}<t<t_{0}$; 
these durations are anyway short enough that the phase fluctuation does not develop 
in practice~\cite{yusupov2010coherent, PhysRevB.103.054109}. 
To study the further evolution of the domain wall for $t>t_0$, however, it is 
crucial to include the two-component nature of $\psi$ as well as the 
thermal fluctuation effect. This is the major focus of this work. 

Secondly, Eq.~(\ref{dw_form}) includes the inertia term, i.e. 
the second-order time-derivative term $\partial_t^2 \psi$, while 
the ``model A" dynamics that we mainly use in this work for $t>t_0$ does not include the inertia term. 
To describe the formation of the domain wall during $t_{\rm pump}<t<t_{0}$, 
the inertia term is crucial; without the inertia, $\psi$ 
in the surface region falls into the $\psi \approx 0$ state and would 
stay there for the rest of the duration.  
To study the further evolution of the domain wall for $t>t_0$, we will mainly study 
the ``model A" dynamics without the $\partial_t^2 \psi$ term for simplicity. 
In Sec.~IV of this supplemental material 
as well as the end of the main text, we investigate the effects of the $\partial_t^2 \psi$ term in the domain wall evolution for $t>t_0$. We argue 
that the inertia term is suppressed 
when $T_0$ is close to $T_{\rm c}$, 
modifying only quantitatively the domain wall evolution for $t>t_0$.  

\section*{\textsl{\NoCaseChange{Supplementary Note 2: }}The first-stage dynamics}
In this section, we will discuss the first-stage dynamics for a general system with and without the $U(1)$ symmetry. We consider the following free energy and dynamics with the thermal noise; 
\begin{align}
    &\frac{1}{\gamma} \partial_t \psi = -\frac{1}{E_{\text{c}}} \frac{\delta F}{\delta \psi^*} + \eta  \label{model_A_sm} \\
    &F = E_{\text{c}} \int d^{D} \bm{r} \ \  \xi_0 ^2  |\nabla \psi|^2 + (|\psi|^2 + \alpha)^2 + 2\beta\psi_2^2. \label{free_energy_sm}\\
    &\langle \eta_{i}(\bm{r},t) \eta_{i^{\prime}}(\bm{r^{\prime}},t^{\prime}) \rangle = \frac{2T}{\gamma E_{\text{c}}} \delta_{ii^{\prime}}\delta(\bm{r}-\bm{r^{\prime}})\delta(t-t^{\prime}) \nonumber   
\end{align} 
Here $\gamma^{-1}$ has dimension of time, $\alpha$, $\beta$ and 
$\psi$ are dimensionless, and $E_c$ has a dimension of energy density. i.e. [Energy][Length]$^{-D}$. The system has the $U(1)$-symmetry for $\beta =0$, where $F$ is invariant under a transformation $\psi\rightarrow \psi^{i\theta}$. The symmetry reduces to $Z_2 \otimes Z_2$ symmetry for $\beta \ne 0$, where $F$ is invariant under the transformations 
$\psi_{1(2)} \rightarrow \pm \psi_{1(2)}$. 
At $t=t_0$, $\psi(\bm{r},t_0) = \psi_0 + \delta\psi$ with an initial 
fluctuation $\delta \psi = \phi_1 + i\phi_2$. The initial fluctuation 
comes from the thermal noise accumulated during the time $t<t_0$. We will 
discuss the nature of the initial fluctuation in the later section of this 
section. By expanding small $\phi_1$ and $\phi_2$, we obtain Eq.~(3) in the main text, 
\begin{align}
     \frac{1}{\gamma} \partial_t \phi_{a} &= -(\hat{L}_{a} - \xi_0 ^2 \nabla_{\bm{R}}^2)\phi_{a} + \eta_{a},
    \label{linear_dynamics_sm}
\end{align}
with $a=1,2$, ${\bm r}\equiv ({\bm R},z)$, $\nabla_{\bm{R}}^2=\nabla^2-\partial_z^2$ is the Laplacian in $D-1$ dimensions. and  $\hat{L}_{1} = -\xi_0 ^2 \partial_z^2 + 2\alpha + 6\psi_0^2$, and $\hat{L}_{2} = -\xi_0 ^2 \partial_z^2 + 2\alpha + 2\beta +  2\psi_0^2$. 
According to Eq.~(\ref{linear_dynamics_sm}), 
eigenvectors of $-\hat{L}_{a}+\xi^2_0 \nabla^2_{\bm R}$ with positive eigenvalues amplify the fluctuations, dominating the first-stage dynamics 
for $t>t_0$. The eigenvectors are given by a product of eigenmodes of $\hat{L}_{a}$ and plane waves $e^{i{\bm K}\cdot{\bm R}}$.
Thus, the amplifying eigenvectors must 
comprise of the eigenmodes of $\hat{L}_a$ with negative eigenvalues. 
 
\subsection{Eigenvalues of the $\hat{L}_{a}$ operators.}
$\hat{L}_1$ and $\hat{L}_2$ take forms of one-dimensional scattering problem with Poschl-Teller potential hole. The scattering problem is defined by  
a following Hamiltonian with a parameter $\lambda$, 


\begin{align}
\hat{H}(\overline{z};\lambda) = - \partial_{\overline{z}}^2 - \frac{\lambda(\lambda - 1)}{\cosh^2(\overline{z})}.
    \label{sech_eig}
\end{align}

Setting $\overline{z}=z/\xi$, 
and $\xi\equiv \xi_0/\sqrt{|\alpha|}$, we obtain $\hat{L}_1 = |\alpha| \hat{H}(\overline{z};3) + 4|\alpha|$,  
$\hat{L}_2 = |\alpha| \hat{H}(\overline{z};2) + 2\beta$. 
The scattering problem has bound-states solutions 
with negative energies as well as continuum states with 
positive energies~\cite{flugge2012practical}. The negative eigenvalues 
are $-(\lambda-1-n)^2 $ for non-negative integer $n$ 
smaller than $\lambda-1$: $0\le n<\lambda-1$.  
Thus, $\hat{L}_1$ has always positive eigenvalues. 
$\hat{L}_2$ can have one negative eigenvalue ($-|\alpha| + 2\beta$) 
with an eigenfunction ($\text{sech}(\overline{z})$) for $\beta<|\alpha|/2$.

\subsection{Solving \texorpdfstring{$C(\bm{R},t) = \langle \varphi (\bm{0},t)\varphi (\bm{R},t) \rangle$}{TEXT}  based on Eq.~(\ref{linear_dynamics_sm}).}

\label{sec_solve_C}

\subsubsection{An equation of motion for \texorpdfstring{$\varphi (\bm{R},t)$}{TEXT}.}
To see how the eigenmode of $\hat{L}_2$ with 
the negative eigenvalue amplifies the fluctuation, 
let us set $\phi_2(\bm{r},t) = \varphi (\bm{R},t)\text{sech}(z/\xi)$ with ${\bm r}=({\bm R},z)$. Eq.~(\ref{linear_dynamics_sm}) in the main text reads 
\begin{align}
    \frac{1}{\gamma} \partial_t \varphi (\bm{R},t) \text{sech}(z/\xi) = (|\alpha| - 2\beta + \xi_0^2 \nabla_{\bm{R}}^2 )\varphi (\bm{R},t) \text{sech}(z/\xi) + \eta_2(\bm{r},t)
\end{align}
By multiplying this by $\text{sech}(z/\xi)$ and taking an integral with $\int \frac{dz}{2\xi}$, we obtain 
\begin{align}
    \frac{1}{\gamma} \partial_t \varphi (\bm{R},t) = (|\alpha|- 2\beta + \xi_0^2 \nabla_{\bm{R}}^2 )\varphi (\bm{R},t) + \eta_\varphi (\bm{R},t).
    \label{linear_dynamics_C}
\end{align}
with $\eta_\varphi (\bm{R},t) \equiv \int\frac{dz}{2\xi} \text{sech}(z/\xi) \eta_2(\bm{r},t)$ and 
\begin{align}
    \langle \eta_\varphi (\bm{R},t) \eta_\varphi (\bm{R^{\prime}},t^{\prime}) \rangle = \int\frac{dz}{2\xi} \text{sech}(z/\xi) \int\frac{dz^{\prime}}{2\xi} \text{sech}(z^{\prime}/\xi) \langle \eta_2(\bm{r},t) \eta_2 (\bm{r^{\prime}},t^{\prime}) \rangle   = \frac{T}{ \gamma  E_{\text{c}} \xi} \delta(\bm{R}-\bm{R^{\prime}})\delta(t-t^{\prime}).
\end{align}


\subsubsection{Derivation of  Eqs.~(4,6) in the main text.}
To analyze the dynamics of $\varphi (\bm{R},t)$ 
for $t>t_0$ from Eq.~(\ref{linear_dynamics_C}), we follow the literature~\cite{Sun.2020_Metastable} and study 
the Fourier transform of Eq.~(\ref{linear_dynamics_C}) with 
$\tilde{\varphi}({\bm K},t) = \frac{1}{l^{D-1}}
\int d^{D-1}{\bm R} \ \ \varphi({\bm R},t) 
e^{-i{\bm K}\cdot{\bm R}}$ and $\tilde{\eta}_{\varphi}({\bm K},t) 
= \frac{1}{l^{D-1}}
\int d^{D-1}{\bm R} \ \ \eta_{\varphi}({\bm R},t) 
e^{-i{\bm K}\cdot{\bm R}}$.  \footnote{In this work, we use the following notation for the Fourier transform.
\begin{align}
    f(x) &= \sum_k \tilde{f}(k)e^{ikx} ; \ \ \ \ \
    \tilde{f}(k) = \frac{1}{l} \int   f(x)e^{-ikx} dx . \nonumber
\end{align} }
\begin{align}
    \frac{1}{\gamma}\partial_t \tilde{\varphi}(\bm{K},t) = (|\alpha| - 2\beta - \xi_0^2 K^2)\tilde{\varphi}(\bm{K},t) + \tilde{\eta}_\varphi (\bm{K},t).
    \label{linear_dynamics_Ck}
\end{align}
Here $l$ is the system size. A general solution of Eq.~(\ref{linear_dynamics_Ck}) is given by 
\begin{align}
    \tilde{\varphi}(\bm{K},t) = \tilde{\varphi}(\bm{K},t_0)e^{S_{K}(t,t_0) } + \gamma\int_{t_0}^{t}  \tilde{\eta}_\varphi (\bm{K},t^{\prime}) e^{S_{K}(t,t^{\prime})} dt^{\prime}
\end{align}
with $S_{K}(t,t^{\prime}) \equiv  \gamma\int_{t^{\prime}}^{t}  (|\alpha| - 2\beta -\xi_0^2 K^2 ) ds$. Note that $\langle \tilde{\varphi}(-\bm{K},t) \tilde{\varphi}
(\bm{K}^{\prime},t) \rangle=\delta_{{\bm K},{\bm K}^{\prime}}D_{K}(t)$. An amplitude of $\tilde{\varphi}(\bm{K},t)$, $D_{K}(t) \equiv \langle \tilde{\varphi}(-\bm{K},t) \tilde{\varphi}(\bm{K},t) \rangle$, is calculated as 
\begin{align}
    D_{K}(t) &= \frac{1}{l^{D-1}} e^{2S_K (t,t_0)} \Big(  l^{D-1} D_K (t_0) + \frac{T\gamma}{E_{\text{c}} \xi} \int_{t_0}^t dt^{\prime} e^{-2S_K (t^{\prime},t_0)} \Big) \\ \nonumber
    &= \frac{1}{l^{D-1}}e^{2S_K (t,t_0)} \Big(  l^{D-1}D_K (t_0) + \frac{T}{2E_{\text{c}} \xi} \frac{1}{|\alpha| - 2\beta - \xi_0^2 K^2} (1-e^{-2S_K (t,t_0)} )   \Big). 
\end{align}
$C({\bm R},t) \equiv \langle \varphi(0,t) \varphi
(\bm{R},t) \rangle$ is given by an inverse Fourier transform of $D_K (t)$, 
\begin{align}
    C(\bm{R},t) &= \sum_{\bm{K}} D_K (t) e^{i\bm{K}\cdot \bm{R}} = l^{D-1}\int\frac{d^{D-1} \bm{K}}{(2\pi)^{D-1}} D_K (t) e^{i\bm{K}\cdot \bm{R}} \nonumber \\
    &= C_1(\bm{R},t) + C_2(\bm{R},t), 
    \label{g_Rt}
\end{align}
where 
\begin{align}
     C_1(\bm{R},t) &\equiv \int\frac{d^{D-1} \bm{K}}{(2\pi)^{D-1}}   l^{D-1} D_K(t_0) e^{2(|\alpha| - 2\beta)\gamma t} e^{-2\xi_0^2 \gamma (t-t_0) K^2 } e^{i\bm{K}\cdot \bm{R}}
     \label{C_1}
\end{align}
\begin{align}
    C_2(\bm{R},t) &\equiv  \int\frac{d^{D-1} \bm{K}}{(2\pi)^{D-1}}    \frac{T}{2E_{\text{c}} \xi} \frac{e^{2S_K(t,t_0)}-1}{(|\alpha| - 2\beta) - \xi_0^2 K^2}  
    e^{i\bm{K}\cdot \bm{R}} 
    \label{C_2}
\end{align}
$C_1$ is dependent on the initial fluctuation at $t=t_0$, 
while $C_2$ is a contribution of the thermal noises that are accumulated during $t_0<t^{\prime}<t$. In the calculation below, we set $t_0=0$ for convenience.

To calculate $C_2$ for larger $R$ and $t$, we follow an approximation 
in Ref.~\onlinecite{Sun.2020_Metastable}, and obtain a simpler analytical expression. 
For $\gamma t \gg 1/(|\alpha|-2\beta)$, the first term in Eq.~(\ref{C_2}) 
dominates over the second term. Also, due to the $e^{-2\xi_0^2 \gamma t K^2}$ factor in the integrand, the integration over $\bm{K}$ converges well within the $\xi_0^2 K^2 \ll |\alpha|-2\beta $ region in the $\gamma t \gg 1/(|\alpha|-2\beta)$ limit. Thus, the $K$-dependence in the denominator becomes unimportant and could be neglected. Thus, we have  
\begin{align}
    C_2 (\bm{R},t) 
    & \approx \int \frac{d^{D-1} \bm{K}}{(2\pi)^{D-1}} 
    \frac{T}{2E_{\text{c}} \xi}\frac{1}{(|\alpha| - 2\beta) - \xi_0^2 K^2}  e^{2(|\alpha| - 2\beta)\gamma t} e^{-2\xi_0^2 \gamma t K^2 } e^{i\bm{K}\cdot \bm{R}} 
    \nonumber \\
    &\approx \frac{T}{2E_{\text{c}} \xi (|\alpha| - 2\beta)}  e^{2(|\alpha| - 2\beta)\gamma t} \int\frac{d^{D-1} \bm{K}}{(2\pi)^{D-1}}    e^{-2\xi_0^2 \gamma t K^2 } e^{i\bm{K}\cdot \bm{R}} \nonumber \\
    &= \frac{T}{2E_{\text{c}} \xi \xi_0^{D-1} (|\alpha| - 2\beta)}  \frac{e^{2 (|\alpha| - 2\beta)\gamma t} } {\left( \sqrt{ 8 \pi \gamma t }\right)^{D-1} }  e^{- \frac{R^2} { 8\xi_0^2 \gamma t} } \nonumber \\
    &= \frac{G}{2\sqrt{|\alpha|}(1 - 2\beta/|\alpha|)}  \frac{e^{2 (|\alpha| - 2\beta)\gamma t} } {\left( \sqrt{ 8 \pi \gamma t }\right)^{D-1} }  e^{- \frac{R^2} { 8\xi_0^2 \gamma t} } \nonumber\\
    &= \frac{\zeta|\alpha|}{2(1 - 2\beta/|\alpha|)}  \frac{e^{2 (|\alpha| - 2\beta)\gamma t} } {\left( \sqrt{ 8 \pi |\alpha| \gamma t }\right)^{D-1} }  e^{- \frac{R^2} { 8\xi_0^2 \gamma t} }. 
    \label{C_2_res}
\end{align} 
Here we introduced two dimensionless quantities 
\begin{align}
G &\equiv  T/(E_{\text{c}}\xi_0^D), \label{G-def} \\ 
\zeta &\equiv G|\alpha|^{D/2 - 2},  \label{zeta-def}
\end{align}
that measures the strength of the thermal fluctuation. Since $t>t_0$ in the calculation above, $T=T_0$ in $G$ and $\zeta$ appeared above. Especially, 
$\zeta$ plays a similar role as the so-called Ginzburg parameter does 
in theories of equilibrium critical phenomenon~\cite{cardy_1996}. A similar 
$C_2 (\bm{R},t)$ appears in 
the literature~\cite{Sun.2020_Metastable}. 

To calculate $C_1$, we need the expression of the initial fluctuation at $t=t_0$, i.e. $D_K(t_0)$. The fluctuation originates from the thermal fluctuation at equilibrium for $t<t_{\rm pump}$. It also evolves during the time domain 
$t_{\rm pump} < t < t_0$. To determine $D_K(t_0)$, we 
calculate the thermal fluctuation at equilibrium for $t<t_{\rm pump}$ first.

$\bullet$ {\bf Thermal fluctuation at equilibrium }: 
Following the ``model A" dynamics, the thermal fluctuation at equilibrium can be also calculated in terms of 
the linearized equation of motion for small imaginary part $\psi_2$ around 
the uniform ground state 
$\psi = -\sqrt{|\alpha|}$;
\begin{align}
     \frac{1}{\gamma} \partial_t \psi_{2} &=  \xi_0 ^2 \nabla^2 
 \psi_2 -2\beta\psi_2  + \eta_{2}. 
\end{align}
The equation can be solved in the momentum space with $\tilde{\psi}_2({\bm k},t) = \frac{1}{l^D}
\int d^{D}{\bm r} \psi_2({\bm r},t) 
e^{-i{\bm k}\cdot{\bm r}}$ and $\tilde{\eta}_{2}({\bm k},t) 
= \frac{1}{l^{D}}
\int d^{D}{\bm r} \eta_{2}({\bm r},t) 
e^{-i{\bm k}\cdot{\bm r}}$,  
\begin{align}
    \frac{1}{\gamma} \partial_t \tilde{\psi}_{2}(\bm{k},t) &=  -(2\beta + \xi_0 ^2 k^2) \tilde{\psi}_{2}(\bm{k},t) + \tilde{\eta}_{2}(\bm{k},t), \nonumber \\
    \tilde{\psi}_{2}(\bm{k},t) &= \tilde{\psi}(\bm{k},-\infty)e^{-\gamma (2\beta + \xi_0 ^2 k^2) \int_{-\infty}^t dt^{\prime}} + \gamma\int_{-\infty}^{t}  \tilde{\eta}_2 (\bm{k},t^{\prime}) e^{-\gamma(2\beta + \xi_0 ^2 k^2)(t-t^{\prime})} dt^{\prime}.
\end{align}
Here we regard that the linear system size along $z$ direction is also given by $l$.  
Taking the correlation at $t=-\infty$ to be zero, we obtain 
\begin{align}
    \langle\tilde{\psi}_2(\bm{k},t) \tilde{\psi}_2(\bm{k^{\prime}},t) \rangle  = \delta
_{\bm{k},-\bm{k^{\prime}}} \frac{T}{ l^D E_{\text{c}} (2\beta + \xi_0 ^2 k^2)  }
\end{align}
and
\begin{align}
    \langle\psi_2(\bm{r},t) \psi_2(\bm{r^{\prime}},t) \rangle &= \sum_{\bm{k},\bm{k^{\prime}}} 
    e^{i\bm{k}\cdot \bm{r}}
    e^{i\bm{k^{\prime}}\cdot \bm{r^{\prime}}}
    \langle\tilde{\psi}_2(\bm{k},t) \tilde{\psi}_2(\bm{k^{\prime}},t) \rangle = \frac{2T\gamma}{E_{\text{c}}} \int\frac{d^D \bm{k}}{(2\pi)^D} e^{i\bm{k}(\bm{r}-\bm{r^{\prime}})} 
    \frac{1}{2\gamma (2\beta + \xi_0 ^2 k^2)}
    \label{goldstone}
\end{align}
From Eq.~(\ref{goldstone}), we obtain
\begin{align}
    \langle\psi_2(\bm{r},t) \psi_2(\bm{r},t) \rangle &= 
     \begin{cases}
    \frac{T}{E_{\text{c}} } \int^{1/a}_{0} \frac{k^2 dk}{2\pi^2} \frac{1}{(2\beta + \xi_0 ^2 k^2) },   & \text{in 3D,} \\
    \frac{T}{E_{\text{c}} } \int^{1/a}_0 \frac{k dk}{2\pi} \frac{1}{(2\beta + \xi_0 ^2 k^2)},              & \text{in 2D.}
    \end{cases}
    \label{psi_2_r}
\end{align}
where $1/a$ stands for a cutoff for the integral in the 
ultraviolet (UV) region. With the UV cutoff, the integral is 
convergent in 3D. Meanwhile, the integral in 2D has an infrared 
(IR) divergence for $\beta=0$. The divergence 
indicates that the fluctuation is so strong in 2D 
that the U(1) symmetry breaking does not occur at 
finite temperature~\cite{cardy_1996, kardar_2007}. 
Thus, we discuss only (i) $\beta \ne 0$ case in 2D, 
and (ii) $\beta=0$ and $\beta\ne 0$ cases in 3D. 

We will next discuss the evolution of the 
fluctuation during $t_{\rm pump}<t<t_0$. 

$\bullet$ {\bf Evolution of the thermal fluctuation during  $t_{\rm pump}<t<t_0$}: 
In the time slot, due to the pump, $\psi_1$ acquires an evolution from $\psi_1(\bm{r},t_{\rm pump})= -1$ to $\psi_1(\bm{r},t_0) = \sqrt{|\alpha|}\tanh(z/\xi) $ as shown in Fig.~\ref{landscape},\ref{dw_res}.  The ``model A" dynamics gives the following linearized equation of motion of $\psi_2$.

\begin{align}
    \frac{1}{\gamma} \partial_t \psi_{2} &=  \xi_0 ^2 \nabla^2 
 \psi_2 -2[\psi_1^2(z,t) + \alpha(z,t) + \beta ] \psi_2  + \eta_{2},
 \label{psi_2_eff}
\end{align}
with 
\begin{align}
    \langle \eta_{2}(\bm{r},t) \eta_{2}(\bm{r^{\prime}},t^{\prime}) \rangle &= \frac{2T(z,t)}{\gamma E_{\text{c}}} \delta(\bm{r}-\bm{r^{\prime}})\delta(t-t^{\prime}) .
\end{align}
Here $\psi_1^2(z,t)$ follows the solution of Eq.~(\ref{dw_form}), $\alpha(z,t) $ and $T(z,t)$ take the form appeared in Eq.~(\ref{dw_form}). Apply the Fourier transform to the $x$-$y$ coordinates, $\psi_2(\bm{r},t) \rightarrow \tilde{\psi}(\bm{K},z,t)$, $\eta_2(\bm{r},t) \rightarrow \tilde{\eta}_2(\bm{K},z,t)$ with the momentum ${\bm K}$ in the $x$-$y$ plane, we obtain the following equation 
\begin{align}
    \frac{1}{\gamma} \tilde{\psi}_2(\bm{K},z,t) = s(\bm{K},z,t)\tilde{\psi}_2(\bm{K},z,t) +  \tilde{\eta}_2(\bm{K},z,t)
\end{align}
with 
\begin{align}
    s(\bm{K},z,t) = -\xi_0^2 K^2 + \tilde{\psi}^{-1}_2(\bm{K},z,t) \xi_0^2 \partial_z^2 \tilde{\psi}_2(\bm{K},z,t) - 2[\psi_1^2(z,t) + \alpha(z,t) + \beta ].
\end{align}

Therefore, we have the following solution for $\tilde{\psi}_2(\bm{K},t_0)$,
\begin{align}
    \tilde{\psi}_2(\bm{K},z,t_0) &= e^{\gamma \int_{t_{\rm pump}}^{t_0} s(\bm{K},z,t^{\prime}) dt^{\prime}} \tilde{\psi}(\bm{K},z,t_{\rm pump}) + \gamma  \int_{t_{\rm pump}}^{t_0}
 e^{\gamma \int_{t^{\prime}}^{t_0} s(\bm{K},z,t^{\prime\prime}) dt^{\prime\prime}} \tilde{\eta}_2(\bm{K},z,t^{\prime}) dt^{\prime} \nonumber \\
  &\equiv  \mathcal{T}(\bm{K},z,t_0,t_{\rm pump})  \tilde{\psi}_2(\bm{K},z,t_{\rm pump}) + \gamma  \int_{t_{\rm pump}}^{t_0} \mathcal{T}(\bm{K},z,t_0,t^{\prime}) 
 \tilde{\eta}_2(\bm{K},z,t^{\prime}) dt^{\prime}
\end{align}
with $\mathcal{T}(\bm{K},z,t_1,t_2) = {\rm exp}[\gamma \int_{t_2}^{t_1} s(\bm{K},z,t) dt]$. Since the formation of the domain wall is a fast process, e.g. $\gamma(t_0-t_{\rm pump}) \sim 1$ (see Fig.~\ref{dw_res}), the $\mathcal{T}$-factor gives a result of order 1  for long-wavelength modes with $\xi_0^2 K^2 \leq 1$. Therefore, for a practical calculation, we neglect the $\mathcal{T}$-factor in later discussion. 

$\bullet$ {\bf Calculation of $D_{K}(t_0)$}: 
To calculate $D_{K}(t_0) \equiv \langle \tilde{\varphi}(-\bm{K},t_0) \tilde{\varphi}(\bm{K},t_0) \rangle$, we read out $\tilde{\varphi}(\bm{K},t_0)$ from 
$\tilde{\psi}_2({\bm K},z,t_0)$ by 
$\tilde{\varphi}(\bm{K},t_0) = \int u(z) \tilde{\psi}_2(\bm{K},z,t_0) dz$ with $u(z) \equiv (2\xi)^{-1} \text{sech}(z/\xi)$ being the mode amplified with time, which is localized around the domain wall. 
\begin{align}
    \tilde{\varphi}(\bm{K},t_0) &\approx \int u(z) \tilde{\psi}_2(\bm{K},z,t_{\rm pump}) dz  + 
     \int u(z) \gamma  \int_{t_{\rm pump}}^{t_0}
 \tilde{\eta}_2(\bm{K},z,t^{\prime}) dt^{\prime} dz \equiv \tilde{\varphi}_a (\bm{K},t_0) + \tilde{\varphi}_b(\bm{K},t_0). \nonumber \\
    D_{K}(t_0)  &\approx \langle \tilde{\varphi}_a (-\bm{K},t_0) \tilde{\varphi}_a (\bm{K},t_0) \rangle + \langle \tilde{\varphi}_b (-\bm{K},t_0) \tilde{\varphi}_b (\bm{K},t_0) \rangle\equiv D_{K,a}(t_0) + D_{K,b}(t_0). 
\end{align}
Here $\tilde{\varphi}_a$ and $\tilde{\varphi}_b$ come from the thermal fluctuation in the equilibrium state for $t<t_{\rm pump}$ and the thermal fluctuation accumulated in $t_{\rm pump} < t < t_0$, respectively.  Now we calculate $D_{K,a}(t_0)$ and $D_{K,b}(t_0)$.

$\bullet \star$ {\bf Calculation of $D_{K,a}(t_0)$}:
\begin{align}
l^{D-1}D_{K,a}(t_0) 
&=\int \frac{dk_z}{2\pi}   \frac{T}{E_{\text{c}}} \frac{1 
}{ 2\beta + \xi_0^2(k_z^2 + K^2)} \!\ |l\tilde{u}(k_z)|^2  \nonumber \\
 &=\int \frac{dk_z}{2\pi}   \frac{T}{E_{\text{c}}} \frac{1 
}{ 2\beta + \xi_0^2(k_z^2 + K^2)} \!\ \Bigg|\int \frac{dz}{2\xi} 
{\rm sech}\big(\frac{z}{\xi}\big) e^{ik_z z} \!\ \Bigg|^2  \nonumber \\
&= \int \frac{dk_z}{2\pi}   \frac{T}{E_{\text{c}}} \frac{\pi^2}{4}\frac{ \text{sech}^2(k_z\xi \pi/2) }{ 2\beta + \xi_0^2(k_z^2 + K^2)} \nonumber \\
&= \frac{T}{E_{\text{c}}\xi|\alpha|} \frac{\pi^2}{4} \frac{1}{\sqrt{(2\beta + \xi_0^2K^2)/|\alpha|}} \int \frac{dq_z}{2\pi}   \frac{ \text{sech}^2(q_z \sqrt{(2\beta + \xi_0^2K^2)/|\alpha|} \pi/2 ) }{ 1 + q_z^2 } 
    \label{DK} 
\end{align}  
with $k_z \xi \equiv q_z\sqrt{(2\beta + \xi_0^2 K^2)/|\alpha|}$. Due to a factor of 
$e^{-2\xi_0^2 \gamma t K^2 }$ in the integrand of 
Eq.~(\ref{C_1}), $C_1({\bm R},t)$ in the limit 
of $\gamma t \gg 1/|\alpha|$ is dominated by 
smaller $K$, and only those $D_{K}(t_0)$ in 
$\xi_0^2K^2 \ll |\alpha|$ matters in the long-time 
behavior of $C_1({\bm R},t)$.
Besides, since $2\beta<|\alpha|$ and 
$\sqrt{(2\beta + \xi_0^2K^2)/|\alpha|} \leq 1$,
the $q_z$-integral in 
Eq.~(\ref{DK}) always gives a result on the order of $1$. See the estimations below).
\begin{align}
    \int   \frac{ \text{sech}^2(q_z)}{1+q_z^2} dq_z \approx 1.41\ \ < \ \ \int   \frac{ \text{sech}^2(q_z\sqrt{(2\beta + \xi_0^2K^2)/|\alpha|} \pi/2 )}{1+q_z^2} dq_z \ \ < \ \ \int   \frac{1}{1+q_z^2} dq_z = \pi.
\end{align}
We approximate it to be $\pi$ for convenience and obtain
\begin{align}
    l^{D-1} D_{K,a}(t_0) \approx \frac{T}{E_{\text{c}}\xi|\alpha|} \frac{\pi^2}{8} \frac{1}{\sqrt{(2\beta + \xi_0^2K^2)/|\alpha|}}. 
    \label{DK_approx}
\end{align}

$\bullet \star$ {\bf Calculation of $D_{K,b}(t_0)$}:
\begin{align}
    l^{D-1} D_{K,b}(t_0) &= \gamma \int_{t_{\rm pump}}^{t_0} dt \ \ \int dz \ \  \frac{2T(z,t)}{ E_{\rm c}} u^2(z) \stackrel{z_{\rm p} \gg \xi}{\approx} \gamma \int_{t_{\rm pump}}^{t_0} dt \ \ \int dz \ \  \frac{2T(0,t)}{ E_{\rm c}} u^2(z)  = \gamma \int_{t_{\rm pump}}^{t_0} dt \ \   \frac{T(0,t)}{ E_{\rm c} \xi } \equiv D_b. 
\end{align}

The calculation gives a constant result independent of $K$.

Substituting $D_K(t_0) = D_{K,a}(t_0)+D_{K,b}(t_0)$ in Eq.~(\ref{C_1}), we obtain $C_1({\bm R},t) = C_{1,a}(\bm{R},t) + C_{1,b}(\bm{R},t)$ with 

\begin{align}
    C_{1,a}(\bm{R},t)  &= \int\frac{d^{D-1} \bm{K}}{(2\pi)^{D-1}}   
    l^{D-1} D_{K,a}(t_0)
    e^{2(|\alpha|-2\beta)\gamma t} e^{-2\xi_0^2 \gamma t K^2 } e^{i\bm{K}\cdot \bm{R}}  \nonumber \\
    &= \frac{T}{E_{\text{c}}\xi_0} e^{2(|\alpha|-2\beta)\gamma t} \frac{\pi^2}{8}  \int\frac{d^{D-1} \bm{K}}{(2\pi)^{D-1}}   
   \frac{1}{\sqrt{(2\beta + \xi_0^2K^2)}}
     e^{-2\xi_0^2 \gamma t K^2 } e^{i\bm{K}\cdot \bm{R}}.
     \label{C_1_general}
\end{align}

\begin{align}
    C_{1,b}(\bm{R},t)  &= \int\frac{d^{D-1} \bm{K}}{(2\pi)^{D-1}}   
    l^{D-1} D_{K,b}(t_0)
    e^{2(|\alpha|-2\beta)\gamma t} e^{-2\xi_0^2 \gamma t K^2 } e^{i\bm{K}\cdot \bm{R}}  \nonumber \\
    &= D_b e^{2(|\alpha|-2\beta)\gamma t} \int\frac{d^{D-1} \bm{K}}{(2\pi)^{D-1}} 
        e^{-2\xi_0^2 \gamma t K^2 } e^{i\bm{K}\cdot \bm{R}} \nonumber \\
    &= \frac{D_b |\alpha|^{D/2-2}}{\sqrt{|\alpha|}\xi_0^{D-1}} \frac{|\alpha|}{\sqrt{8\pi|\alpha|\gamma t}^{D-1}} e^{2(|\alpha|-2\beta)\gamma t} e^{-\frac{R^2}{8\xi_0^2 \gamma t}}.
    \label{C_1_b}
\end{align}
By the definition of $D_b$, the factor $\frac{D_b |\alpha|^{D/2-2}}{\sqrt{|\alpha|}\xi_0^{D-1}} \sim \zeta$. The integration of $C_{1,a}(\bm{R},t)$ could be evaluated in some relevant limits. In the following, we discuss $C_{1,a}(\bm{R},t)$ in these limits. 

$\bullet$ {\bf $\gamma t \gg 1/\beta $ limit}: 
In this limit, 
the $K$ integral is dominated by those $K$ in $\xi_0^2K^2 \ll \beta$ due to a factor of 
$e^{-2\xi_0^2 \gamma t K^2}$ in the integrand of 
Eq.~(\ref{C_1_general}). Thus, we can estimate the 
integral as follows
\begin{align}
    C_{1,a}(\bm{R},t) &\approx  \frac{T}{E_{\text{c}}\xi_0} e^{2(|\alpha|-2\beta)\gamma t} \frac{\pi^2}{8}  \int\frac{d^{D-1} \bm{K}}{(2\pi)^{D-1}}   
   \frac{1}{\sqrt{2\beta }}
     e^{-2\xi_0^2 \gamma t K^2 } e^{i\bm{K}\cdot \bm{R}} \nonumber \\
    &=  G e^{2(|\alpha|-2\beta)\gamma t} \frac{\pi^2}{8} \frac{1}{\sqrt{2\beta}} \frac{1}{\sqrt{8\pi\gamma t}^{D-1}} e^{-\frac{R^2}{8\xi_0^2\gamma t}} \nonumber \\
    &\approx \zeta|\alpha| \frac{\sqrt{|\alpha|/\beta}}{\sqrt{8\pi|\alpha|\gamma t}^{D-1}} e^{2(|\alpha|-2\beta)\gamma t} e^{-\frac{R^2}{8\xi_0^2 \gamma t}}.
    \label{C_1_res}
\end{align}
Combining with Eqs.~(\ref{C_2_res}, \ref{C_1_b}), 
we obtain Eq.~(6) in the main text.  

$\bullet$ {\bf U(1) symmetric case 
in the 3D system ($\beta=0$ and $D=3$)}: 
The 3D ${\bm K}$-integral in 
Eq.~(\ref{C_1_general}) can be exactly evaluated in the polar coordinate with the Bessel functions,  
\begin{align}
    C_{1,a}(\bm{R},t) &=  \frac{T}{E_{\text{c}}\xi_0^2} e^{2|\alpha|\gamma t} \frac{\pi^2}{8}  \int_0^{+\infty} \frac{dK}{2\pi}  e^{-2\xi_0^2 \gamma t K^2 } J_0(KR) = 
   \frac{T}{E_{\text{c}}\xi_0^2} e^{2|\alpha|\gamma t} \frac{\pi^2}{16} \frac{1}{\sqrt{8\pi\xi_0^2\gamma t}} I_0(\frac{R^2}{16\xi_0^2 \gamma t}) e^{-\frac{R^2}{16\xi_0^2 \gamma t}} \nonumber \\
 &\approx G e^{2|\alpha|\gamma t} \frac{1}{\sqrt{8\pi\gamma t}} I_0(\frac{R^2}{16\xi_0^2 \gamma t}) e^{-\frac{R^2}{16\xi_0^2 \gamma t}} 
 = \zeta|\alpha| e^{2|\alpha|\gamma t} \frac{1}{\sqrt{8\pi|\alpha|\gamma t}} I_0(\frac{R^2}{16\xi_0^2 \gamma t}) e^{-\frac{R^2}{16\xi_0^2 \gamma t}}. 
    \label{C_1_I0}
\end{align}
Here $J_0(x)$ is the Bessel function of the first kind, and $I_0(x)$ is the modified Bessel function of the first kind. A comparison between Eq.~(\ref{C_1_I0}) and Eq.~(\ref{C_2_res}) with $D=3$ and $\beta=0$ 
shows that $C_{1,a}({\bm R},t)$ always dominates over $C_{1,b}({\bm R},t)$ and $C_{2}({\bm R},t)$ in the long-time limit, leading to Eq.~(4) in the main text. From Eq.~(\ref{C_1_I0}), 
$\lim_{x\rightarrow 0} I_0(x) = 1$, and 
$\lim_{x\rightarrow +\infty} = e^x / \sqrt{2\pi x}$, we can see that 
the spatial dependence of the correlation in the long-time limit 
($\gamma t\gg 1$) shows a crossover from an exponential decay to 
a power-law decay; 
\begin{align}
C_{1,a}({\bm R},t) \propto 
\begin{cases}
    e^{-R^2/16\xi_0^2 \gamma t},   & \text{for} \!\ \!\  
    R\ll  4\xi_0\sqrt{\gamma t} \\ 
    \frac{4\xi_0\sqrt{\gamma t}}{R},       & \text{for} \!\ \!\ 
    R\gg 4\xi_0\sqrt{\gamma t}. 
\end{cases}
\label{crossover}
\end{align}
The crossover is consistent with 
the power-law spatial correlation $\langle \psi_{2}(\bm{r}) \psi_{2}(\bm{r^{\prime}}) \rangle \sim 1/|\bm{r} - \bm{r^{\prime}}|$ at 
long distances in the initial equilibrium fluctuation in the $3D$ systems. 

\subsection{Determine the crossover time $t_{\text{c}}$.}
When $|\psi_2({\bm r},t)|^2 = C(0,t){\rm sech}^2(z/\xi)$ 
becomes on the 
same order as $|\psi_1({\bm r},t)|^2=|\alpha|\tanh^2(z/\xi)$, 
the first-stage dynamics 
crossover into the second-stage dynamics. The crossover time scale 
$t_c$ is defined as $C({\bm 0},t_c)=C_{1,a}({\bm 0},t)+C_{1,b}({\bm 0},t)+C_{2}({\bm 0},t) \sim C_{1,a}({\bm 0},t)+2C_{2}({\bm 0},t) =|\alpha|$.
Note that 
the integrals in Eq.~(\ref{C_1_general}) can be calculated exactly at $\bm{R}=0$ for both 2D and 3D cases, 
 \begin{align}
    C_{1,a}(0,t) = 
    \begin{cases}
    \zeta |\alpha| e^{2|\alpha|\gamma (t-t_0)} \frac{\pi\sqrt{\pi}}{32}\frac{1}{\sqrt{2|\alpha|\gamma (t-t_0)}} (1-\text{erf}(\sqrt{4\beta\gamma (t-t_0)})),   & \text{in 3D systems,} \\
    \zeta |\alpha| e^{2(|\alpha|-\beta)\gamma (t-t_0)}\frac{\pi}{16}K_0(2\beta\gamma (t-t_0)),              & \text{in 2D systems.}
\end{cases}
\label{C_1(0,t)}
\end{align}
Here $\text{erf}(x)$ and $K_0(x)$ are the error function and the modified Bessel function of the second kind, respectively. For the long-time limit ($\gamma |t-t_0| \gg 1/(|\alpha|-2\beta)$), $C_2(0,t)$ is evaluated in Eq.~(\ref{C_2_res}).


In the general case, we can use Eq.~(\ref{C_2_res},\ref{C_1(0,t)}) to calculate $C(0,t_{\text{c}})$ and determine $t_{\text{c}}$ by $C(0,t_{\text{c}})=|\alpha|$. The results of $\gamma t_{\text{c}}$ mentioned in the main text are approximated results in the limits discussed.

\section*{\textsl{\NoCaseChange{Supplementary Note 3: }}The second-stage dynamics}
When $t$ reaches the crossover time $t_c$, randomly distributed positive-$\psi_2$ and negative-$\psi_2$ domains are located on
the $z=0$ interface. Since $\psi$ is a continuous function of $\bm{r}$, the boundary between the positive-$\psi_2$ and negative-$\psi_2$ domains is nothing but the zero of $\psi_2(\bm{R},z=0)$. Furthermore, $\psi_1(\bm{R},z=0)$ remains to be $0$. As a result, $|\psi|=0$ on the boundary of the domains on the $z=0$ interface, which suggests the existence of topological defects. Writing $\psi = |\psi|e^{i\theta}$, the profile of $\theta$ around these defects could be analyzed as follows. On the interface, $\theta(\bm{R},z=0)$ takes the value of $\pi/2$ and $3\pi/2$ in the positive-$\psi_2$ and negative-$\psi_2$ domains. Far away from the interface, the boundary condition at $z=\pm \infty$ set $\psi_1 = \pm \sqrt{|\alpha|}$ and $\psi_2=0$, i.e. $\theta = 0/\pi$ at $z=+\infty/-\infty$. 

\begin{figure}[h]
    \centering
    \includegraphics[width= 7in]{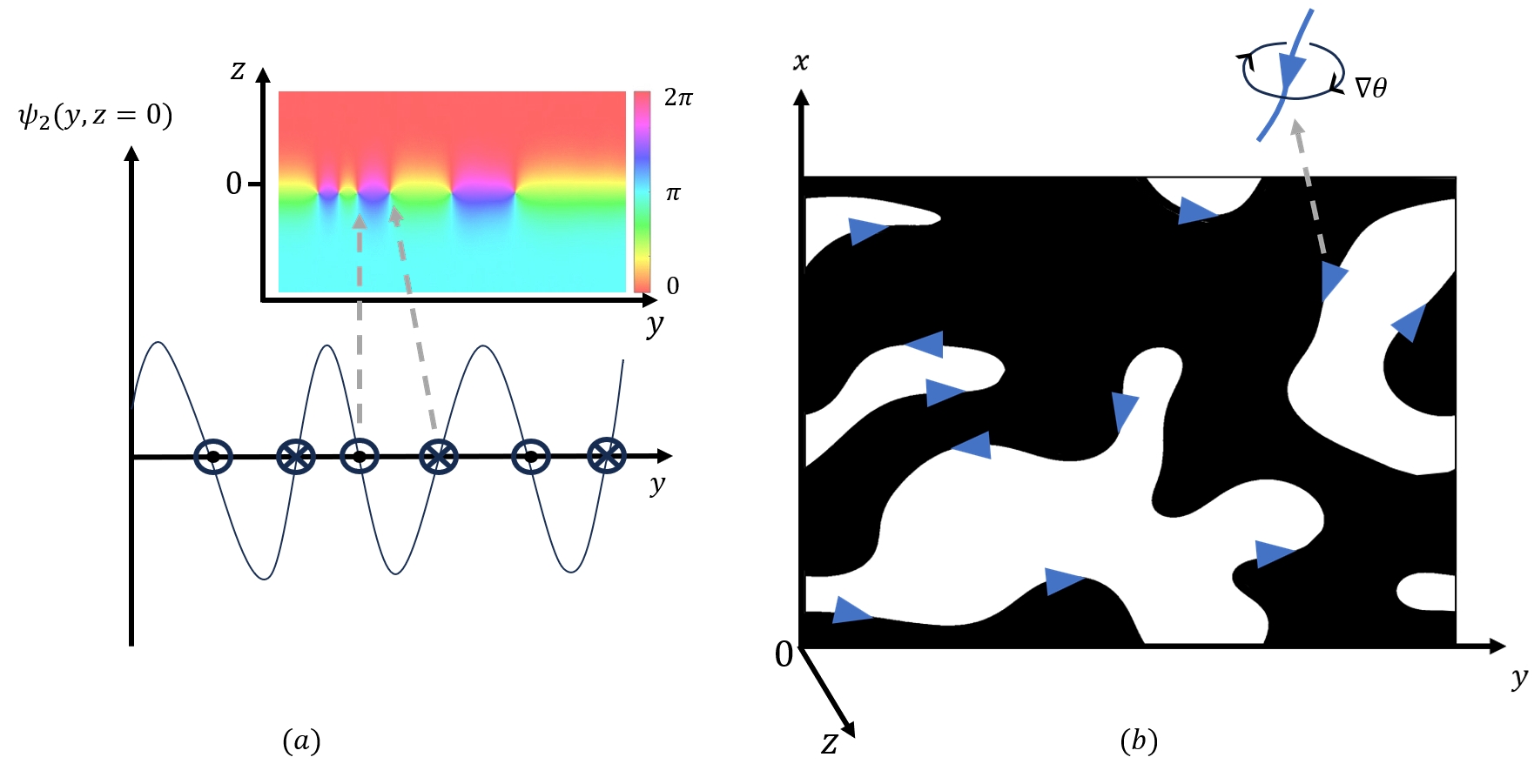} 
    \caption{  Schematic plot of the profile of $\psi_2$ after the first-stage dynamics on the $z=0$ interface. (a) On the 1D interface of 2D systems, the nodes of $\psi_2$ generate a vortex-antivortex-pair array denoted by $\otimes/\odot$. The inset is the corresponding profile of $\theta$. (b) On the 2D interface of 3D systems, $\psi_2$ are positive/negative in the black/white regions. The borders between the black and white regions, where $\psi=0$, generate vortex strings. The vortex strings have specific orientations marked by the blue arrows. The orientations are defined by a right-hand rule concerning the direction of $\nabla \theta$ around the strings, illustrated in the inset.  }
    \label{psi2_profile}
\end{figure}

Based on the analysis above, on the 1D interface of 2D systems, a clockwise closed path around a zero of $\psi_2$ obtains a phase winding $2\pi$($-2\pi$) if the slope $\partial_y \psi_2(y,z=0)$ is negative(positive). Therefore, a zero of $\psi_2$ with a negative(positive) $\partial_y \psi_2$ is nothing but a vortex(antivortex). Since the positive-$\psi_2$ and negative-$\psi_2$ domains appear alternatively along the $y$ axis, the vortex and antivortex also appear alternatively. In other words, a vortex(antivortex) has to lie in between two neighboring antivortex(vortex). See Fig.~(\ref{psi2_profile})(a) for a schematic plot of the $\psi_2(y,z=0)$ function, the $\theta(y,z)$ profile and the vortex/antivortex. On the 2D interface of 3D systems, the zeros of $\psi_2$ form 1D lines called vortex strings. The orientations of the vortex strings are defined by a right-hand rule with respect to the direction of $\nabla \theta$ around the strings, as shown in \fig{psi2_profile}(b). In any intersecting 2D plane parallel to $z$, $\theta$ distributes in a similar way as in the 2D systems, and vortex/antivortex appears alternatively on the $z=0$ interface in the plane. Since the positive-$\psi_2$ and negative-$\psi_2$ domains are closed, the line of zeros of $\psi_2$, which is nothing but the boundary between the domains, are closed loops. Therefore, the vortex strings form loops.


In the second stage dynamics of 2D systems, the vortex moves under an attractive force from 
neighboring antivortices, a frictional force from the damping term $\gamma^{-1}\partial_t \psi$ and Langevin force from the thermal noise $\eta$ in the ``model A" dynamics. Due to the attractive force, 
the neighboring vortex and antivortex move toward each other and annihilate. The annihilation processes 
lead to a coarsening phenomenon of the topological defects in the 
interface~\cite{bray2002theory}. 
In the second stage dynamics of 3D systems, the vortex string contracts under a tension corresponding to the local curvature, a frictional force from the damping term $\gamma^{-1}\partial_t \psi$ and Langevin force from the thermal noise $\eta$ in the ``model A" dynamics. It contracts and annihilates, leading to a coarsening phenomenon. In this section, we derive an equation of motion of the defects,  
and analyze quantitatively the coarsening process in the second stage dynamics.

\setcounter{subsection}{0}
\subsection{Spatial profile of the topological defect without the $U(1)$ symmetry ($\beta\ne 0$).}
\label{secIIIA}
We first determine a spatial profile of the defect to derive the equation of motion of a topological defect from the ``model A" dynamics. 
For the $U(1)$-symmetric case ($\beta=0$), the spatial profile of the vortex or vortex string 
is well known~\cite{bray2002theory}. In this section, we will determine the 
spatial profile of the vortex without the $U(1)$ symmetry ($\beta \ne 0$). 
Especially, we will clarify the spatial profile far from the vortex core,  
and the profile near the core. In the following, we will focus on a 2D spatial profile of a vortex pinned in the 1D interface of the 2D systems. 
A vortex loop in the 3D systems can be regarded as a stack of the 
2D spatial profile of the vortex along the third direction. We also limit ourselves to the `nearly U(1) symmetric' case ($\beta\ll |\alpha|$) for analytical convenience.

The spatial profile of a single vortex forms a local minimum of the Ginzburg-Landau free energy functional of $\psi({\bm r})=|\psi(\bm{r})| e^{i\theta({\bm r})}$, i.e. Eq.~(\ref{free_energy_sm}). In the limit of $\beta\ll |\alpha|$, 
it is a good approximation to neglect the amplitude dynamics and fix the amplitude $\psi=\sqrt{|\alpha|}$, and study the free energy 
functional of $\theta({\bm r})$ with ${\bm r}\equiv (y,z)$; 
\begin{align}
    F &= E_{\text{c}}|\alpha| \int d^{2} \bm{r} \ \  \xi_0 ^2  |\nabla \theta|^2 + 2\beta \sin^2 \theta.
    \label{free_energy_phase}
\end{align} 
Or equivalently, $\delta F /\delta \theta =0$; 
\begin{align}
    \xi_0^2 \nabla^2 \theta = \beta \sin ( 2\theta ). 
    \label{sg}
\end{align}
With the boundary condition $\lim_{z\rightarrow +\infty} \theta_0 = 0$, $\lim_{z\rightarrow -\infty} \theta_0 = \pi$,
the equation leads to the following two solutions, 
\begin{align}
    \theta_0^{\pm}(z) = \text{arg}(\sinh (\sqrt{2\beta}z/\xi_0) \pm i)
\end{align}
These 
solutions are nothing but the 
Sine-Gordon solitons. The soliton has a finite  energy per unit length along the $y$-direction, 
\begin{align}
    \sigma_0 = E_{\text{c}} |\alpha| \int dz \ \ \xi_0^2|\nabla\theta_0^{\pm}|^2 + 2\beta\sin^2\theta_0^{\pm}  = 4 E_{\text{c}}\xi_0 |\alpha|\sqrt{2\beta} . 
\end{align}
In the second-stage dynamics, $\theta(y,z)=\theta^{+}_0(z)$ and $\theta(y,z)=\theta^{-}_0(z)$ appears alternatively along the $y$-direction, connected by vortices.
Consider a vortex at $(y,z)=(0,0)$. The 2D spatial profile of the vortex 
can be described by combining the two soliton solutions at $y=0$;
$\theta(y,z)=\theta_0^{+}(z)$ for $y\rightarrow +\infty$ and $\theta(y,z)=\theta_0^{-}(z)$ for $y\rightarrow -\infty$~\cite{Eto2023-ky}. 
Such spatial profile of the vortex is generally given by 
\begin{eqnarray}
\theta(y,z) = 
\begin{cases}
\theta^+_0(z) + s(y,z)  & {\rm for} \ \ \!\ y>0, \\ 
\theta^{-}_0(z) + s(y,z) & {\rm for} \ \ \!\ y<0. \\
\end{cases} 
\label{ansatz}
\end{eqnarray}

Note that at $y$ approaches $\pm\infty$, the vortex profile approaches the Sine-Gordon soliton profile due to small but finite $\beta$. At $|z|$ approaches $+\infty(-\infty)$, $\theta = 0(\pi)$. Therefore, $s$ in Eq.~(\ref{ansatz}) satisfies the boundary condition
$\lim_{y\rightarrow \pm\infty} s(\bm{r}) = 0$. $s$ and $\lim_{z\rightarrow \pm\infty} s(\bm{r}) = 0$. 
Thus, we can determine a form of $s(y,z)$ for the larger $|y|$ by substituting $\theta(\bm{r}) = \theta_0(z) + s(\bm{r})$ into Eq.~(\ref{sg}) and linearizing the equation in small $s$;  
\begin{align}
    \xi_0^2 \nabla^2 s = 2\beta \cos (2\theta_0) s.  
    \label{s_eq}
\end{align}
The linearized equation is decomposed into an operator in $y$ and an operator 
in $z$, 
\begin{align}
    \frac{\xi_0^2}{2\beta} \partial_y^2 s(y,z) = [-\frac{\xi_0^{2}}{2\beta}\partial_z^2 - \frac{2}{\cosh^2(\sqrt{2\beta} z /\xi_0)} + 1 ]s(y,z) 
    \equiv 
    [H(z^{\prime};2) + 1]s(y,z),
\end{align}
with $z^{\prime}\equiv \sqrt{2\beta} z /\xi_0$. $H(z^{\prime};\lambda)$ is nothing but the 1D  Hamiltonian with the Poschl-Teller potential hole mentioned in Eq.~(\ref{sech_eig}). Thus, 
in terms of the eigenmodes of $H(z^{\prime};2)$, $s(y,z)$ can be expanded as
\begin{align}
s(y,z) = c_0 h_0(z^{\prime}) + 
\int dk^{\prime} \!\ c(k^{\prime}) \!\ 
 e^{-\sqrt{{k^{\prime}}^2+1}  |y^{\prime}|} 
 h(k^{\prime},z^{\prime}). 
 \label{s_sol}
\end{align}
with $k^{\prime}\equiv \xi_0 k /\sqrt{2\beta}$, $y^{\prime} = y\sqrt{2\beta}/\xi_0$, $z^{\prime}=z\sqrt{2\beta}/\xi_0$,  
and $c^*(k^{\prime})=c(-k^{\prime})$. Here 
$h_0(z^{\prime}) \propto {\rm sech}(z^{\prime})$ is the 
bound-state eigenmode of $H(z^{\prime},2)$ 
with negative eigenenergy $-1$. 
$h(k^{\prime},z^{\prime}) \propto \big(1+i/(k^{\prime})  
\tanh(z^{\prime})\big) e^{ik^{\prime} z^{\prime}}$ 
is a continuum eigenstate of $H(z^{\prime},2)$ 
with positive eigenenergy $(k^{\prime})^2$. 
From the boundary condition 
at $y=\pm \infty$, $c_0=0$.
$c(k^{\prime})$ is chosen to satisfy the boundary condition $ \lim_{z\rightarrow \pm\infty} s =0$.
The energy for the Sine-Gordon soliton with a single vortex 
is given by 
\begin{align}
E(R_{\rm d}) = 
\int^{R_{\rm d} /2}_{-R_{\rm d} /2} dy \!\ \sigma(y), \nonumber 
\end{align} 
with 
\begin{align}
\sigma(y) = E_c |\alpha| 
\int dz \ \ \xi_0^2|\nabla(\theta_0^{\pm} + s)|^2 + 2\beta\sin^2(\theta_0^{\pm}+s).
\nonumber   
\end{align}
 Here $R_{\rm d}$ is a size of the vortex along the 1D interface, and
$\theta^{+}_0$ and $\theta^{-}_0$ in the right hand side are 
for an energy density per length, $\sigma(y)$, 
in the $y>0$ and $y<0$ regions, respectively. 
Since the size of the vortex is constrained by its 
neighboring antivortex, whose 2D spatial profile 
takes the same form as Eq.~(\ref{ansatz}) with $y\rightarrow -y$, 
$R_{\rm d}$ can be also regarded as a distance between the two defects 
along the 1D interface. 

For large $|y|$, the energy density per length, $\sigma(y)$,
can be further evaluated by an expansion in small 
$s(y,z)$. The zeroth order term in $s$ is the energy density of the domain wall without the vortices, $\sigma_0$. The first-order term in $s$ vanishes because the soliton solution $\theta^{\pm}_0$ is a local energy minimum of the free energy. The second-order term gives the energy density of a vortex per length for large $|y|$;
\begin{align}
    \sigma(y) &\approx \sigma_0 + E_{\text{c}} |\alpha|\int dz \ \ \xi_0^2(\partial_y s)^2 + \xi_0^2(\partial_z s)^2 + 2s^2\beta\cos(2\theta_0) \nonumber \\
    &= \sigma_0 + E_{\text{c}} |\alpha|\int dz \ \ \xi_0^2(\partial_y s)^2 + \xi_0^2\partial_z(s\partial_z s) - \xi_0^2( s \partial_z^2 s) + 2s^2\beta\cos(2\theta_0) \nonumber \\
   & = \sigma_0 + E_{\text{c}} |\alpha|\int dz \ \ \xi_0^2(\partial_y s)^2 + \xi_0^2(s\partial_y^2 s) +  \xi_0^2\partial_z(s\partial_z s)  \nonumber \\
   &= \sigma_0 + 
   2 E_{\rm c}|\alpha| \xi_0\sqrt{2\beta} \int |c(k^{\prime})|^2
   ( k^{\prime 2} + 1)e^{-2\sqrt{k^{\prime 2}+2} |y^{\prime}|} \ dk^{\prime} +  E_{\text{c}}|\alpha| \xi_0^2 (s\partial_z s |^{z=+\infty}_{z=-\infty} ) \nonumber \\
   &= \sigma_0 + 
   2 E_{\rm c}|\alpha| \xi_0\sqrt{2\beta} \int |c(k^{\prime})|^2
   ( k^{\prime 2} + 1)e^{-2\sqrt{k^{\prime 2}+2} |y|\sqrt{2\beta}/\xi_0} \ dk^{\prime} 
   \label{wall_energy}
\end{align}
From the second line to the third line, we use Eq.~(\ref{s_eq}). 
From the third line to the fourth line, we use 
Eq.~(\ref{s_sol}) and an orthonormal condition of 
$h(k^{\prime},z^{\prime})$; $\int dz^{\prime} h(k^{\prime}_1,z^{\prime}) h^*(k^{\prime}_2,z^{\prime}) 
= \delta(k^{\prime}_1-k^{\prime}_2)$. From the fourth line to the 
fifth line, we use the boundary condition 
$ \lim_{z\rightarrow \pm\infty} s = 0$. 

The free energy can be decomposed into an energy of 
the vortex, $E_d(R_{\rm d})$, and the energy of the 
soliton, $\sigma_0 R_{\rm d}$, 
\begin{align}
E(R_{\rm d}) = \sigma_0 R_{\rm d} + \int^{R_{\rm d}/2}_{-R_{\rm d}/2} dy \ \ [\sigma(y) -\sigma_0] 
\equiv \sigma_0 R_{\rm d} + E_{d}(R_{\rm d}). 
\end{align}
By Eq.~(\ref{wall_energy}), $\sigma(y)-\sigma_0$ decays exponentially for larger $|y|$. Therefore, the energy of the vortex is convergent in the larger $R_{\rm d}$ limit;
$\lim_{R_{\rm d}\rightarrow +\infty} E_{\text{d}}(R_{\rm d}) = E_0 \propto E_{\text{c}}|\alpha|\xi_0^2$. 

Since $R_{\rm d}$ can also be regarded as 
the distance between vortex and antivortex along the 1D interface, 
$dE_{\rm d}/dR_{\rm d}$ gives an estimate of an action-reaction force 
between the two defects. The force is attractive and it is 
dependent on the distance $R_{\rm d}$. For larger $R_{\rm d}$ with 
$R_{\rm d} \gg \xi_0/\sqrt{2\beta}$, the magnitude of the force decays exponentially, 
\begin{align}
    F_{\rm d} = \frac{dE_{\text{d}}}{dR_{\rm d}} = \sigma(R_{\rm d}/2) \sim  E_{\rm c} |\alpha| \xi_0 \sqrt{2\beta} e^{-R_{\rm d}\sqrt{2\beta}/\xi_0}.
    \label{dEdL}
\end{align}

We have discussed the energy and attraction force of the vortex far away 
from the vortex core, $R_{\rm d} \gg \xi_0/\sqrt{2\beta}$. Next, we will 
discuss the energy and force of the vortex near the vortex core 
$R_{\rm d}\ll \xi_0/\sqrt{2\beta}$. Let us rewrite Eq.~(\ref{free_energy_phase}) in the polar coordinates,
\begin{align}
    F &= E_{\text{c}}|\alpha| \int^{R_{\rm d}/2}_0 rdr \int^{2\pi}_{0} d\phi \ \ 
    \Big[  \xi^2_0 
    (\frac{\partial \theta}{\partial r})^2 +  \frac{\xi^2_0}{r^2}(\frac{\partial \theta}{\partial \phi})^2 + 2 \beta \sin^2 \theta \Big] 
    \equiv E_{\text{c}}|\alpha| \int^{R_{\rm d}/2}_0 dr \sigma(r).
\end{align}
For $r \ll \xi_0/\sqrt{2\beta}$, the vortex profile approaches the U(1) symmetric vortex, where 
$\sigma(r)$ is dominated by the gradient term;
\begin{align}
\sigma(r) \equiv \int^{2\pi}_{0} d\phi \ \ 
    \frac{\xi^2_0}{r}\Big[  r^2
    (\frac{\partial \theta}{\partial r})^2 + (\frac{\partial \theta}{\partial \phi})^2 + \frac{2\beta r^2}{\xi^2_0} \sin^2 \theta \Big] \approx \int^{2\pi}_{0} d\phi \ \ 
    \frac{\xi^2_0}{r}\Big[  r^2
    (\frac{\partial \theta}{\partial r})^2 + (\frac{\partial \theta}{\partial \phi})^2 \Big]. 
\end{align}
Thus, for $R_{\rm d}\ll \xi_0/\sqrt{2\beta}$, 
$E_d(R_{\rm d}) \approx E_{\text{c}}|\alpha| \int d^{2} \bm{r} \!\ \xi_0 ^2  |\nabla \theta|^2 \simeq 
|\alpha|E_{\text{c}} \xi_0^2 \ln(R_{\rm d}/\xi)$. Here the coherence length $\xi$ plays the role of the core size of the vortices. 
Thereby, the attractive force decays with a power law 
for $R_{\rm d}\ll \xi_0/\sqrt{2\beta}$; 
\begin{align}
F_{\text{d}}(R_{\rm d}) =  \frac{dE_{\rm d}}{dR_{\rm d}} \sim  \frac{\xi_0^2 |\alpha| E_{\text{c}}}{R_{\rm d}}, \label{dEdL2}
\end{align}
which is the same as the $U(1)$-symmetric case.

\begin{figure}[h]
    \centering
    \includegraphics[width= 4in]{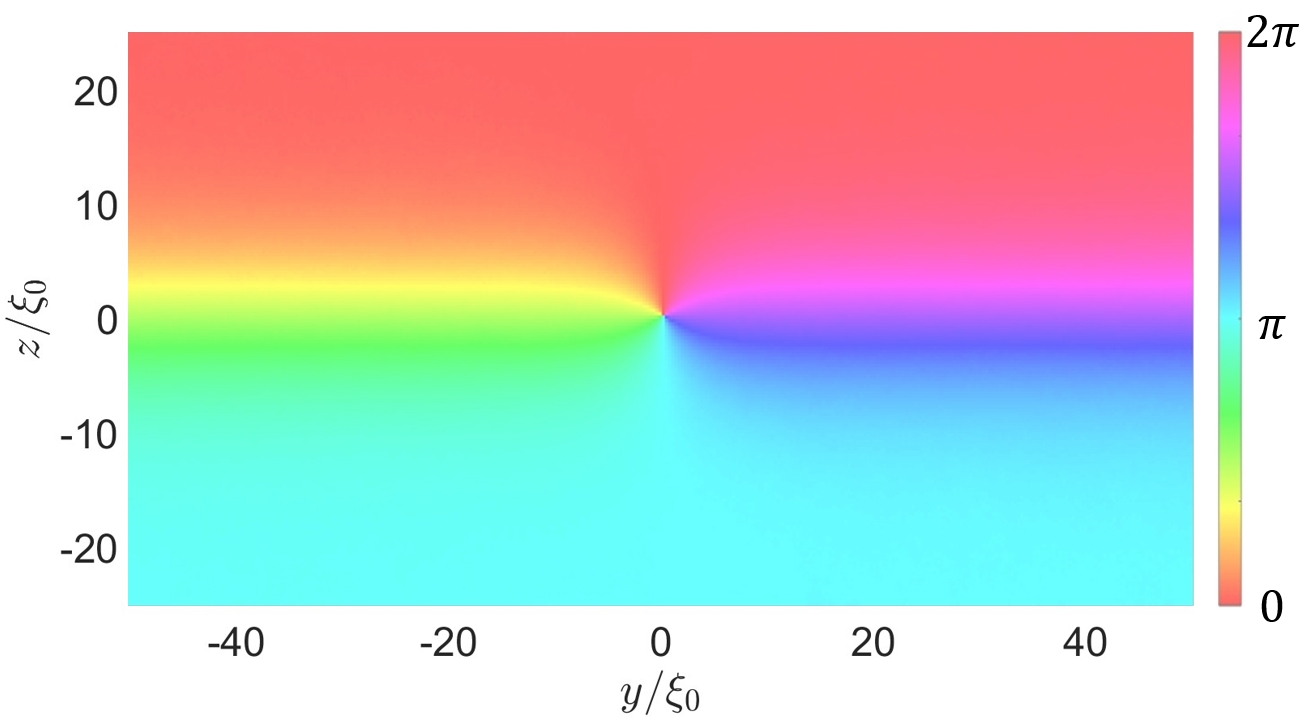} 
    \caption{Numerical solution of a vortex of the free energy Eq.~(\ref{free_energy_sm}) for the 2D spatial 
    profile of the phase with a vortex core at $(y,z)=(0,0)$. The numerical solution is obtained with $|\alpha|=-1$, $\beta = 0.02$. i.e. $\xi_0/\sqrt{2\beta} = 5\xi_0$.}
    \label{Z2_vortex}
\end{figure}

The 2D spatial profile of the vortex discussed so far has a characteristic length scale, $\xi_0/\sqrt{2\beta}$, which separates the long-length regime with the exponential decay of the attractive force, and the short-length regime with the power law of the force. We confirmed that numerical solutions of the vortex are consistent with these analytical results in the two limits; Fig.~\ref{Z2_vortex}, verifying the conclusion above. In Fig.~\ref{Z2_vortex}, we can see that the defect behaves like a circular symmetric vortex in the shorter length regions $|y| \ll \xi_0/\sqrt{2\beta}$, while in the long-length region $|y| \gg \xi_0/\sqrt{2\beta}$, it behaves as a Sine-Gordon soliton 
with a translation symmetry along the $y$-direction.

In the 2D interface of the 3D systems, the vortex forms a vortex string, and the cross-section of the string contains the vortices discussed above. For a vortex-string loop with diameter $\sim R_{\text{d}}$, its energy is estimated as $\sim R_{\text{d}}E_{\rm d}(R_{\rm d})$ where $E_{\rm d}(R_{\rm d})$ takes the expression obtained above.

\subsection{Force balance of the defects}
In this section, we start from the 2D systems and derive an equation of motion for a single vortex from the ``model A dynamics". The equation of motion becomes a force balance equation 
among the attractive force between vortex and antivortex, frictional and Langevin forces. When a pair of vortex and antivortex meet each other due to the 
attractive force, they annihilate. Events of the pair annihilations keep enlarging the mean distance between vortex and antivortex in the whole interface, leading to a coarsening phenomenon~\cite{PhysRevE.47.1525, PhysRevA.46.7765,PhysRevA.42.5865}.  From the force balance equation, we derive an equation of motion of the mean distance and estimate a growth law of the mean distance (correlation length) in the coarsening process. In 3D systems, a similar thing happens and the topological defects are vortex strings. They contract and annihilate driven by the tension that stems from the local curvature~\cite{10.1143/PTP.78.237, PhysRevA.45.657}, leading to a coarsening phenomenon. The analytical methods used in this section are mainly adopted from a review paper by  Bray~\cite{bray2002theory}. 
\subsubsection{2D systems}
\begin{figure}[h]
    \centering
    \includegraphics[width= 5in]{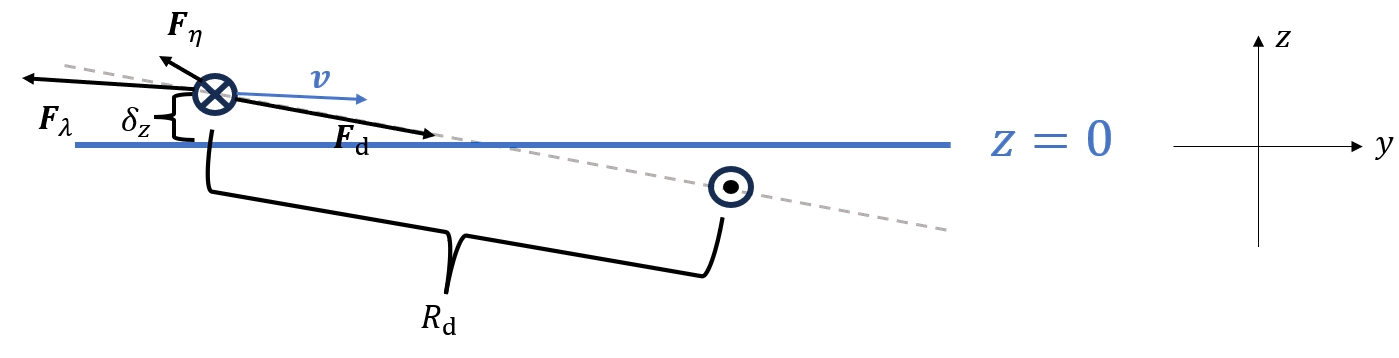} 
    \caption{A schematic picture of the force balance relation for a vortex moving with a velocity $\bm{v}$ around the interface at $z=0$. An attractive force $\bm{F}_{\text{d}}$ is along the grey dashed line, which connects the vortex ($\otimes$) and its neighboring antivortex $(\odot)$ with a distance $R_{\rm d}$. $\delta_z$ is a displacement of the vortex from $z=0$. We consider the $\delta_z/R_{\rm d} \rightarrow 0$ limit. Thus, the direction of $\bm{F}_{\rm d}$ is nearly along the interface.} 
    \label{sm_force_2d}
\end{figure}

Let us begin with the 2D system. Consider that a vortex core 
moves with a velocity $\bm{v}(t)$, represented by the $\otimes$ symbol in Fig.~(\ref{sm_force_2d}). The spatial profile of such a moving vortex is given by $\psi(\bm{r},t) = f(\bm{r}-\int^t \bm{v}(t^{\prime}) dt^{\prime})$, where $f(\bm{r})$ is given by a defect at ${\bm r}=(0,0)$ in Eq.~(\ref{ansatz}); $f({\bm r}) \equiv \sqrt{|\alpha|}{\rm exp}[i\theta({\bm r})]$. $\psi(\bm{r},t) $ satisfies the dynamical equation, 
\begin{align}
&\frac{1}{\gamma^2_0} \partial^2_t \psi 
+ \frac{1}{\gamma} \partial_t \psi = -\frac{\delta F}{\delta\psi^*} 
+ \eta,  \label{gl1} 
\end{align}
where $F[\psi]$ takes the form of Eq.~(\ref{free_energy_sm}). Here for the usage of the later section, we also added the inertia term $\gamma_0^{-2} \partial^2_t \psi$ based on the ``model A" dynamics in Eq.~(\ref{model_A_sm}) and derived a general 
equation of motion for the vortex. 

For a moving vortex with $\psi(\bm{r},t) = f(\bm{r}-\int^t \bm{v}(t^{\prime}) dt^{\prime})$,  its distance from the neighboring antivortex could be regarded as its size $R_{\rm d}$. In the second-stage dynamics, the vortex moves toward the neighboring antivortex under the attraction force and $R_{\rm d}$ gets smaller with time going by. Thus, its energy $E_{\rm d}$ changes with the following energy-dissipation rate
\begin{align}
    \frac{dE_{\rm d}}{dt} = \frac{dE_{\rm d}}{dR_{\rm d}} \frac{dR_{\rm d}}{dt}  = -\bm{F}_{\rm d} \cdot \bm{v}.
    \label{work_2d}
\end{align}

Here $\bm{F}_{\rm d}$ is the attraction force from the neighboring antivortex with a magnitude $F_{\rm d} = dE_{\rm d}/dR_{\rm d}$, as illustrated in Fig.~\ref{sm_force_2d}. We consider the vortices are nearly confined in the interface, i.e. the distance between the vortices are much larger than their displacements from the $z=0$ interface, which is illustrated by the $\delta_z / R_{\rm d} \rightarrow 0$ limit in Fig.~\ref{sm_force_2d}. Therefore, $\bm{F}_{\rm d}$ is along the $y$-direction, i.e. ${F}_{{\rm d},z}=0$ and $F_{{\rm d},y}=F_{\rm d} $.
Since $\bm{v}$ is measured from the frame of the neighboring antivortex, $v_y = -dR_{\rm d}/dt$ and $-\bm{F}_{\rm d} \cdot \bm{v} = -F_{\rm d,y} v_y = F_{\rm d} dR_{\rm d}/dt $, yielding Eq.~(\ref{work_2d}). The asymptotic forms of $F_{\rm d}$ in the two limits, $R_{\rm d}\gg \xi_0/\sqrt{2\beta}$ and $R_{\rm d}\ll \xi_0/\sqrt{2\beta}$, are discussed in Eqs.~(\ref{dEdL},\ref{dEdL2}).



According to Eq.~(\ref{gl1}), the 
energy-dissipation rate can also be given by the inertia term of the vortex, frictional force, and Langevin force acting on the vortex,  
\begin{align}
    \frac{dE_{\text{d}}}{dt} &= \int d^2 \bm{r} \ \ \frac{\delta F}{\delta \psi } \frac{\partial \psi}{\partial t} + \frac{\delta F}{\delta \psi^* } \frac{\partial \psi^*}{\partial t} \nonumber \\
    &= \int d^2 \bm{r} \ \ \Big( - \frac{E_{\rm c}}{\gamma^2_0} \frac{\partial^2 \psi^*}{\partial t^2}
    -\frac{E_{\text{c}}}{\gamma} \frac{\partial \psi^* }{\partial t} + E_{\text{c}} \eta^* \Big) \!\  \frac{\partial \psi}{\partial t} + 
    \Big( - \frac{E_{\rm c}}{\gamma^2_0} \frac{\partial^2 \psi}{\partial t^2}
    -\frac{E_{\text{c}}}{\gamma} \frac{\partial \psi}{\partial t} + E_{\text{c}} \eta \Big) \!\ \frac{\partial \psi^*}{\partial t} \nonumber \\
    &=-\sum_{i,j,k = y,z}\int d^2 \bm{r} \ \ \Big(
    -\frac{E_c}{\gamma^2_0} 
    (\partial_i\partial_k f^*) {\bm v}_i {\bm v}_k 
    + \frac{E_c}{\gamma^2_0} (\partial_i f^*) \frac{\partial {\bm v}_i}{\partial t} + \frac{E_{\text{c}}}{\gamma}  (\partial_i  f^*) \bm{v}_i + E_{\text{c}} \eta^*\Big) \!\ (\partial_j  f) \bm{v}_j \nonumber \\
    &\hspace{3cm} + \Big(
    -\frac{E_c}{\gamma^2_0} 
    (\partial_i\partial_k f) {\bm v}_i {\bm v}_k 
    + \frac{E_c}{\gamma^2_0} (\partial_i f) \frac{\partial {\bm v}_i}{\partial t} + 
    \frac{E_{\text{c}}}{\gamma} (\partial_i f) \bm{v}_i + E_{\text{c}} \eta\Big) \!\ (\partial_j  f^*) \bm{v}_j \nonumber \\
    &\equiv ( \bm{F}_{\lambda} - \bm{F}_{\eta} - {\bm A} ) \cdot \bm{v}.
    \label{dissip_2d}
\end{align}
Here ${\bm A}=\sum_i m_i a_i 
\hat{i}$ with ${\bm a}= d{\bm v}/dt$, is the inertia term of the vortex~\cite{PhysRevLett.68.1216}, 
$\bm{F}_{\lambda} = -\sum_i \lambda_i v_i \hat{i}$  is the frictional force, and $\bm{F}_{\eta}$ is the Langevin force. $m_i$ and $\lambda_i$ are the inertia mass and the friction coefficient in the $i$-direction, respectively. The inertia mass and 
friction coefficients and the Langevin force are given 
as follows;
\begin{align}
    m_i & = \frac{E_{\text{c}}}{\gamma^2_0}  \int d^2 \bm{r} \ \ 2|\partial_i f |^2, \\
    \lambda_{i} &= \frac{E_{\text{c}}}{\gamma}  \int d^2 \bm{r} \ \ 2|\partial_i f  |^2,  \\
    \bm{F}_{\eta}  &= E_{\text{c}} \int d^2 \bm{r} \ \ (\eta^* \nabla  f + \eta \nabla  f^*).
\end{align}
From the third line to the fourth line in Eq.~(\ref{dissip_2d}), we 
set $\int d^2{\bm r} (\partial_i \partial_k f^*) (\partial_j f) + {\rm c.c.}=0$, because $f({\bm r})=-f(-{\bm r})$. 


Equating Eq.~(\ref{dissip_2d}) with Eq.~(\ref{work_2d}), 
we obtain a general equation of motion
for the vortex under the force from its neighboring antivortex, 
\begin{align}
 m_y \frac{d^2 y}{dt^2} 
+ \lambda_y \frac{dy}{dt} &= F_{{\rm d},y} - F_{\eta,y}, \nonumber \\ 
 m_z  \frac{d^2 z}{dt^2} 
+ \lambda_z \frac{dz}{dt} &=  -
F_{\eta,z}. 
\label{eom_2d}
\end{align}
For the ``model A" dynamics without the inertia term ($m_i=0$), 
the equation of motion does not have the acceleration term, 
leading to the force balance equation;

\begin{align} 
\lambda_y \frac{dy}{dt} = F_{{\rm d},y} - F_{\eta,y},  \ \ 
\lambda_z \frac{dz}{dt} = - F_{\eta,z}. \label{balance}
\end{align}
Using $F_{\rm d,z}=0$ and $F_{\rm d,y} = F_{\rm d}$,  the force balance equation is solved for the coordinate of the vortex core,
\begin{align}
     z(t) &=  \int dt  \ \ \lambda_z ^{-1}  F_{\eta,z} \nonumber \\
     y(t) &= \int dt \ \ \lambda_y ^{-1}    (F_{\rm d} - F_{\eta,y} ). 
     \label{Brownian}
\end{align}
The mean displacement is 
\begin{align}
    \langle z(t) \rangle &= 0 ; \ \  \langle y(t) \rangle = \int dt \ \ \lambda_y ^{-1}    F_{\text{d}} .
    \label{mean_xy}
\end{align}

Eq.~(\ref{mean_xy}) leads to a conclusion: the mean velocity $v(R_{\text{d}})$ of a vortex in a contracting vortex-antivortex pair with distance $R_{\text{d}}$ reads 
\begin{align}
    v_y (R_{\text{d}}) = \frac{d\langle y(t) \rangle }{dt} = -\frac{dR_{\text{d}}}{dt} = \lambda_y^{-1}(R_{\text{d}})F_{\text{d}}(R_{\text{d}}).
    \label{vR}
\end{align}
For a coarsening system, a single typical time-dependent correlation length $L(t)$, which is, physically, the typical inter-defect distance or the defect size, characterizes the system. Thus, the typical velocity reads $ v(R_{\rm d} \sim L)$~\cite{bray2002theory, PhysRevA.46.7765, PhysRevE.47.1525, PhysRevE.47.1525}. Equating it to the rate of change of $L$ gives the following Eq.~(\ref{estimate_sm}), an equation to obtain the growth exponent $p$ of the growing length $L(t) \propto t^p$ ~\cite{bray2002theory, PhysRevA.46.7765, PhysRevE.47.1525, PhysRevE.47.1525}.  
\begin{align}
     \frac{dL}{dt} = \lambda_y^{-1}(L) F_{\text{d}}(L).
    \label{estimate_sm}
\end{align}
In Sec. \ref{SI_B3}, a formal of Eq.~(\ref{estimate_sm}) from Eq.~(\ref{vR}) is presented.  Eq.~(\ref{estimate_sm}) is referred to as Eq.~(5) in the main text, where $\lambda$ denotes $\lambda_y$ for simplicity.

To solve Eq.~(\ref{estimate_sm}), $\lambda_i(L)$ and $F_{\text{d}}(L)$ need to be evaluated. In $L \ll \xi_0/\sqrt{2\beta}$ limit, the defects behave like circular vortices in the $U(1)$-symmetric case. Therefore, $E_{\text{d}} (L) \sim E_{\text{c}} |\alpha| \xi_0^2 \ln (L/\xi)$, $F_{\text{d}}(L) = |dE_{\text{d}} /dL| = E_{\text{c}} |\alpha| \xi_0^2 /L$.
For $\lambda_y$, notice that for $L \ll \xi_0/\sqrt{2\beta}$,
\begin{align}
    \lambda_{i=y,z} &  \approx \frac{E_{\text{c}}|\alpha|}{\gamma}  \int d^2 \bm{r} \ \ 2|\nabla  \theta \cdot \hat{i} |^2 =  \frac{E_{\text{c}}|\alpha|}{\gamma}  \int d^2 \bm{r} \ \ |\nabla  \theta|^2 = \frac{E_{\text{d}}}{\gamma \xi_0^2} \sim  \frac{E_{\text{c}} |\alpha|}{\gamma} \ln (L/\xi) .
    \label{lambda_i_expr}
\end{align}
Thus, 
\begin{align}
     L(t) \sim  \xi_0 \sqrt{\frac{\gamma t} { \ln (|\alpha|\gamma t )} }.
     \label{Lt_tlnt}
\end{align}
The scaling is the same as the 2D $U(1)$-symmetric systems after a global thermal quench~\cite{bray2002theory}.

In the $L \gg \xi_0/\sqrt{2\beta}$ limit, the typical attraction force reads $F_{\text{d}} (L)  \sim E_{\text{c}} |\alpha| \xi_0 e^{-L\sqrt{2\beta}/\xi_0} $ according to Eq.~(\ref{dEdL}).  For $\lambda_y $, 
\begin{align}
    \lambda_y &= \frac{E_{\text{c}} |\alpha|}{\gamma} \int d^2 \bm{r} \ \ 2(\partial_y \theta)^2 = \frac{2E_{\text{c}}|\alpha|}{\gamma}  \int_{-L/2}^{L/2} dy  \int dz\ \   (\partial_y s)^2 .
   \label{lambda_2d_largeL}
\end{align}
In the $|y| \rightarrow \infty$ limit, by Eq.~(\ref{s_sol}), the integrand in the integral of $y$ approaches the following asymptotic form, 
\begin{align}
    \int dz \ (\partial_y s)^2 = \frac{\sqrt{2\beta}}{\xi_0} \int dk^{\prime} |c(k^{\prime})|^2 (k^{\prime 2} + 1) e^{-\sqrt{k^{\prime 2} +1} |y|\sqrt{2\beta}/\xi_0} .
\end{align}
As $|y|$ gets larger, the integrand show an exponential decay. Therefore, the integral of $y$ is convergent for $L\rightarrow +\infty$. In other words, for larger $L$, the leading-order contribution of $\lambda_y(L)$ is a $L$-independent constant $\lambda_y(+\infty) 
\equiv \lambda_0 \sim E_{\text{c}}|\alpha|/\gamma$. Thus, from Eq.~(\ref{estimate_sm}, \ref{dEdL}), 
\begin{align}
    \frac{dL}{dt} = \lambda_y^{-1} F_{\text{d}} \sim \gamma \xi_0 \sqrt{2\beta} e^{-L\sqrt{2\beta}/\xi_0} 
\end{align}
i.e. 
\begin{align}
    L \propto \frac{\xi_0}{\sqrt{2\beta}} \ln (2\beta\gamma t ).
    \label{Lt_lnt}
\end{align}

$\bullet$ {\bf Defect displacement in the $z$-direction and stability of the interface}: The analysis above assumes that the vortices are confined within the interface. In fact, from Eq.~(\ref{balance}), the vortices acquire a weak Brownian-like motion along the $z$-direction due to the thermal noise. Write the equation of motion in the $z$-direction in the following dimensionless form, 
\begin{align}
    \frac{d(z/\xi_0)}{d(\gamma t)} = H_z .
    \label{Brownian_z_2d}
\end{align}
Here $H_z = -\lambda_z^{-1}  F_{\eta,z}/(\xi_0 \gamma)$, whose correlator reads
\begin{align}
    \langle H_z (t) H_z(t^{\prime})  \rangle = \frac{2T}{\gamma \xi_0^2} \lambda_z^{-1}(R_{\rm d}) \delta(\gamma t - \gamma t^{\prime}).
    \label{Hz_lambda_2d}
\end{align}
Here $R_{\rm d}$ is the vortex size. By the expression of $\lambda_i$ in \equa{lambda_i_expr}, $\lambda_i = E_{\rm c}|\alpha|\gamma^{-1}h_i(R_{\rm d})$ where $h_i(R_{\rm d})=\int d^2 \bm{r} \ |\partial_i  \theta|^2$ is dimensionless. Therefore, 
\begin{align}
	\langle H_z (t) H_z(t^{\prime})  \rangle = \frac{2T}{E_{\rm c} \xi_0^2 |\alpha|} h_z^{-1}(R_{\rm d}) \delta(\gamma t - \gamma t^{\prime}) = 2\zeta h_z^{-1}(R_{\rm d}) \delta(\gamma t - \gamma t^{\prime}).
	\label{Hz_2d}
\end{align}
Replacing $R_{\rm d}$ in \equa{Hz_2d} by the typical defect size $L$, we obtain Eq.~(9) in the main text. \equa{Brownian_z_2d} and \equa{Hz_2d} imply the defect motion in the $z$-direction is negligible because $\zeta$ is small in consideration of this paper. Thus, the interface is stable. 

For the $L \ll \xi_0/\sqrt{2\beta}$ case with a nonzero $\beta$, one has $\lambda_z = \lambda_y$ and $h_z(L) = h_y(L) \sim \ln(L/\xi)$, while it is hard to evaluate $h_z(L)$ for the other cases. In the $L \ll \xi_0/\sqrt{2\beta}$ case, after replacing $R_{\rm d}$ in Eqs.(\ref{Hz_lambda_2d},\ref{Hz_2d}) by its typical value $L$ , we could solve the following scaling of $\langle z^2(L) \rangle$ 
\begin{align}
    \langle z^2(L(t)) \rangle =& 2T \int^t \lambda_z^{-1}(L(t^{\prime})) \ dt^{\prime} = 2T \int^t \lambda_y^{-1}(L(t^{\prime})) \ dt^{\prime}  \nonumber \\
    \stackrel{\text{By eq.}~(\ref{estimate_sm})}{=} & 2T \int^L F_{\rm d}^{-1}(L^{\prime}) \ dL^{\prime} = 2T \int^L \frac{L^{\prime}}{E_{\rm c} |\alpha| \xi_0^2} \ dL^{\prime} \approx  \zeta L^2 . 
    \label{zL_2d}
\end{align}
i.e. $\sqrt{\langle z^2(L) \rangle} \approx \sqrt{\zeta} L \ll L$.  $\sqrt{\langle z^2(L) \rangle}$ may physically represent the typical value of the defects' displacement in the $z$-direction when the typical inter-defect distance is $L$. In that way, $\sqrt{\langle z^2(L) \rangle} = \sqrt{\zeta} L \ll L$ means the interface is stable. For other cases, quantitative evaluation of $\sqrt{\langle z^2(L) \rangle}$ is hard to achieve since $\lambda_z (L)$ and $h_z(L)$ are hard to evaluate. 

\subsubsection{3D systems} 
\begin{figure}[h]
	\centering
	\includegraphics[width= 4in]{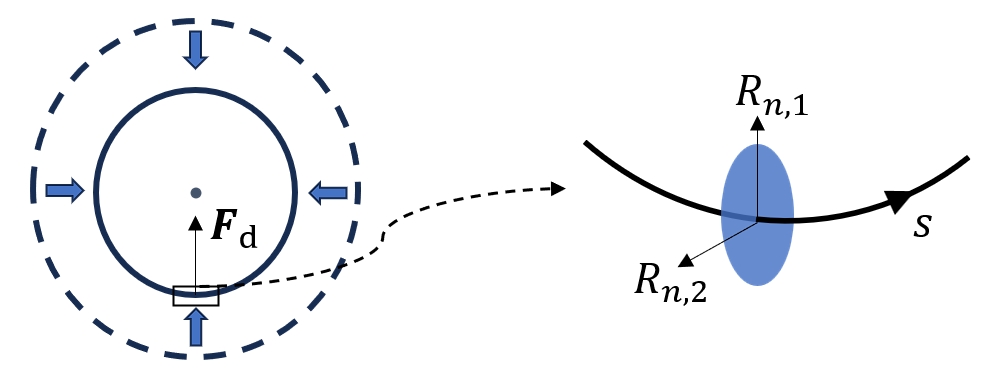} 
	\caption{A circular vortex string loop contracts under the driving of the force $\bm{F}_{\rm d}$, which points to the center of the circle. The local motion of the string is confined in the local $R_{n,1}-R_{n,2}$ plane normal to the string, which is marked by blue. }
	\label{sm_force_3d}
\end{figure}

In 3D systems, vortex string loops contract and $\psi(\bm{r},t)$ satisfies \equa{gl1}. For convenience, 
we separate the coordinates by $\bm{r} = (\bm{R}_n,s)$ where $\bm{R}_n$ is the local coordinate of the intersecting plane normal to the string and $s$ is the coordinate along the string. The motion of the string is characterized by the ansatz $\psi(\bm{r}=(\bm{R}_n,s),t) = f(\bm{R}_n-\int^t \bm{v}(s,t^{\prime}) dt^{\prime})$. Here $f(\bm{R}_n)$ is the order-parameter profile of a point defect in the 2D plane spanned by $\bm{R}_n$. As illustrated in \fig{sm_force_3d}, the ansatz implies that $\bm{v}(s,t)$ is confined in the 2D plane spanned by $\bm{R}_n$. 

To further simplify the problem, we idealize the shape of the vortex-string loop to be circular with a diameter $ R_{\text{d}}$. The energy of such a vortex-string loop is $E_{\text{s}} =   R_{\rm d} E_{\text{d}} (R_{\text{d}})$, where a constant $\pi$ factor is omitted. Here $E_{\text{d}}(R_{\rm d})$ is the energy per unit length of the string. $E_{\text{d}}$ equals the energy of point vortices in the 2D intersecting plane of the string, whose expression is given in Sec.~\ref{secIIIA}. Since the vortex string loop is contracting, $R_{\rm d}$ as well as $E_{\text{s}} =  R_{\rm d} E_{\text{d}} (R_{\text{d}})$ decrease with time. The energy dissipation rate of the vortex string loop reads
\begin{align}
    \frac{d E_{\text{s}}}{dt} = \frac{d E_{\text{s}}}{dR_{\text{d}}} \frac{dR_{\text{d}}}{dt} = \int_0^{R_{\rm d}} ds\ \ -\bm{F_{\rm d}}(s) \cdot \bm{v}(s).
    \label{work_3d}
\end{align}
Here $\bm{F_{\rm d}}(s)$ is a force acting on per unit length of the string, which comes from the string tension stem from the local curvature. $\bm{F_{\rm d}}(s)$ points to the center of the circle and drives the loop to contract, as shown in \fig{sm_force_3d}. Since the loop is circular, the magnitude of $\bm{F_{\rm d}}$ is independent on $s$ with the following expression, 
\begin{align}
	F_{\text{d}} = \frac{1}{R_{\text{d}}} \frac{dE_{\rm s}}{dR_{\text{d}}} = \frac{E_{\text{d}}(R_{\text{d}})}{R_{\text{d}}} + \frac{dE_{\text{d}}(R_{\text{d}})}{dR_{\text{d}}}.
	\label{Fd_3d}
\end{align}
In our theory, the strings are nearly confined in the interface, i.e. the displacement of the strings from the $z=0$ interface is negligible compared to $R_{\rm d}$. Therefore, the direction of $\bm{F_{\rm d}}(s)$ is nearly confined in the $x-y$ plane. 

Consider an infinitely small segment of the string with length $\epsilon$ located at $s=s_0$. Set its energy as $E_{\epsilon}$. \equa{work_3d} impies that the energy dissipation rate of the small segment reads
\begin{align}
	\frac{dE_{\epsilon}}{dt} =\int_{s_0}^{s_0 + \epsilon} ds\ \ -\bm{F_{\rm d}}(s) \cdot \bm{v}(s)= -\epsilon\bm{F}_{\rm d}(s_0) \cdot \bm{v}(s_0).
	\label{work_epsilon}
\end{align}
Moreover, by \equa{gl1} and the ansatz of $\psi(\bm{r}=(\bm{R}_n,s_0),t) = f(\bm{R}_n-\int^t \bm{v}(s_0,t^{\prime}) dt^{\prime})$, the energy dissipation rate of the small segment can be given by the inertia term, and friction and Langevin forces acting on the small segment. 
\begin{align}
    \frac{dE_{\epsilon} }{dt } &= \int_{s_0}^{s_0+\epsilon} ds \int d^2 \bm{R}_n \ \ \frac{\delta F}{\delta \psi } \frac{\partial \psi}{\partial t} + \frac{\delta F}{\delta \psi^* } \frac{\partial \psi^*}{\partial t} \nonumber \\
    &=  \int_{s_0}^{s_0+\epsilon} ds\int d^2 \bm{R}_n \ \ \Big( - \frac{E_{\rm c}}{\gamma^2_0} \frac{\partial^2 \psi^*}{\partial t^2}
    -\frac{E_{\text{c}}}{\gamma} \frac{\partial \psi^* }{\partial t} + E_{\text{c}} \eta^* \Big) \!\  \frac{\partial \psi}{\partial t} + 
    \Big( - \frac{E_{\rm c}}{\gamma^2_0} \frac{\partial^2 \psi}{\partial t^2}
    -\frac{E_{\text{c}}}{\gamma} \frac{\partial \psi}{\partial t} + E_{\text{c}} \eta \Big) \!\ \frac{\partial \psi^*}{\partial t}  \nonumber \\
    &=- \sum_{i,j,k=R_{n,1},R_{n,2}}\int_{s_0}^{s_0+\epsilon} ds\int d^2 \bm{R}_n \ \ \Big(
    -\frac{E_c}{\gamma^2_0} 
    (\partial_i\partial_k f^*) {\bm v}_i {\bm v}_k 
    + \frac{E_c}{\gamma^2_0} (\partial_i f^*) \frac{\partial {\bm v}_i}{\partial t} + \frac{E_{\text{c}}}{\gamma}  (\partial_i  f^*) \bm{v}_i + E_{\text{c}} \eta^*\Big) \!\ (\partial_j  f) \bm{v}_j \nonumber \\
    &\hspace{5cm} + \Big(
    -\frac{E_c}{\gamma^2_0} 
    (\partial_i\partial_k f) {\bm v}_i {\bm v}_k 
    + \frac{E_c}{\gamma^2_0} (\partial_i f) \frac{\partial {\bm v}_i}{\partial t} + 
    \frac{E_{\text{c}}}{\gamma} (\partial_i f) \bm{v}_i + E_{\text{c}} \eta\Big) \!\ (\partial_j  f^*) \bm{v}_j \nonumber \\
    &= ( \epsilon\bm{F}_{\lambda} - \bm{F}_{\eta} - \epsilon{\bm A} ) \cdot \bm{v}(s_0). 
    \label{dissip_3d}
\end{align}
Here $\epsilon{\bm A}=\sum_{i=y,z} \epsilon m_i a_i 
\hat{i}$ with ${\bm a}= d{\bm v}(s_0)/dt$, is the inertia term for the small segment. 
$\epsilon\bm{F}_{\lambda} = -\sum_{i=y,z} \epsilon\lambda_i v_i(s_0) \hat{i}$ is the frictional force. $\bm{F}_{\eta}$ is the Langevin force. $m_i$ and $\lambda_i$ are the inertia mass and the friction coefficient in the $i$-direction, respectively. The inertia mass, friction coefficients, and the Langevin force are given as follows;
\begin{align}
    m_i & = \frac{E_{\text{c}}}{\gamma^2_0}  \int d^2 \bm{R_n} \ \ 2|\partial_i  f |^2, \\
    \lambda_{i} &= \frac{E_{\text{c}}}{\gamma}  \int d^2 \bm{R_n} \ \ 2|\partial_i  f |^2,  \\
    \bm{F}_{\eta} &= E_{\text{c}}  \int_{s_0}^{s_0+\epsilon} ds\int d^2 \bm{R}_n \ \ (\eta^* \nabla_n  f + \eta \nabla_n  f^*).
    \label{F_eta_3d} .
\end{align}
Here $\nabla_n = \partial / \partial \bm{R}_n$. Without loss of generality, we set the tangent of the string at $s=s_0$ is along the $x$-direction and $\bm{R}_n = (y,z)$. Therefore, $\bm{F}_{\rm d} = F_{\rm d} \hat{y}$. Equating \equa{work_epsilon} and \equa{dissip_3d}, we obtain an equation of motion for the small segment of the string under the driving force $\bm{F}_{\rm d}$.
\begin{align}
    m_y \frac{d^2 y}{dt^2} 
+ \lambda_y \frac{dy}{dt} &= F_{{\rm d},y} - \epsilon^{-1} F_{\eta,y}, \nonumber \\ 
 m_z  \frac{d^2 z}{dt^2} 
+ \lambda_z \frac{dz}{dt} &= 
- \epsilon^{-1} F_{\eta,z}.
\label{eom_3d}
\end{align}
For the ``model A" dynamics without the inertia term ($m_i=0$), we obtain the force-balance equation
\begin{align}
    \lambda_y \frac{dy}{dt} = F_{{\rm d},y} - \epsilon^{-1}F_{\eta,y},  \ \ 
\lambda_z \frac{dz}{dt} = - \epsilon^{-1}F_{\eta,z}.
\label{balance_3d}
\end{align}
The force 
balance equation is solved for the coordinate of the vortex core,  
\begin{align}
     z(t) &=  \int dt  \ \ -\epsilon^{-1}\lambda_z ^{-1}  F_{\eta,z} \nonumber \\
     y(t) &= \int dt \ \ \lambda_y ^{-1}    (-\epsilon^{-1}F_{\eta,y} + F_{\text{d}} ). 
     \label{Brownian_3d}
\end{align}
The mean displacement is
\begin{align}
    \langle z(t) \rangle &= 0 ; \ \  \langle y(t) \rangle = \int dt \ \ \lambda_y ^{-1}    F_{\text{d}} .
    \label{mean_xy_3d}
\end{align}
Eq.~(\ref{mean_xy_3d}) leads to a conclusion: that the mean velocity of the string motion reads $v(R_{\text{d}}) = -dR_{\rm d}/dt = \lambda_y ^{-1}(R_{\text{d}}) F_{\text{d}}(R_{\text{d}})$. Similar to the analysis for the 2D systems, taking $R_{\text{d}}$ by its typical value $L$ and equating $dL/dt = v(L)$ gives Eq.~(\ref{estimate_sm}) ~\cite{bray2002theory}. See Sec. \ref{SI_B3} for a more formal derivation.

To solve Eq.~(\ref{estimate_sm}), $\lambda_i(L)$ and $F_{\text{d}}(L)$ need to be evaluated. In the presence of the $U(1)$-symmetry ($\beta=0$) or in the $L\ll \xi_0/\sqrt{2\beta}$ limit when $\beta \ne 0$, $E_{\text{d}} (L) \propto E_{\text{c}} |\alpha| \xi_0^2 \ln( L/\xi)$ and the leading order contribution of $F_{\text{d}}(L)$ is the $E_{\text{d}}(L)/L$ term in Eq.~(\ref{Fd_3d}).  Therefore, 
\begin{align}
	F_{\text{d}} &\sim E_{\text{c}}|\alpha|\xi_0^2 \frac{1}{L} \ln(\frac{L}{\xi}).
\end{align}
For $\lambda_{i=y,z}$, as shown in the analysis of 2D systems, 
\begin{align}
    \lambda_{z/y} &\sim \frac{E_{\text{c}} |\alpha|}{\gamma} \ln (\frac{L}{\xi}). 
\end{align}
Thus, from Eq.~(\ref{estimate_sm}), we obtain
\begin{align}
    L (t) \sim \xi_0 \sqrt{\gamma t}.
\end{align}

In the $L\gg \xi_0/\sqrt{2\beta}$ limit, $E_{\rm d}(L) \sim E_{\rm c}|\alpha|\xi_0^2$ according to the calculation in Sec.~\ref{secIIIA}. The leading order contribution of $F_{\text{d}}(L)$ is the $E_{\text{d}}(L)/L$ term in Eq.~(\ref{Fd_3d}). Therefore
\begin{align}
	F_{\text{d}} &\sim E_{\text{c}}|\alpha|\xi_0^2 \frac{1}{L}.
\end{align}
For $\lambda_y$, according to the calculation in \equa{lambda_2d_largeL}, 
\begin{align}
    \lambda_{y} &= \lambda_0 \propto E_{\text{c}} |\alpha|/\gamma 
\end{align}
Therefore, from Eq.~(\ref{estimate_sm}), we obtain
\begin{align}
    L (t) \sim \xi_0 \sqrt{\gamma t}.
\end{align}
The analysis reveals a discrepancy between 2D and 3D systems. In 2D
systems with $\beta \ne 0$, as $L$ gets larger, the scaling of $L(t)$ undergoes a crossover from a diffusive behavior to a subdiffusive behavior. See Eq.~(\ref{Lt_tlnt}, \ref{Lt_lnt}). However, in 3D systems, $L(t)$ always shows a diffusive behavior. 

The discrepancy could be explained in the following way. In 2D systems, as shown above, the $L$-dependence of the attraction force $F_{\text{d}}$ shows a $ 1/L$ (long range) to $ e^{-L\sqrt{2\beta}/\xi_0}$ (short range) crossover as $L$ goes across the scale  $\xi_0/\sqrt{2\beta}$. The crossover of $F_{\text{d}}$ makes the motion and annihilation of vortices slower with time going by, leading to the diffusive-to-subdiffusive crossover of the coarsening dynamics. In 3D systems, the dominant contribution of the driving force $F_{\text{d}}(L)$ always comes from $E_{\rm d}(L)/L$, the first term in Eq.~(\ref{Fd_3d}), which is from the derivative of the string-loop diameter $L$ rather than $E_{\rm d}(L)$. (Physically, it means that the driving force $F_{\text{d}}(L)$ is dominated by the tension of the string.) Moreover, the $L$-dependence of $E_{\text{d}}(L)$ is the same as $\lambda_y(L)$ in both $L\ll\xi_0/\sqrt{2\beta}$ and $L\gg\xi_0/\sqrt{2\beta}$ limits. Therefore, we always obtain $dL/dt \sim \gamma \xi_0^2 /L$, leading to the result $L(t)\sim \xi_0\sqrt{\gamma t}$ . 

$\bullet$ {\bf Defect displacement in the $z$-direction and stability of the interface}: Similar to the analysis for 2D systems, the discussion above assumes the strings are confined within the 2D interface. In fact, the string-loop could acquire a weak Brownian-like motion in the $z$-direction due to the thermal noise.  Consider the center-of-mass motion in the $z$-direction of a macroscopical segment of the string loop with diameter $R_{\rm d}$. The length of the segment is $\sim R_{\rm d}$.  The equation of motion is derived by simply replacing $\epsilon$ by $ R_{\rm d}$ in Eq.~(\ref{dissip_3d}, \ref{F_eta_3d}, \ref{balance_3d}), where $R_{\rm d}$ is the diameter of the loop. Written in the dimensionless form, the equation of motion reads
\begin{align}
	 \frac{d(z/\xi_0)}{d(\gamma t)} = H_z
	 \label{Brownian_z_3d}
\end{align}
Here 
\begin{align}
  H_z = -(\xi_0 \gamma)^{-1}(R_{\rm d} \lambda_z)^{-1}E_{\text{c}}  \int_0^{ R_{\rm d}} ds\int d^2 \bm{R}_n \ \ (\eta^* \partial_z  f + \eta \partial_z  f^*).
\end{align}
$H_z$ satisfies
\begin{align}
	\langle H_z (t) H_z (t^{\prime})  \rangle =  \frac{2T}{ R_{\rm d} \xi_0^2 \gamma } \lambda_z^{-1}(R_{\rm d}) \delta(\gamma t - \gamma t^{\prime}).
  \label{Hz_lambda_3d}
\end{align}
By the expression of $\lambda_i$ in \equa{F_eta_3d}, $\lambda_i = E_{\rm c}|\alpha|\gamma^{-1}h_i(R_{\rm d})$ where $h_i(R_{\rm d})=\int d^2 \bm{R}_n \ |\partial_i  \theta|^2$ is dimensionless. Therefore, 
\begin{align}
	\langle H_z (t) H_z(t^{\prime})  \rangle &= \frac{2T}{ R_{\rm d} \xi_0^2 E_{\rm c} |\alpha| } h_z^{-1} \delta(\gamma t - \gamma t^{\prime}) = \frac{2T}{ \xi_0^3 E_{\rm c} \sqrt{|\alpha|} }  \Big(\frac{R_{\rm d}}{\xi}h_z(R_{\rm d})\Big)^{-1} \delta(\gamma t - \gamma t^{\prime}) \nonumber \\
    &= 2\zeta  \Big(\frac{R_{\rm d}}{\xi}h_z(R_{\rm d})\Big)^{-1}  \delta(\gamma t - \gamma t^{\prime}).
	\label{Hz_3d}
\end{align}

Replacing $R_{\rm d}$ in \equa{Hz_3d} by the typical defect size $L$, we obtain Eq.~(9) in the main text. \equa{Brownian_z_3d} and \equa{Hz_3d} imply the defect motion in the $z$-direction is negligible because $\zeta$ is small in consideration of this paper. Thus, the interface is stable. 

For the $U(1)$ symmetric case as well as the $L \ll \xi_0/\sqrt{2\beta}$ case with a nonzero $\beta$, one has $\lambda_z = \lambda_y$ and $h_z(L) = h_y(L) \sim \ln(L/\xi)$, while it is hard to evaluate $h_z(L)$ for the other cases. In the $L \ll \xi_0/\sqrt{2\beta}$ case, after replacing $R_{\rm d}$ in Eqs.(\ref{Hz_lambda_3d},\ref{Hz_3d}) by its typical value $L$ , we could solve the following scaling of $\langle z^2(L) \rangle$ 
\begin{align}
    \langle z^2(L(t)) \rangle =& 2T \int^t L^{-1}(t^{\prime}) \lambda_z^{-1}(L(t^{\prime})) \ dt^{\prime} = 2T \int^t L^{-1}(t^{\prime}) \lambda_y^{-1}(L(t^{\prime})) \ dt^{\prime}  \nonumber \\
    \stackrel{\text{By eq.}~(\ref{estimate_sm})}{=} & 2T \int^L L^{\prime -1} F_{\rm d}^{-1}(L^{\prime}) \ dL^{\prime} = 2T \int^L_{2\xi} \frac{1}{E_{\rm c} |\alpha| \xi_0^2 \ln(L^{\prime}/\xi)} \ dL^{\prime} \nonumber \\
    =& 2\zeta \xi^2 \int_{2}^{L/\xi} \frac{1}{\ln (L^{\prime}/\xi) } d(L^{\prime}/\xi) \stackrel{L \gg \xi}{=} 2\zeta  \frac{\xi L}{\ln (L/\xi) }  
    \label{zL_3d}
\end{align}
Here we set a lower limit of $2\xi$ in the integration over $L^{\prime}$ to prevent the unphysical singularity. Eq.~\eqref{zL_3d} gives $\sqrt{\langle z^2(L) \rangle} \ll L$.  $\sqrt{\langle z^2(L) \rangle}$ may physically represent the typical value of the defects' displacement in the $z$-direction when the typical inter-defect distance is $L$. In that way, $\sqrt{\langle z^2(L) \rangle} \ll L$ means the interface is stable. For other cases, quantitative evaluation of $\sqrt{\langle z^2(L) \rangle}$ is hard to achieve since $\lambda_z (L)$ and $h_z(L)$ are hard to evaluate. 

\subsubsection{Derivation of Eq.~\eqref{estimate_sm} from Eq.~\eqref{vR}. \label{SI_B3}}
This section discusses how Eq.~(\ref{estimate_sm}) is derived From Eq.~(\ref{vR}). 
The interface contains random domains of size $R_{\text{d}}$. $R_{\text{d}}$ satisfies a probability distribution function $p(R_{\text{d}})$ with $\int_0^{+\infty} p(R_{\text{d}}) dR_{\text{d}} = 1$. Suppose the coarsening dynamics is self-similar with a single length scale $L(t)$, we obtain $p(R_{\text{d}}) = g(R_{\text{d}}/L) / L$ where $g$ is a function satisfying $\int_0^{+\infty} g(x) dx = 1$. 

Eq.~(\ref{vR}) implies that the typical velocity of the defect motion depends on the defect size (domain size). Statistically, when the typical correlation length is $L(t)$, the typical velocity of the defect's motion reads $v(L)=  F_{\text{d}}(L)/\lambda_y(L)$. Therefore, within a time $dt$, the number of the annihilated domains reads
\begin{align}
    dN_{\text{d}} = -N_{\text{d}} \int_0^{v(L)dt} p(R_{\text{d}})dR_{\text{d}} = -N_{\text{d}} \int_0^{v(L)dt/L} g(x)dx.
    \label{anni_Nd}
\end{align}
Here $N_{\text{d}}$ denotes the number of domains. Suppose the system size is $l$. In the 1D interface of 2D systems, $N_{\text{d}} \approx  l/L$. In the 2D interface of 3D systems, $N_{\text{d}} \approx l^2/L^2$. Obviously, in both cases, $dN_{\text{d}} / N_{\text{d}} \approx -dL/L$. Thus, Eq.~(\ref{anni_Nd}) implies 
\begin{align}
    \frac{dL}{L} \approx g(0) \frac{v(L)}{L}  dt,
\end{align}
for $v(L)dt/L \ll 1$. Neglecting the dimensionless constant $g(0)$, we derive Eq.~(\ref{estimate_sm}).

\section*{\textsl{\NoCaseChange{Supplementary Note 4: }}Effects of the $\partial_t^2 \psi$ term}
\setcounter{subsection}{0}
\subsection{Suppression of the $\partial_t^2 \psi$ term when the temperature is near $T_{\rm c}$.}
\begin{figure}[h]
	\centering
	\includegraphics[width= 5in]{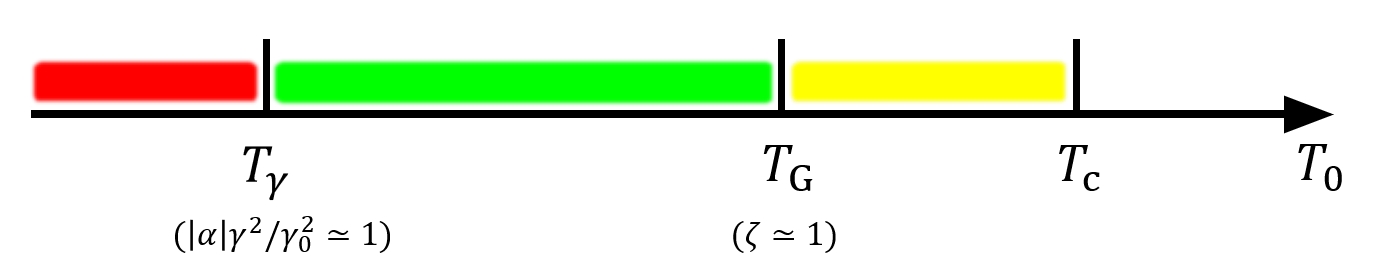} 
	\caption{Valid criteria of the analysis in this paper is marked by green.}
	\label{valid}
\end{figure}

The ``model A" dynamics used in Eq.~(\ref{model_A_sm}) neglects the inertia term, the second-order time-derivative $\partial_t^2 \psi$. In fact, as mentioned in Eq.~(\ref{dw_form}), the inertia term is crucial to form a domain wall. In this section, we discuss the effect of the inertia term and analyze how it modifies our results. With the inertia term, the dynamical equation reads
\begin{align}
    \frac{1}{\gamma_0^2}\partial^2_t \psi + \frac{1}{\gamma} \partial_t \psi = -\frac{1}{E_{\text{c}}} \frac{\delta F}{\delta \psi^*} + \eta = \xi_0^2 \nabla^2 \psi - 2\alpha\psi -2\beta i \psi_2 - 2|\psi|^2 \psi + \eta.
    \label{model_A2}
\end{align}
Here $\gamma_0$ is a parameter controlling the strength of the inertia. Typically, one has $\gamma,\, \gamma_0\sim T$, $\alpha \sim (T-T_{\text{c}})/T_{\text{c}}$ for pure electronic orders arising from Fermi surface nesting, see Ref.~\cite{Mitra.2018} and Appendix~E of Ref.~\cite{Sun.2020_SC}. The following scaling analysis shows that when the temperature $T$ is close to $T_{\text{c}}$, the inertia term is effectively suppressed. Eq.~(\ref{model_A2}) is invariant under the following rescaling;
\begin{align}
    \alpha &= \bar{\alpha} b^2,  \ \ \beta = \bar{\beta} b^2,  \ \ E_{\text{c}} = \bar{E_{\text{c}}} b^{-4}, \ \  \psi= \bar{\psi} b, \nonumber \\ \ \ \eta &=  \bar{\eta} b^3, 
    \ \ \xi_0 =  \bar{\xi_0} b, \ \ \gamma = \bar{\gamma} b^{-2} , \ \ \gamma_0 = \bar{\gamma_{0}} b^{-1}.
    \label{rescaling}
\end{align}
The rescaled equation reads,
\begin{align}
    \frac{1}{\bar{\gamma_{0}}^2}\partial^2_t \bar{\psi} + \frac{1}{\bar{\gamma}} \partial_t \bar{\psi} = \bar{\xi_0}^2 \nabla^2 \bar{\psi} - 2\bar{\alpha}\bar{\psi} -2\bar{\beta} i \bar{\psi_2} - 2|\bar{\psi}|^2 \bar{\psi} + \bar{\eta}.
    \label{model_A2_rescale}
\end{align}
with the noise correlator $\langle \bar{\eta}_{i}(\bm{r},t) \bar{\eta}_{i^{\prime}}(\bm{r^{\prime}},t^{\prime}) \rangle = 2T(\bar{\gamma} \bar{E_{\text{c}}})^{-1} \delta_{ii^{\prime}}\delta(\bm{r}-\bm{r^{\prime}})\delta(t-t^{\prime})$. We could set $\bar{\alpha} = -1$. If $T$ is close to $T_{\text{c}}$, $|\alpha|$ and $b=\sqrt{|\alpha|}$ will be small. Thus, $\bar{\gamma}^2 / \bar{\gamma_0}^2=b^2 \gamma^2/\gamma_0^2$ is suppressed to be small. Since we discuss the dynamics happening after $t_0$, as long as $T(t>t_0)=T_0$ is near $T_{\text{c}}$, the inertia term is much weaker than the damping term and our analysis using the ``model A" dynamics is valid. The condition $\bar{\gamma}^2 / \bar{\gamma_0}^2 = |\alpha| \gamma^2/\gamma_0^2 = 1$ gives a lower critical temperature $T_{\gamma}$, below which the inertia term is non-negligible. Moreover, another condition $\zeta=G|\alpha|^{D/2 -2} \geq 1$ sets an upper critical temperature $T_{\rm G}$, with $T<T_{\rm G}$ being the Ginzburg criterion in the conventional theory of critical phenomenon~\cite{cardy_1996}. If $T_{\rm G} < T_0 < T_{\rm c}$, the thermal fluctuation is strong and the mean-field solution of $\psi=-\sqrt{\alpha}$ for $t<t_{\rm pump}$ is invalid and the domain wall will not form. The successful formation of the domain wall in the CDW materials proves those systems are inside the Ginzburg criterion. In summary, the valid criterion of the analysis in this paper is $T_{\gamma} < T_0 < T_{\rm G}$, shown by Fig.~\ref{valid}.

Note that the inertia term is suppressed for $t>t_0$ by $T_0$. While within $t_{\rm pump} < t< t_{\rm off}$, the pump lifts the temperature to $T_{\rm H}$ and $\alpha \sim (T_{\rm H}-T_{\rm c})/T_{\rm c}$ may be larger, making the inertia term significant enough to form the domain wall. 

\subsection{Quantitative modifications of the first and second stage dynamics by the $\partial_t^2 \psi$ term.}
In the following discussion, we discuss how the weak inertia modifies our results quantitatively for $t>t_0$. 

$\bullet$ {\bf First-stage dynamics}. For the first-stage dynamics, we calculate the correlation $C(\bm{R},t) = C_1(\bm{R},t) + C_2(\bm{R},t)$ defines in Eq.~(\ref{g_Rt}, \ref{C_1}, \ref{C_2}). A calculation is shown in Sec.~\ref{derive_C_inertia}. $C_2 (\bm{R},t)$ is evaluated in the $\gamma t \gg 1/|\alpha|$ limit, 
\begin{align}
    C_2 (\bm{R},t)  &=  \frac{\zeta|\alpha|}{2\kappa^2 c/|\alpha|} \frac{e^{2c\gamma t} } {\left( \sqrt{ 8  \pi |\alpha| \gamma t/\kappa }\right)^{D-1} }  e^{- \frac{R^2} { 8\xi_0^2 \gamma t/\kappa} }.  
    \label{C_2_inertia}
\end{align}
Besides, $C_1(\bm{R},t)$ is given by the following integral.
\begin{align}
   C_1(\bm{R},t) &= e^{2c\gamma t}\int\frac{d^{D-1} \bm{K}}{(2\pi)^{D-1}}   l^{D-1} D_K(0) e^{-2\xi_0^2\gamma t K^2 /\kappa} e^{i\bm{K}\cdot \bm{R}} 
   \label{C_1_inertia}
\end{align}
Here $\kappa = \sqrt{1+4(|\alpha|-2\beta)\gamma^2/\gamma_0^2}, c = 2^{-1}(\kappa-1)\gamma_0^2/\gamma^2  $.

Compared with Eq.~(\ref{C_2_res}, \ref{C_1}), Eq.~(\ref{C_2_inertia} ,\ref{C_1_inertia}) only contains corrections of the coefficients. Espectially, the time-dependent correlation length $\xi_0\sqrt{\gamma t}$ shown in the results of Eq.~(\ref{C_2_res}, \ref{C_1_b}, \ref{C_1_res}, \ref{C_1_I0}) is renormalized to $\xi_0\sqrt{\gamma t}/\sqrt{\kappa}$.  

$\bullet$ {\bf Second-stage dynamics}. For the second-stage dynamics, the equations of motion with the inertia term are given in Eqs.~(\ref{eom_2d},\ref{eom_3d}) for 2D and 3D systems. have nonzero $m_i$. An average of Eqs.~(\ref{eom_2d},\ref{eom_3d}) over the thermal noise $\eta$ gives the equation of motion
\begin{align}
    m_y \frac{d^2 \langle y \rangle }{dt^2} 
+ \lambda_y \frac{d\langle y \rangle }{dt} &= F_{{\rm d},y}  \nonumber \\
    m_z \frac{d^2 \langle z \rangle }{dt^2} 
+ \lambda_z \frac{d\langle z \rangle }{dt} &= 0.
\end{align}
Therefore, Eq.~(\ref{vR}) is modified to
\begin{align}
    m_y\frac{d v_y(R_{\rm d})}{dt} +  \lambda_y v_y(R_{\rm d}) &=    F_{\text{d}}(R_{\rm d}).
    \label{vR2}
\end{align}
Taking $v_y(R_{\rm d}\sim L) = dL/dt$, Eq.~(\ref{estimate_sm}) is modified to
\begin{align}
    m_y\frac{d^2 L}{dt^2} +  \lambda_y\frac{dL}{dt} &=    F_{\text{d}}(L).
    \label{estimate_sm_2}
\end{align}

Eq.~(\ref{estimate_sm_2}) is the Eq.~(10) in the main text, where $\lambda$, $m$ in the main text denote $\lambda_y,m_y$ here for simplicity. Note that all the results of $L(t)$ satisfying Eq.~(\ref{estimate_sm}) obtained in this work satisfy Eq.~(\ref{estimate_sm_2}) in the
long time limit, at which time the term $m_y(L)d^2L /dt^2$ is much smaller than the other two terms.


\subsection{Derivation of Eq.~(\ref{C_2_inertia}, \ref{C_1_inertia}).} 
\label{derive_C_inertia}
In the derivation in this section, we use the rescaled notation used in the dynamical equation Eq.~(\ref{model_A_rescale}) with $\bar{\gamma}^2/\bar{\gamma}_0^2 \ll 1$. Moreover, we drop the bar symbol on the top of the characters for convenience. 

In the first-stage dynamics, the dynamical equation Eq.~(\ref{linear_dynamics_sm}) with an inertia term reads
\begin{align}
   \frac{1}{\gamma_0^2}\partial^2_t \phi_{1/2} + \frac{1}{\gamma} \partial_t \phi_{1/2} &= -(\hat{L}_{1/2} - \xi_0 ^2 \nabla_{\bm{R}}^2)\phi_{1/2} + \eta_{1/2}.
    \label{linear_dynamics_2}
\end{align}
Here $\gamma^2/\gamma_0^2 \ll 1$. Only the modes $\phi_{2}^{\bm{K}}(\bm{r}) = \text{sech}(z/\xi)e^{i\bm{K}\cdot\bm{R}}$ with $K<\sqrt{|\alpha|-2\beta}/\xi_0$ will be exponentially amplified with time. Writing $\phi_2 (\bm{r}) = \varphi (\bm{R},t)\text{sech}(z/\xi)$, Eq.~(\ref{linear_dynamics_C},  
 \ref{linear_dynamics_Ck}) are modified to be
\begin{align}
    \frac{1}{\gamma_0^2} \partial_t^2 \varphi (\bm{R},t) + \frac{1}{\gamma} \partial_t \varphi (\bm{R},t) &= (|\alpha|- 2\beta + \xi_0^2 \nabla_{\bm{R}}^2 )\varphi (\bm{R},t) + \eta_\varphi (\bm{R},t), \nonumber \\
    \frac{1}{\gamma_0^2} \partial_t^2 \tilde{\varphi}(\bm{K},t) + \frac{1}{\gamma}\partial_t \tilde{\varphi}(\bm{K},t) &= (|\alpha| - 2\beta - \xi_0^2 K^2)\tilde{\varphi}(\bm{K},t) + \tilde{\eta}_\varphi (\bm{K},t).
    \label{linear_dynamics_C2}
\end{align}
Construct the following two functions, 
\begin{align}
    A_{1}(\bm{K},t) &=  \frac{\gamma^2}{\gamma_0^2} \frac{1}{\gamma}\partial_t \tilde{\varphi}(\bm{K},t) +   [\frac{\gamma^2}{\gamma_0^2}a_1(\bm{K})+1] \tilde{\varphi}(\bm{K},t), \nonumber \\
    A_{2}(\bm{K},t) &=  \frac{\gamma^2}{\gamma_0^2} \frac{1}{\gamma}\partial_t \tilde{\varphi}(\bm{K},t) +   [\frac{\gamma^2}{\gamma_0^2}a_2(\bm{K})+1] \tilde{\varphi}(\bm{K},t),
    \label{A1A2}
\end{align}
with 
\begin{align}
    a_1(\bm{K}) = \frac{-1+\Delta(K)}{2 } \frac{\gamma_0^2}{\gamma^2} ,\nonumber \\
    a_2(\bm{K}) = \frac{-1-\Delta(K)}{2 } \frac{\gamma_0^2}{\gamma^2} .
\end{align}
Here $\Delta(K) = \sqrt{1+4(|\alpha| - 2\beta - \xi_0^2 K^2)\gamma^2 / \gamma_0^2 }$. From Eq.~(\ref{linear_dynamics_C2}), $A_{1/2}$ satisfies
\begin{align}
    \frac{1}{\gamma } \partial_t A_{1/2} = a_{1/2}(\bm{K}) A_{1/2} + \tilde{\eta}_\varphi (\bm{K},t) .
    \label{A_eq}
\end{align}
The solution reads
\begin{align}
    A_{1/2}(\bm{K},t) = A_{1/2}(\bm{K},t_0)e^{\gamma\int_{t_0}^{t}  a_{1/2}(K) dt^{\prime} } + \gamma\int_{t_0}^{t}  \tilde{\eta}_\varphi (\bm{K},t^{\prime}) e^{\gamma\int_{t^{\prime}}^{t}  a_{1/2}(K) dt^{\prime\prime}} dt^{\prime}.
\end{align}
Therefore, the function $A_{1}(\bm{K},t)$ will be exponentially amplified with time if  $K<\sqrt{|\alpha|-2\beta}/\xi_0$ since $a_1(\bm{K}) > 0$, while $A_{2}(\bm{K},t)$ will exponentially decay with time since $a_2(\bm{K}) < 0$. Notice that the decay of $A_{2}(\bm{K},t)$ is extremely rapid, since $a_2(\bm{K})$ is large with $a_2(\bm{K}) \approx \gamma_0^2/\gamma^2$ in the $\gamma^2/\gamma_0^2 \ll 1$ limit. Therefore, by the relation
\begin{align}
    A_1(\bm{K},t) - A_2(\bm{K},t) = \Delta(K)\tilde{\varphi}(\bm{K},t),
    \label{AADelta}
\end{align}
$A_1(\bm{K},t)$ could be approximated by $
    A_1(\bm{K},t)  \approx \Delta(K)\tilde{\varphi}(\bm{K},t).$ for $t>t_0$.
Thereby, from Eq.~(\ref{A_eq}), we obtain
\begin{align}
    \frac{1}{\gamma } \partial_t \tilde{\varphi}(\bm{K},t) = a_{1/2}(\bm{K}) \tilde{\varphi}(\bm{K},t) + \frac{\tilde{\eta}_\varphi (\bm{K},t)}{\Delta(K)} ,
\end{align}
\begin{align}
    \tilde{\varphi}(\bm{K},t) = \tilde{\varphi}(\bm{K},t_1)e^{U_K(t,t_0)} + \gamma\int_{t_0}^{t}  \frac{\tilde{\eta}_\varphi (\bm{K},t^{\prime})}{\Delta(K)} e^{U_K(t,t^{\prime})} dt^{\prime}
\end{align}
Here $U_K(t_1,t_2) = \gamma \int_{t_2}^{t_1} a_1(K) dt$. Define the amplitude $D_{K}(t) = \langle \tilde{\varphi}(-\bm{K},t) \tilde{\varphi}(\bm{K},t) \rangle$. We obtain 
\begin{align}
    D_{K}(t) &= \frac{1}{l^{D-1}} e^{2U_K (t,t_0)} \Big(  l^{D-1} D_K(t_0) + \frac{T\gamma}{E_{\text{c}} \xi} \frac{1}{\Delta^2(K)} \int_{t_0}^t dt^{\prime} e^{-2U_K(t^{\prime},t_0)} \Big) \\ \nonumber
    &= \frac{1}{l^{D-1}}e^{2U_K (t,t_0)} \Big(  l^{D-1}D_K(t_0) + \frac{T}{2E_{\text{c}} \xi} \frac{1}{\Delta^2(K)}  \frac{1}{a_1(K)} (1-e^{-2U_K(t,t_0)} )   \Big). 
\end{align}
Here $l$ is the system size. In the following derivation, we set $t_0=0$ for convenience. Apply the inverse Fourier transform to $D_K (t)$ to obtain the function $C(\bm{R},t)$.
\begin{align}
    C(\bm{R},t) &= \sum_{\bm{K}} D_K (t) e^{i\bm{K}\cdot \bm{R}} \equiv l^{D-1}\int\frac{d^{D-1} \bm{K}}{(2\pi)^{D-1}} D_K (t) e^{i\bm{K}\cdot \bm{R}} \nonumber \\
    &= C_1(\bm{R},t) + C_2(\bm{R},t) ,
    \label{g_Rt_2}
\end{align}
with
\begin{align}
     C_1(\bm{R},t) &= \int\frac{d^{D-1} \bm{K}}{(2\pi)^{D-1}}   l^{D-1} D_K(0) e^{2U_K (t,0)} e^{i\bm{K}\cdot \bm{R}} 
     \label{C_1_2}
\end{align}
\begin{align}
    C_2(\bm{R},t) &=  \int\frac{d^{D-1} \bm{K}}{(2\pi)^{D-1}}    \frac{T\sqrt{|\alpha|}}{2E_{\text{c}} \xi_0} \frac{1}{\Delta^2 (K)} \frac{1-e^{-2U_K(t,0)}}{a_1(K)}  e^{2U_K (t,0)} e^{i\bm{K}\cdot \bm{R}} .
    \label{C_2_2}
\end{align}
To make the calculation of $C_1(\bm{R},t)$ and $C_2(\bm{R},t)$ practical, we approximate $a_1(K)$ by the following Taylor expansion at $K=0$, which is suitable for large $\gamma(t-t_0)$ when the long-wavelength mode dominates.
\begin{align}
    e^{2U_K(t,0)} \approx e^{2c\gamma t - 2\xi_0^2 \gamma t K^2/ \kappa}
\end{align}
Here $\kappa =\Delta(0)= \sqrt{1+4(|\alpha|-2\beta)\gamma^2/\gamma_0^2}, c = a_1(0) = 2^{-1}(\kappa-1)\gamma_0^2/\gamma^2  $.

We apply the approximation used in the derivation of Eq.~(\ref{C_2_res}), in the $\gamma t \gg 1/|\alpha|$ limit, we integrate over $\bm{K}$ with the integrand $e^{-2\xi_0^2 \gamma t K^2 /b} e^{i\bm{K}\cdot \bm{R}}$ and omit the $K$-dependence of other factors in Eq.~(\ref{C_2_2}), i.e.
\begin{align}
    C_2 (\bm{R},t) & \approx\frac{T\sqrt{|\alpha|}}{2E_{\text{c}} \xi_0}\frac{1}{\kappa^2 c  }  e^{2c\gamma t} \int\frac{d^{D-1} \bm{K}}{(2\pi)^{D-1}}    e^{-2\xi_0^2 \gamma t K^2/\kappa } e^{i\bm{K}\cdot \bm{R}} \nonumber \\
    &= \frac{T}{E_{\rm c}\xi_0^D} \frac{\sqrt{|\alpha|}}{2\kappa^2 c}  \frac{e^{2c\gamma t} } {\left( \sqrt{ 8 \pi \gamma t/\kappa }\right)^{D-1} }  e^{- \frac{R^2} { 8\xi_0^2 \gamma t/\kappa} } \nonumber \\
    &=  \frac{\zeta|\alpha|}{2\kappa^2 c/|\alpha|} \frac{e^{2c\gamma t} } {\left( \sqrt{ 8  \pi |\alpha| \gamma t/\kappa }\right)^{D-1} }  e^{- \frac{R^2} { 8\xi_0^2 \gamma t/\kappa} }.
    \label{C_2_res_2}
\end{align}  
$C_1$ is given by the following integration.
\begin{align}
     C_1(\bm{R},t) &= e^{2c\gamma t}\int\frac{d^{D-1} \bm{K}}{(2\pi)^{D-1}}   l^{D-1} D_K(0) e^{-2\xi_0^2\gamma t K^2 /\kappa} e^{i\bm{K}\cdot \bm{R}} 
     \label{C_1_2_int}
\end{align}
Note that Eq.~(\ref{C_2_res_2}, \ref{C_1_2_int}) are written by the notation used in Eq.~(\ref{model_A_rescale}) with the bar symbol on the top of the characters being left out. They could be rescaled back to the notation used in Eq.~(\ref{model_A2}) to give Eq.~(\ref{C_2_inertia}, \ref{C_1_inertia}). 

\section*{\textsl{\NoCaseChange{Supplementary Note 5: }}Coupled three-component field in 3D systems.}
\setcounter{subsection}{0}
\subsection{Free energy and first-stage dynamics}
As mentioned in the last discussion part in the main text, for a theory of coupled three-component field in 3D systems, similar dynamics could happen. For example, we consider the following free energy and ``model A" dynamics
\begin{align}
    &F =  E_{\text{c}} \int d^{3} \bm{r} \ \  f_0(\psi) +  f_1(\Psi,\Psi^*) + 2g|\Psi|^2 \psi^2 .  
    \label{free_energy_couple}  \\
    &\frac{1}{\gamma} \partial_t \Psi = -\frac{1}{E_{\text{c}}} \frac{\delta F}{\delta \Psi^*} + \eta \nonumber \\
    &\frac{1}{\gamma^{\prime}} \partial_t \psi = -\frac{1}{2E_{\text{c}}} \frac{\delta F}{\delta \psi} + \eta^{\prime} \nonumber \\
    & \langle \eta_{i}(\bm{r},t) \eta_{i^{\prime}}(\bm{r^{\prime}},t^{\prime}) \rangle = \frac{2T}{\gamma E_{\text{c}}} \delta_{ii^{\prime}}\delta(\bm{r}-\bm{r^{\prime}})\delta(t-t^{\prime}) \nonumber \\
    & \langle \eta^{\prime}(\bm{r},t) \eta^{\prime}(\bm{r^{\prime}},t^{\prime}) \rangle = \frac{2T}{\gamma^{\prime} E_{\text{c}}} \delta(\bm{r}-\bm{r^{\prime}})\delta(t-t^{\prime}) \nonumber
\end{align} 
with
\begin{align}
    f_0(\psi_1) &= \xi_0^2  |\nabla \psi|^2 + (\psi^2 + \alpha)^2 \nonumber \\
    f_1(\Psi) &=   \xi_1^2 |\nabla \Psi|^2 + (|\Psi|^2 + \chi )^2. \nonumber \\
\end{align}
Here $ \alpha < \chi <  0$, $g>0$. $\psi$ is real. and $\Psi$ is complex with $\Psi = \Psi_1 + i\Psi_2$. 

The theory describes a $Z_2$ real-scalar field competing with a $U(1)$ complex-scalar field. Physically, it could effectively describe some CDW-superconductor competing systems~\cite{PhysRevLett.118.106405,joe2014emergence}. In this case, $\psi$ represents the amplitude of the commensurate CDW order and its phase fluctuations can be ignored. The CDW free energy in Eq.~(\ref{model_A_sm}) with $\beta \gg |\alpha|$ belongs to this case. $\Psi$ can represent the order parameter of the superconducting order, and $f_1(\Psi)$ has the $U(1)$ symmetry $\Psi \rightarrow \Psi e^{i\theta}$.

The symmetry of the theory is $Z_2 \otimes U(1)$. Its mean-field ground state reads $\psi = \pm \sqrt{|\alpha|}$. By the optical pump, a domain wall configuration $ \psi_0(\bm{r}) = \sqrt{|\alpha|}\tanh\frac{z}{\xi }$  could be generated by the mechanism mentioned in Sec.~\ref{dw_form_intro}. Similar to the analysis applied in the first-stage dynamics, we expand the fluctuation of $\Psi$ to the linear term and obtain
\begin{align}
    \frac{1}{\gamma}\partial_t \Psi = -(\hat{M} -  \xi_1 ^2 \nabla_{\bm{R}}^2)\Psi + \eta.
    \label{linear_couple}
\end{align}
with
\begin{align}
    \hat{M} &= - \xi_1^2 \partial_z^2 + 2\chi + 2g\psi_0^2 \nonumber \\
    &= - \xi_1^2 \partial_z^2 - 2g|\alpha|\frac{1}{\cosh^2 (z/\xi)} + 2g|\alpha| + 2\chi \nonumber \\
    &=  \xi_1^2[-\partial_z^2 -\frac{1}{\xi^2}\frac{2g \xi_0^2/\xi_1^2 }{\cosh^2(x/\xi)}] + 2g|\alpha| + 2\chi.
\end{align}
Based on the result of the eigenvalue problem in Eq.~(\ref{sech_eig}), the lowest eigenvalue of $\hat{M}$, which is denoted by $m_0$, reads
\begin{align}
    m_0 = - \frac{ \xi_1^2}{ \xi^2 }(\frac{\sqrt{1+8g\xi_0^2/\xi_1^2}-1}{2})^2 + 2g|\alpha| + 2\chi.
\end{align}
If $m_0 < 0$, from Eq.~(\ref{linear_couple}), the domain wall is unstable and the first-stage dynamics, the growth of $\Psi$ around the interface, could happen. 

\subsection{The second-stage dynamics}
After the first-stage dynamics, topological defects are generated and a second-stage dynamics will happen. Basically, the following two situations may appear:  \\

1. Only one component of $\Psi$ grows up around the interface. In that case, the systems go back to the case of the two-component parameter theory we discuss above. The defects are string defects. The coarsening should be diffusive. \\

2. Both two components of $\Psi$ grows up around the interface. In that case, the defects are point defects called monopoles. \\

For the last case, we present a brief discussion of the monopole spatial profile and the corresponding coarsening dynamics. We separate the coordinate by $\bm{r} = (z,\bm{R}) = (z,R\cos \phi, R\sin \phi)$ and write $\Psi = \rho e^{i\theta}$. Suppose a monopole solution with a rotational symmetry around the line of $\bm{R}=0$, i.e. $\rho(\bm{R}=0)=0$. The monopole solution satisfies
\begin{align}
    \frac{\delta F}{\delta \psi} &= -2\xi_0^2 \nabla^2 \psi + 4\alpha \psi + 4\psi^3 + 4g\rho^2 \psi = 0 \nonumber \\
    \frac{\delta F}{\delta \rho} &= -2  \xi_1^2 \nabla^2 \rho + 2  \xi_1^2 \rho |\nabla \theta|^2 + 4\chi\rho + 4\rho^3 + 4g\psi^2 \rho = 0 \nonumber \\
    \frac{\delta F}{\delta \theta}&= -2  \xi_1^2 \rho^2 \nabla^2 \theta = 0,
    \label{monopole_saddle}
\end{align}
with boundary conditions 
\begin{align}
    \lim_{z \rightarrow \pm \infty} \psi &= \pm \sqrt{|\alpha|} \nonumber \\
    \lim_{z \rightarrow \pm \infty} \rho &= 0 \nonumber \\
    \oint  d\bm{R} \ \ \theta(z,\bm{R}) &= \pm 2\pi. 
    \label{monopole_cond}
\end{align}
In the asymptotic regions with $|\bm{R}-\bm{R}_0| \rightarrow +\infty$, the following solution
\begin{align}
    \rho(\bm{r}) &= \rho_0 (z) , \nonumber \\
    \theta &= \pm\phi, \nonumber \\
    \psi(\bm{r}) &= \psi_0(z),
    \label{monopole_sol}
\end{align}
satisfies Eq.~(\ref{monopole_saddle}, \ref{monopole_cond}), as long as the the solution of $\psi_0(z) $ and $ \rho_0(z)$ with respect to the following equation exists.
\begin{align}
     -2\xi_0^2 \partial_z^2 \psi_0 + 4\alpha \psi_0 + 4\psi_0^3 + 4g\rho_0^2 \psi_0 &= 0, \nonumber \\
     -2 \xi_1^2 \partial_z^2 \rho_0 + 4\chi\rho_0 + 4\rho_0^3 + 4g\psi_0^2 \rho_0 &= 0,
\end{align}
with the boundary condition $\lim_{z \rightarrow \pm \infty} \psi_0(z) = \pm \sqrt{|\alpha|} $, $
\lim_{z \rightarrow \pm \infty} \rho_0(z) = 0$. \\
Substituting $\rho_0(z)$, $\psi_0(z)$ into Eq.~(\ref{free_energy_couple}), we can separate the free energy by
\begin{align}
    F = E_{\rm d} + E_0 \nonumber
\end{align}
with
\begin{align}
    E_{\rm d} &=  E_{\text{c}}\int d^2 \bm{R} \ \ \xi_1^2 \Gamma |\nabla_{\bm{R}} \theta|^2 , \ \ \ \ \ \Gamma =  \int dz \ \ \rho_0^2(z) \nonumber \\
    E_0 &= E_{\text{c}} \int d^2\bm{R} \int dz \ \ \xi_0^2 ( \psi_0^{\prime}(z) )^2 + (\psi_0^2 + \alpha)^2 + \xi_1^2 ( \rho_0^{\prime}(z) )^2 + (\rho_0^2 + \chi)^2 + 2g\rho_0^2 \psi_0^2.
\end{align}
Here $E_0$ is the energy of the solution $\rho_0(z)$ and $\psi_0(z)$, 
$E_{\rm d}$ is the energy per monopole, which has the same form as the energy per vortex of a 2D system with the $U(1)$ symmetry. Therefore, in this case, the coarsening dynamics has the same scaling behavior as the 2D vortex system, where the typical length scale $L(t)$ scales as $L(t) \sim \xi_0 \sqrt{\frac{\gamma t}{\ln (|\alpha|\gamma t)}}$~\cite{bray2002theory}. 

\section*{\textsl{\NoCaseChange{Supplementary Note 6: }}Details of the numerical simulation}
In Fig.~3 in the main text, the result of the numerical simulation of the scaling of $L(t)$ in the second stage dynamics was shown. In the simulation, the ``model A" dynamics were rescaled in the same way as in Eq.~(\ref{rescaling}, \ref{model_A2_rescale}) with $b=\sqrt{|\alpha|}$, $\bar{\alpha} = -1$. We further define $\tau = |\alpha|\gamma t$, $\tilde{\bm{r}} = \bm{r}/\xi$ and $\tilde{\nabla} = \partial/\partial \tilde{\bm{r}}$ and write the dynamical equation in the following form;
\begin{align}
     \frac{\partial \bar{\psi}}{\partial \tau}  =  \tilde{\nabla}^2 \bar{\psi} + 2\bar{\psi} -2\frac{\beta}{|\alpha|} i \bar{\psi_2} - 2|\bar{\psi}|^2 \bar{\psi} + \bar{\eta}, 
    \label{model_A_rescale}
\end{align}
Here $\bar{\psi} = \psi/\sqrt{|\alpha|}$, $\bar{\eta} = \eta|\alpha|^{-3/2}$. The rescaled noise correaltor reads $\langle \bar{\eta}_{i}(\tilde{\bm{r}},\tau) \bar{\eta}_{i^{\prime}}(\tilde{\bm{r}}^{\prime},\tau^{\prime}) \rangle = 2\zeta \delta_{ii^{\prime}}\delta(\tilde{\bm{r}}-\tilde{\bm{r}}^{\prime})\delta(\tau - \tau^{\prime})$. In this equation, there are only two dimensionless parameters, $\zeta$ and $\beta/|\alpha|$. We discretize the space and time by the steps $d\tilde{r}_{j=x,y,z} = 0.5$ and $d\tau = 0.03$ in the program and use the following Euler method to update $\bar{\psi}$ in the time evolution;
\begin{align}
    \bar{\psi}(\tilde{\bm{r}}, \tau + d\tau) - \bar{\psi}(\tilde{\bm{r}}, \tau) =  \Big( \tilde{\nabla}^2 \bar{\psi}(\tilde{\bm{r}}, \tau) + 2\bar{\psi}(\tilde{\bm{r}}, \tau) -2\frac{\beta}{|\alpha|} i \bar{\psi_2}(\tilde{\bm{r}}, \tau) - 2|\bar{\psi}(\tilde{\bm{r}}, \tau)|^2 \bar{\psi}(\tilde{\bm{r}}, \tau) \Big) d\tau + \bar{\eta}(\tilde{\bm{r}},\tau) d\tau.
\end{align}
Here $\bar{\eta}(\tilde{\bm{r}},\tau) d\tau$ are generated by $\bar{\eta}_i(\tilde{\bm{r}},\tau) d\tau = \sqrt{\frac{2\zeta d\tau}{\Pi_j dr_j}} n_i(\tilde{r},\tau)$ with $2\zeta = 0.0001$ with $n_i(\tilde{r},\tau)$ being generated from independent standard normal distribution for each discretized $\tilde{\bm{r}}$ and $\tau$.  

The initial condition of Eq.~(\ref{model_A_rescale}) in the program is simply set as  $\bar{\psi}(\tilde{\bm{r}},\tau=0) = \tanh{\tilde{z}}$. Note that the initial fluctuation $\delta\bar{\psi}(\tilde{\bm{r}})$ is neglected in the program because 
it is hard to generate the initial fluctuation, which is a random function of $\bm{r}$ satisfying a specific statistical correlation function. The simplification is appropriate since the simulation is to fit the scaling of $L(t)$ in the second-stage dynamics, which is independent of the initial condition information.

The following method evaluates the length scale $L(t)$ in the simulation. In 2D systems, we count a 
number $N$ of vortices in the $y$-$z$ plane and define $L=L_y/ \langle N \rangle$ with $\langle N \rangle$ being the average of $N$ over ten independent runs, where $L_y$ is the system size along $y$. In 3D systems, we count a number $N$ of vortices in the $y$-$z$ planes and calculate $\langle N \rangle$ by an average of $N$ over each discrete $x$ and independent runs to obtain $L = L_y/\langle N\rangle$. To count the number of vortices, we extract the function $P(y) = 1-|\bar{\psi}(x,y,z=0)|^2$ for each $x$. A peak of $P(y)$ with a height larger than 0.6 is identified as a vortex(or antivortex). $N$ equals the number of peaks.


In the main text, \revise{the average results over ten independent simulations are shown}. Here we list the results of the fitted value of $p$ for each simulation in Table.~\ref{Table_p} to show that our simulation size is large enough to reduce the statistical error.

\begin{table}[!ht]
\centering
    \begin{tabular}{|l|l|l|l|l|l|l|l|l|l|l|}
    \hline
        $D=3$. Run number: & 1 & 2 & 3 & 4 & 5 & 6 & 7 & 8 & 9 & 10 \\ \hline
        $\beta/|\alpha|=0$ & 0.45365 & 0.45123 & 0.45414 & 0.44534 & 0.45779 & 0.46566 & 0.46025 & 0.44490 & 0.44960 & 0.45672 \\ \hline
        $\beta/|\alpha|=0.1$  & 0.45910 & 0.44516 & 0.45337 & 0.46073 & 0.45576 & 0.45869 & 0.46217 & 0.45613 & 0.45594 & 0.44143 \\ \hline
    \end{tabular}
    
    \centering
    \begin{tabular}{|l|l|l|l|l|l|l|l|l|l|l|}
    \hline
        $D=2$. Run number: & 1 & 2 & 3 & 4 & 5 & 6 & 7 & 8 & 9 & 10 \\ \hline
        $\beta/|\alpha|=0.001$ & 0.27497 & 0.26555 & 0.30045 & 0.30077 & 0.27175 & 0.29875 & 0.28527 & 0.27797 & 0.28693 & 0.27962 \\ \hline
        $\beta/|\alpha|=0.1$ & 0.19070 & 0.16815 & 0.18172 & 0.17463 & 0.17473 & 0.16713 & 0.17261 & 0.16296 & 0.18281 & 0.17346 \\ \hline
    \end{tabular}

    \caption{ Linear fit results of the  values of $p = \frac{\log_{10}(L/\xi)}{\log_{10}(|\alpha|\gamma t)}$ for ten runs of the program in $D=3$ and $D=2$. The parameter set is the same as mentioned in the caption of Fig.~(3) in the main text. }
    \label{Table_p}
\end{table}

\section*{\textsl{\NoCaseChange{Supplementary Note 7: }}Microscopic derivation of the free energy model in CDW systems.}
\revise{ This section gives an electron-phonon coupling model to derive the CDW free energy in general spatial dimensions. The derivation is similar to the theory of CDW in one dimension~\cite{PhysRevLett.31.462}. The Hamiltonian reads
\begin{align}
    H-\mu N &= \sum_{\bm{p} \sigma} \epsilon_{\bm{p}} c^{\dagger}_{\bm{p}\sigma}c_{\bm{p}\sigma} + \sum_{\bm{q}} \omega_{\bm{q}} b^{\dagger}_{\bm{q}} b_{\bm{q}} + \frac{1}{l^{D/2}} \sum_{\bm{q},\sigma,\bm{q}} g_{\bm{q}}b_{\bm{q}} c^{\dagger}_{\bm{p+q}\sigma} c_{\bm{p}\sigma} + cc. 
\end{align}
Here $\epsilon_{\bm{p}}, \omega_{\bm{q}}$ are the dispersions of electron and phonon. $l$ and $D$ are the system size and spatial dimension. $g_{\bm{q}}$ is the electron-phonon coupling strength. Defining the order parameter as $\Delta_{\bm{q}} \equiv  g_{\bm{q}} b_{\bm{q}}$, the Hamiltonian reads 
\begin{align}
    H - \mu N &= \sum_{\bm{p} \sigma} \epsilon_{\bm{p}} c^{\dagger}_{\bm{p}\sigma}c_{\bm{p}\sigma} + \sum_{\bm{q}} \frac{\omega_{\bm{q}}}{g_{\bm{q}}^2} \Delta^{\dagger}_{\bm{q}} \Delta_{\bm{q}} + \frac{1}{l^{D/2}} \sum_{\bm{q},\sigma,\bm{q}} \Delta_{\bm{q}} c^{\dagger}_{\bm{p+q}\sigma} c_{\bm{p}\sigma} + cc. 
\end{align}
In this notation, $c_{\bm{q},\sigma}$ and $b_{\bm{q}}$ are dimensionless, $\epsilon_{\bm{p}}$ and $\omega_{\bm{q}}$ carry dimension $[t^{-1}]$, $g_{\bm{q}}$ carries dimension $[l^{D/2} t^{-1}]$. The action reads 
\begin{align}
   S &= \int_0^{\beta \equiv 1/T} d\tau \ \ \sum_{\bm{p} \sigma} c^{\dagger}_{\bm{p}\sigma} (\partial_\tau + \epsilon_{\bm{p}}) c_{\bm{p}\sigma}  + \sum_{\bm{q}} \frac{\omega_{\bm{q}}}{g_{\bm{q}}^2} \Delta^{\dagger}_{\bm{q}} \Delta_{\bm{q}} + \frac{1}{l^{D/2}} \sum_{\bm{q},\sigma,\bm{q}} \Delta_{\bm{q}} c^{\dagger}_{\bm{p+q}\sigma} c_{\bm{p}\sigma} + cc.  \nonumber \\
   &= \sum_{\bm{q}} \beta\frac{\omega_{\bm{q}}}{g_{\bm{q}}^2} \Delta^{\dagger}_{\bm{q}} \Delta_{\bm{q}} + \sum_{i\omega_n} \ \ \sum_{\bm{p} \sigma} c^{\dagger}_{\bm{p}\sigma}(i\omega_n) (-i\omega_n + \epsilon_{\bm{p}}) c_{\bm{p}\sigma}(i\omega_n)  + \frac{1}{l^{D/2}} \sum_{\bm{q},\sigma,\bm{q}} \Delta_{\bm{q}} c^{\dagger}_{\bm{p+q}\sigma}(i\omega_n) c_{\bm{p}\sigma}(i\omega_n) + cc. 
\end{align}
\vspace{0.3cm}
Here the $\Delta^{\dagger}_{\bm{q}} \partial_\tau \Delta_{\bm{q}}$ term is not shown since the derivation is to obtain the mean-field free energy of $\Delta_{\bm{q}}$ which does not depend on $\tau$. For a given Fermi surface, a nesting vector 
$\bm{Q}$ is preferred for the CDW. Integrating out the fermions, we obtain the effective action for $\Delta_{\bm{Q}}$.
\begin{align}
    S_Q &= \beta\frac{\omega_{\bm{Q}}}{g_{\bm{Q}}^2} |\Delta_{\bm{Q}}|^2  -\sum_{\bm{p},\sigma,i \omega_n}\ln ( 1- \frac{1}{l^D}\frac{|\Delta_{\bm{Q}}|^2}{(-i\omega_n + \epsilon_{\bm{p}})(-i
    \omega_n + \epsilon_{\bm{p+Q}})} )   \nonumber \\
    &\approx  \beta\frac{\omega_{\bm{Q}}}{g_{\bm{Q}}^2} |\Delta_{\bm{Q}}|^2 + \sum_{\bm{p},i\omega_n}  \frac{2}{l^D} \frac{|\Delta_{\bm{Q}}|^2}{(-i\omega_n + \epsilon_{\bm{p}})(-i
    \omega_n + \epsilon_{\bm{p+Q}})}  + \sum_{\bm{p},i\omega_n} \frac{1}{l^{2D}}  \frac{|\Delta_{\bm{Q}}|^4}{(-i\omega_n + \epsilon_{\bm{p}})^2(-i
    \omega_n + \epsilon_{\bm{p+Q}})^2}
\end{align}
Note that the summation over $\epsilon_{\bm{p}}$ is restricted by the characteristic energy of the phonon scattering $W$. i.e. The summation only contains momenta $\bm{p}$ with $|\epsilon_{\bm{p}}| < W$ and $|\epsilon_{\bm{p}+\bm{Q}}| < W$. Other momenta are neglected since they do not contribute to the nesting. The fermi-surface-nesting induced CDW has a quasi-one-dimensional nature and we approximate $\epsilon_{\bm{p}} \approx -\epsilon_{\bm{p} + \bm{Q}}$ in the effective action. Meanwhile, we define the density of states for electrons contributing to the nesting as $\nu$ satisfying $l^{-D} \sum_{\bm{p}} \equiv  \int_{-W}^{W} \nu d\epsilon $. As an estimation, $\nu \sim p_{\rm F}^D/E_{\rm F}$, where $p_{\rm F}$ and $E_{\rm F}$ denote the fermi momentum and fermi energy. Taking the summation over $i\omega_n$, one obtains
\begin{align}
    S_Q/\beta &= (\frac{\omega_{\bm{Q}}}{g_{\bm{Q}}^2}  - 2 \nu \int_{-W}^{W} d  \frac{1-2n_{\rm F}(\epsilon)}{2\epsilon}) |\Delta_{\bm{Q}}|^2  + \frac{1}{\beta l^D} \nu \sum_{i\omega_n}\int_{W}^{W} d\epsilon  
      \frac{1}{(\omega_n^2 + \epsilon^2)^2} |\Delta_{\bm{Q}}|^4 \nonumber \\
      & \approx  (\frac{\omega_{\bm{Q}}}{g_{\bm{Q}}^2}  - 2 \nu \int_{-W}^{W} d\epsilon   \frac{1-2n_{\rm F}(\epsilon)}{2\epsilon}) |\Delta_{\bm{Q}}|^2  + \frac{\pi}{2 \beta l^D} \nu \sum_{i\omega_n} 
      \frac{1}{|\omega_n|^3 } |\Delta_{\bm{Q}}|^4 \nonumber \\
      &= (\frac{\omega_{\bm{Q}}}{g_{\bm{Q}}^2}  - 2 \nu \int_{-W}^{W} d\epsilon   \frac{1-2n_{\rm F}(\epsilon)}{2\epsilon}) |\Delta_{\bm{Q}}|^2  + \frac{1}{ l^D} \frac{\beta^2}{\pi^2} \frac{7}{8} \nu \zeta(3) |\Delta_{\bm{Q}}|^4 \nonumber \\
      &\equiv a(\bm{Q},T) |\Delta_{\bm{Q}}|^2  + b(\bm{Q},T)|\Delta_{\bm{Q}}|^4.
      \label{S_Q}
\end{align}
Here $n_{\rm F}$ is the Fermi distribution, and $\zeta(s) = \sum_{n=1}^{+\infty}1/n^{s}$ is the Riemann zeta function. The equation $a(\bm{Q},T) = 0$ sets the critical temperature $T_{\rm c}$ with~\cite{nagaosa2013quantum, PhysRevLett.31.462}
\begin{align}
    T_{\rm c} = 1.14 W e^{-\frac{\omega_Q}{2\nu g_Q^2}}.
\end{align}
In the vicinity of $T_{\rm c}$, $a(\bm{Q},T) \approx 2\nu (T-T_{\rm c})/T_{\rm c}$ was obtained in the literature~\cite{nagaosa2013quantum, PhysRevLett.31.462}. For a general momentum around $\bm{Q}$, the coefficient of $|\Delta_{\bm{q}+\bm{Q}}|^2$ reads 
\begin{align}
    a(\bm{q}+\bm{Q},T) &= \frac{\omega_{\bm{q+Q}}}{g_{\bm{q+Q}}^2}  + \frac{1}{\beta}\sum_{\bm{p},i\omega_n}  \frac{2}{l^D} \frac{1}{(-i\omega_n + \epsilon_{\bm{p}})(-i
    \omega_n + \epsilon_{\bm{p+Q+q}})} \nonumber \\
    &= \frac{\omega_{\bm{q+Q}}}{g_{\bm{q+Q}}^2}  + \frac{1}{\beta} \sum_{\bm{p}}  \frac{2}{l^D} \ \    \frac{n_{\rm F}(\epsilon_{\bm{p}}) - n_{\rm F}(\epsilon_{\bm{p}+\bm{q}+\bm{Q}})}{\epsilon_{\bm{p}} - \epsilon_{\bm{p}+\bm{q}+\bm{Q}}}  \nonumber \\
    &\approx \frac{\omega_{\bm{q+Q}}}{g_{\bm{q+Q}}^2}  + \frac{1}{\beta} \sum_{\bm{p}}  \frac{2}{l^D} \ \  \frac{n_{\rm F}(\epsilon_{\bm{p}}) - n_{\rm F}(-\epsilon_{\bm{p}+\bm{q}})}{\epsilon_{\bm{p}} + \epsilon_{\bm{p}+\bm{q}}}  \nonumber \\
    & \approx a(\bm{Q},T) +  \nu \sum_{i=x,y,z} \xi_{0,i}^2 q_i^2 .
    \label{a_q}
\end{align}
Without loss of generality, we set the nesting vector $\bm{Q} = Q\hat{y}$, which breaks the rotational symmetry. Therefore, the system is generally anisotropic with $\xi_{0,y} \ne \xi_{0,x/z}$. Neglecting the momentum dependence of the first term $\omega_{\bm{q+Q}}/g_{\bm{q+Q}}^2$, $\xi_{0,i}$ is derived from the second term, which has the same form as the Lindhard function in conventional Bardeen–Cooper–Schrieffer (BCS) superconductors. Following a calculation in BCS superconductors~\cite{nagaosa2013quantum}, we estimate $\xi_{0,y}^2 \approx 7\zeta(3)(v_{\rm F} \beta)^2/24\pi^2 $ where $v_{\rm F}$ is the Fermi velocity along the nesting direction. $\xi_{0,x/z}$ depend on the detailed band structure and we cannot obtain a simple expression for them. To derive the free energy model in the main text, we define a dimensionless quantity $\psi(\bm{r})$ and write the CDW order parameter as
\begin{align}
    \Delta(\bm{r}^{\prime}) = \sqrt{\pi}\psi(\bm{r}^{\prime}) e^{i\bm{Q}\cdot \bm{r}^{\prime}} T_{\rm c}.
\end{align}
Combine Eq.~(\ref{S_Q}, \ref{a_q}) and notice $7\zeta(3) \approx 8$, we obtain the  free energy for $\psi(\bm{r})$ around $T_{\rm c}$
\begin{align}
    F \equiv S/\beta &\approx \nu T^2_{\rm c} \Bigg[ \int d^D\bm{r^{\prime}} \ \   \sum_i \pi |\xi_{0,i} \partial_i \psi|^2 + 2 \pi \frac{T-T_{\rm c}}{T_{\rm c}} |\psi|^2 + |\psi|^4  \Bigg]  \nonumber \\
    &= \nu T^2_{\rm c} \Bigg[ \int d^D\bm{r}  \ \  \xi_0^2 |\nabla \psi|^2 + 2 \pi \frac{T-T_{\rm c}}{T_{\rm c}} |\psi|^2 + |\psi|^4  \Bigg].
\end{align}
Here $\bm{r}$ is the rescaled spatial coordinate of the real space $\bm{r^{\prime}}$ with $r_i = r^{\prime}_i\xi_0/\xi_{0,i}$. $\xi_0 = (\pi \Pi_{i=1}^{D} \xi_{0,i})^{1/D}$ is the averaged bare coherence length over the spatial dimensions. Comparing with the free energy in Eq.~(2) of the main text, we obtain $E_{\rm c} = \nu T^2_{\rm c}$, $\alpha = \pi (T-T_{\rm c})/T_{\rm c}$. If the system does not show a strong anisotropy of $\xi_{0,i}$ in different dimensions, we could approximate $\xi_0 \approx  \pi \xi_{0,y} \approx v_{\rm F} \beta = v_{\rm F}/T_{\rm c}$ around $T=T_{\rm c}$. Therefore, around $T=T_{\rm c}$, the quantities defined in Eq.~(\ref{G-def}, \ref{zeta-def}) read $G \sim T_{\rm c}^{D-1}/E_{\rm F}^{D-1} $ and $\zeta \sim G|(T-T_{\rm c})/T_{\rm c}|^{D/2-2}$. The microscopic derivation and parameter estimation are similar in spin-density waves, excitonic insulators, and BCS superconductors~\cite{RevModPhys.66.1, nagaosa2013quantum,  PhysRev.158.462, kozlov1965metal}. \\
$\bm{Q}$ is along the in-plane direction (normal to the sample surface) for typical materials in experimental measurements~\cite{duan2021optical}.  If $\bm{Q}$ is commensurate to the lattice structure, the ions tend to lock the phase of $\psi$ to reduce the Coulomb energy, producing a $Z_2$ term $2\beta\psi_2^2$ with $\beta > 0$.  }

\section*{\textsl{\NoCaseChange{Supplementary Note 8: }}How does a domain wall in the $Z_2$ symmetric case with $\beta > |\alpha|/2$ decay in real experiments?}
\revise{ In the ideal model considered in this manuscript, the domain wall is flat, and the sample's surface is infinitely away from the domain wall.  In this case, as mentioned in the main text, the domain wall is stable and will not decay for $\beta > |\alpha|/2$ in the $Z_2$ case. However, in real experiments, the pump pulse is a light spot acting on the surface of the bulk material. It only heats the sample with a finite range, roughly equal to the spot size ($\sim 10^2 \mu m$). Therefore, the generated domain wall is not exactly flat but has a curvature. Such a domain wall will shrink to decrease the free energy, as shown in Fig.~\ref{dw_decay}.   
\begin{figure}
    \centering
    \includegraphics[width= 4in]{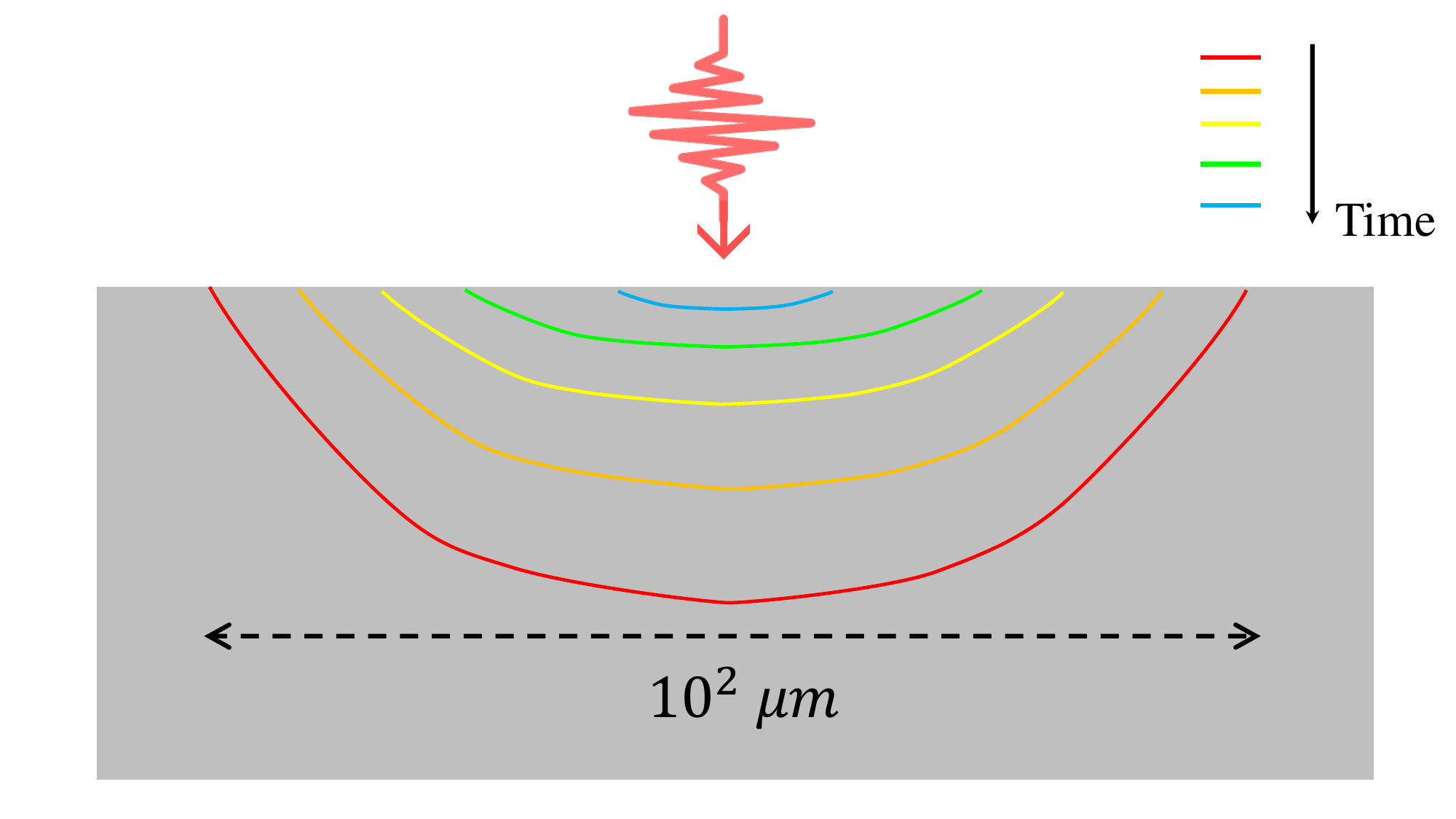} 
    \caption{\revise{The pump-induced domain wall and its decay in the $Z_2$ case with $\beta > |\alpha|/2$. The gray region denotes the sample. The red curve denotes the pump-induced domain wall with a size $\sim 10^2 \mu m$. With time going by, its shape evolves from the red curve to the orange, yellow, green, and blue curves and disappears.} }  
    \label{dw_decay}
  \end{figure}
}

\revise{
Note that this decay also happens in experiments for other cases discussed in the paper ($U(1)$ or with a weak $Z_2$ term). However, the length scale of this decay dynamics is much larger than the length scale of the dynamics proposed in our paper ($\sim 10-100$ nm). Therefore, the ideal model we use is effective for our proposal.   }

\bibliography{supp}